\documentclass[fleqn,usenatbib]{mnras}

\usepackage{newtxtext,newtxmath}

\usepackage[T1]{fontenc}
\usepackage{ae,aecompl}
\usepackage{subfigure}

\usepackage{graphicx}	
\usepackage{amsmath}	
\usepackage{amssymb}	
\usepackage[dvipsnames]{xcolor}

\usepackage{soul}
\usepackage{tabularx,ragged2e}
\usepackage{booktabs,array,times}
\usepackage[export]{adjustbox}

\def\CIVdblt{{\rm C~}\kern 0.1em{\sc iv}~$\lambda\lambda 1548, 1550$}
\def\NVdblt{{\rm N~}\kern 0.1em{\sc v}~$\lambda\lambda 1238, 1242$}
\def\OVIdblt{{\rm O~}\kern 0.1em{\sc vi}~$ 1031, 1037$}
\def\SVIdblt{{\rm S~}\kern 0.1em{\sc vi}~$ 933, 944$}
\def\SiIVdblt{{\rm Si~}\kern 0.1em{\sc iv}~$\lambda\lambda1394, 1403$}
\def\MgIIdblt{{\rm Mg~}\kern 0.1em{\sc ii}~$\lambda\lambda2796, 2803$}
\def\NeVIIIdblt{{\rm Ne~}\kern 0.1em{\sc viii}~$\lambda\lambda770, 780$}
\def\NeV{\hbox{{\rm Ne~}\kern 0.1em{\sc v}}}
\def\NeVI{\hbox{{\rm Ne~}\kern 0.1em{\sc vi}}}
\def\NeVIII{\hbox{{\rm Ne~}\kern 0.1em{\sc viii}}}
\def\OII{\hbox{{\rm O~}\kern 0.1em{\sc ii}}}
\def\OIII{\hbox{{\rm O~}\kern 0.1em{\sc iii}}}
\def\OIV{\hbox{{\rm O~}\kern 0.1em{\sc iv}}}
\def\OV{\hbox{{\rm O~}\kern 0.1em{\sc v}}}
\def\OVI{\hbox{{\rm O~}\kern 0.1em{\sc vi}}}
\def\OVII{\hbox{{\rm O~}\kern 0.1em{\sc vii}}}
\def\OVIII{\hbox{{\rm O~}\kern 0.1em{\sc viii}}}
\def\NI{\hbox{{\rm N~}\kern 0.1em{\sc i}}}
\def\NII{\hbox{{\rm N~}\kern 0.1em{\sc ii}}}
\def\NIII{\hbox{{\rm N~}\kern 0.1em{\sc iii}}}
\def\NIV{\hbox{{\rm N~}\kern 0.1em{\sc iv}}}
\def\NV{\hbox{{\rm N~}\kern 0.1em{\sc v}}}
\def\NVII{\hbox{{\rm N~}\kern 0.1em{\sc vii}}}
\def\CII{\hbox{{\rm C~}\kern 0.1em{\sc ii}}}
\def\CIII{\hbox{{\rm C~}\kern 0.1em{sc iii}}}
\def\SiII{\hbox{{\rm Si~}\kern 0.1em{\sc ii}}}
\def\SiIII{\hbox{{\rm Si~}\kern 0.1em{\sc iii}}}
\def\SiIV{\hbox{{\rm Si~}\kern 0.1em{\sc iv}}}
\def\SIV{\hbox{{\rm S~}\kern 0.1em{\sc iv}}}
\def\SV{\hbox{{\rm S~}\kern 0.1em{\sc v}}}
\def\SVI{\hbox{{\rm S~}\kern 0.1em{\sc vi}}}
\def\SiI{\hbox{{\rm Si~}\kern 0.1em{\sc i}}}
\def\PII{\hbox{{\rm P~}\kern 0.1em{\sc ii}}}
\def\AlII{\hbox{{\rm Al~}\kern 0.1em{\sc ii}}}
\def\AlIII{\hbox{{\rm Al~}\kern 0.1em{\sc iii}}}
\def\CaI{\hbox{{\rm Ca~}\kern 0.1em{\sc i}}}
\def\CaII{\hbox{{\rm Ca~}\kern 0.1em{\sc ii}}}
\def\CrII{\hbox{{\rm Cr~}\kern 0.1em{\sc ii}}}
\def\CII{\hbox{{\rm C~}\kern 0.1em{\sc ii}}}
\def\CIII{\hbox{{\rm C~}\kern 0.1em{\sc iii}}}
\def\CIV{\hbox{{\rm C~}\kern 0.1em{\sc iv}}}
\def\CV{\hbox{{\rm C}\kern 0.1em{\sc v}}}
\def\MgX{\hbox{{\rm Mg}\kern 0.1em{\sc x}}}
\def\MgII{\hbox{{\rm Mg}\kern 0.1em{\sc ii}}}
\def\FeII{\hbox{{\rm C~}\kern 0.1em{\sc ii}}}
\def\FeIII{\hbox{{\rm C~}\kern 0.1em{\sc iii}}}
\def\H{\hbox{{\rm H~}}}
\def\HI{\hbox{{\rm H~}\kern 0.1em{\sc i}}}
\def\HeI{\hbox{{\rm He~}\kern 0.1em{\sc i}}}
\def\HII{\hbox{{\rm H~}\kern 0.1em{\sc ii}}}
\def\Lya{\hbox{{\rm Ly}\kern 0.1em$\alpha$}}
\def\Lyb{\hbox{{\rm Ly}\kern 0.1em$\beta$}}
\def\Lyg{\hbox{{\rm Ly}\kern 0.1em$\gamma$}}
\def\Lyth{\hbox{{\rm Ly}\kern 0.1em$\theta$}}
\def\Lyfive{\hbox{{\rm Ly}\kern 0.1em$5$}}
\def\Lysix{\hbox{{\rm Ly}\kern 0.1em$6$}}
\def\Lyseven{\hbox{{\rm Ly}\kern 0.1em$7$}}
\def\Lyeight{\hbox{{\rm Ly}\kern 0.1em$8$}}
\def\Lynine{\hbox{{\rm Ly}\kern 0.1em$9$}}
\def\Lyten{\hbox{{\rm Ly}\kern 0.1em$10$}}
\def\MnII{\hbox{{\rm Mn~}\kern 0.1em{\sc ii}}}
\def\kms{\hbox{km~s$^{-1}$}}
\def\cmsq{\hbox{cm$^{-2}$}}
\def\cc{\hbox{cm$^{-3}$}}

\newcommand{\angstrom}{\mbox{\normalfont\AA}}

\title[{\OIII} to {\OVI} absorbing systems towards PG~$1522+101$]{Physical Conditions of Five {\OVI} Absorption Systems Towards PG~$1522+101$}

\author[Sankar et al. 2020]{Sriram Sankar$^{1}$\thanks{E-mail: sriram10sankar@gmail.com}, Anand Narayanan$^{2}$\thanks{E-mail: anand@iist.com}, Blair D Savage$^{3}$, Vikram Khaire$^{4}$, \newauthor Benjamin E Rosenwasser$^{3}$, Jane Charlton$^{5}$, and Bart P Wakker$^{3}$  \\
$^{1}$Department of Mechanical Engineering, Federal Institute of Science And Technology, Ernakulam 683577, Kerala, INDIA\\
$^{2}$Department of Earth and Space Sciences, Indian Institute of Space Science \& Technology, Thiruvananthapuram 695547, Kerala, INDIA\\
$^{3}$Department of Astronomy, University of Wisconsin - Madison, Madison, WI\\
$^{4}$Department of Physics, University of California, Santa Barbara 93106, California, USA\\
$^{5}$Department of Astronomy \& Astrophysics, The Pennsylvania State University, 525 Davey Laboratory University Park, PA, 16802, USA\\
}

\date{Accepted 2020 August 21. Received 2020 August 09; in original form 2020 May 18}

\pubyear{2020}

\begin{document}
\label{firstpage}
\pagerange{\pageref{firstpage}--\pageref{lastpage}}
\maketitle

\begin{abstract}
We present the analysis of five {\OVI} absorbers identified across a redshift path of $z \sim (0.6 - 1.3)$ towards the background quasar PG~$1522+101$ with information on five consecutive ionization stages of oxygen from {\OII} to {\OVI}. The combined $HST$ and $Keck$ spectra cover UV, redshifted EUV, and optical transitions from a multitude of ions spanning ionization energies in the range of $\sim (13 - 300)$~eV. Low ionization ({\CII}, {\OII}, {\SiII}, {\MgII}) and very high ionization species ({\NeVIII}, {\MgX}) are non-detections in all the absorbers. Three of the absorbers have coverage of {\HeI}, in one of which it is a $> 3 \sigma$ detection. The kinematic structures of these absorbers are extracted from {\CIV} detected in $HIRES$ spectra. The farthest absorber in our sample also contains the detections of {\NeV} and {\NeVI}. Assuming co-spatial absorbing components, the ionization models show the medium to be multiphased with small-scale density-temperature inhomogeneities that are sometimes kinematically unresolved. In two of the absorbers, there is an explicit indication of the presence of a warm gas phase ($T \gtrsim 10^5$~K) traced by {\OVI}. In the remaining absorbers, the column densities of the ions are consistent with a non-uniform photoionized medium. The sub-solar [C/O] relative abundances inferred for the absorbers point at enrichment from massive Type II supernovae. Despite metal enrichment, the inferred wide range for [O/H] $\sim$~[$-2.1, +0.2$] amongst the absorbers along with their anti-correlation with the observed {\HI} suggest poor small-scale mixing of metals with hydrogen in the regions surrounding galaxies and the IGM. 
\end{abstract}




\section{Introduction}

Quasar absorption line observations remain one of the most sensitive tools for examining the diffuse gaseous component of the universe. At low redshifts ($z \lesssim 1$) such studies are particularly challenging as the gas beyond galaxies is inhomogeneously distributed over a broad temperature - density range. The circumgalactic medium (CGM) is one of the major reservoirs of such gas with $L^*$ galaxies possessing halo baryonic masses of $\gtrsim 10^{10}$~M$_{\odot}$ and collectively accounting for $\sim 25$\% of the cosmic baryon budget at $z \sim 0$ (\citealt{peeples_budget_2014}, \citealt{shull_tracing_2014}, \citealt{keeney_characterizing_2017}, \citealt{hafen_origins_2019}). Absorption systems featuring ions of low and high ionization energies show the CGM to be a richly multiphased region whose ionization, metallicity, and absorption velocity spreads induced by temperature and turbulence, span a wide range of values (\citealt{richter_neutral_2011}, \citealt{werk_cos-halos_2013}, \citealt{fox_cosuves_2014}, \citealt{liang_mining_2014}, \citealt{nuza_distribution_2014}, \citealt{richter_hstcos_2016}, \citealt{tumlinson_circumgalactic_2017}). 

The multiphase properties are many a times brought out by the diversity of metal lines featured in spectroscopic observations. For example, the presence of a diffuse warm ($T \gtrsim 10^5$~K) component to the CGM is known through absorption from high ions such as {\OVI} and {\NeVIII}, whereas concurrent detections of lower ionization species such as {\CII} and {\SiII} reveals much cooler $T \lesssim 10^4$~K CGM gas phases in the same environment (\citealt{savage_multiphase_2011}, \citealt{narayanan_cosmic_2012}, \citealt{pachat_detection_2017}, \citealt{narayanan_detection_2018}, \citealt{pradeep_solar_2020}). The scale height of these low and high ions also serve to define a boundary to the chemically enriched CGM within the dark matter envelope of galaxies (\citealt{liang_mining_2014}, \citealt{liang_column_2016}, (See also Project AMIGA - \citealt{lehner_project_2020}). The metallicities inferred from the low and high ions and their kinematic properties have also served as a tools to probe gas flows in and out of galaxies (\citealt{keres_how_2005}, \citealt{lilly_gas_2013}, \citealt{kacprzak_relationship_2019}). The properties of the CGM are also explored in great detail by hydrodynamic simulations. The recent emphasis in this direction has been in reproducing the wide scatter seen in cloud-to-cloud metallicities within the CGM, and also the differences in physical sizes of gas of various ionizations (\citealt{rauch_small-scale_2001}, \citealt{rauch_small-scale_2001-1}, \citealt{lopez_metal_2005}, \citealt{misawa_spectroscopy_2013}, \citealt{muzahid_probing_2014}, \citealt{churchill_direct_2015}, \citealt{mccourt_characteristic_2018}). Given the complex kinematics, phase structure, and patchy metal distributions with the CGM, these simulations require resolutions of gas structures at sub-kiloparsec scales and the inclusion of sub-grid scale physical processes such as radiative cooling, stellar and AGN feedback, and accretion of gas in cold streams, to match the observed statistical properties of CGM absorption (\citealt{schaye_eagle_2015}, \citealt{oppenheimer_bimodality_2016}, \citealt{oppenheimer_flickering_2018}, \citealt{peeples_figuring_2019}, \citealt{suresh_zooming_2019}, \citealt{hummels_impact_2019}).

The gas beyond the extended halos of galaxies also possess a complex multiphase structure. A significant mass density of baryons in the intergalactic medium (IGM) is in the cool ($T \lesssim 10^4$~K) and diffuse phase, with a roughly equal amount at warm-hot ($10^5 \lesssim T \lesssim 10^7$~K) temperatures (e.g., \citealt{martizzi_baryons_2019}). The cool clouds heated through photoionization and with neutral fractions of $f_{\HI} = N(\HI) / N(\H) \sim  10^{-4}$ is probed well by the discrete {\Lya} forest absorption features seen in quasar spectra (\citealt{shull_baryon_2012}). Compared to this cool phase, resolving the phase structure of the warm-hot gas in the IGM has been challenging. Here again the observational strategy has been to focus on absorbers that feature {\OVI}, {\NeVIII}, and BLAs (thermally broadened {\Lya} with $b(\HI) > 40$~{\kms}). A general challenge in this case, especially with {\OVI} absorbers, is in discriminating photoionized gas from hotter collisionally ionized gas when the phases are kinetmaically overlaid. In such cases, detailed spectral line analysis and ionization modeling of the ions of different energy states help to segregate the neutral and low ionization phases from the high ionization gas traced by {\OVI} (\citealt{narayanan_cosmic_2010}, \citealt{narayanan_cosmic_2011}, \citealt{pachat_pair_2016}, \citealt{narayanan_detection_2018}). 

The intracluster medium (ICM), which dominates the baryonic mass in galaxy clusters, is also a strongly multiphased environment. Emission and absorption in X-rays are typically the probes for studying the $T \gtrsim 10^7$~K fully ionized ICM plasma. Quasar absorption lines, on the other hand, aid in detecting phases of the ICM that are cooler and denser, but with mass fractions comparable to that of the hot gas, especially in regions beyond the cluster virial radius (\citealt{emerick_warm_2015}, \citealt{butsky_ultraviolet_2019}). Observations of such $T \sim 10^4 - 10^5$~K ICM gas is a means to study important physical processes such as ram-pressure and tidal stripping of cluster galaxies, stellar and AGN feedback that push interstellar gas out into the ICM, as well as cold gas inflows through cosmic web filaments that penetrate into the cluster medium (\citealt{conselice_nature_2001}, \citealt{ehlert_ripping_2013}, \citealt{zinger_role_2016}, \citealt{pradeep_detection_2019}, \citealt{manuwal_c_2019}). 

While {\HI} and ionized metal transitions in the near and far-UV have been efficient in revealing diffuse gas with T~$\sim 10^{4} - 10^{6}$~K, the detection of hotter baryons (T~$\sim 10^{6} - 10^{7}$~K) have relied on X-ray emission studies of gas around individual massive galaxies (M$^{*} \sim 10^{11}$ M$_{\sun}$), and also stacking X-ray measurements to detect diffuse emission further along the galactocentric radius (e.g., \citealt{anderson_extended_2013}, \citealt{singh_x-ray_2018}). Such studies show that the hot phase of the CGM accounts for only $\sim 8 - 10 \%$ of the expected cosmic baryon fraction (\citealt{li_baryon_2018}). The baryon deficit is an indication that the CGM is multiphased with some significant baryonic mass entrenched in cool/warm gas phases corresponding to accretion and feedback from star formation and AGN activity, although this in itself is not adequate to explain the shortfall in baryon fraction (\citealt{anderson_hot_2010}). Dispersion measurements of millisecond duration Fast Radio Bursts (FRBs) is an emerging technique that can provide an ionization model independent account of the baryonic mass retained in the CGM and IGM. \citet{macquart_census_2020} recently reported an independent measurement of the cosmic baryon density, $\Omega_{b} = 0.051^{+0.021}_{-0.025} h_{70}$ (95\% confidence) using the dispersion of a small sample of localized FRBs and thus confirming the existence of highly ionized gas and solving the missing baryons problem. Although a larger sample of localized FRBs are required to make more robust measurements and studies await the data from current and forthcoming facilities like Australian Square Kilometre Array Pathfinder (ASKAP), the Deep Synoptic Array (DSA) etc. (e.g., \citealt{mcquinn_locating_2014}, \citealt{shull_dispersion_2018}, \citealt{prochaska_probing_2019}, \citealt{ravi_measuring_2019}, \citealt{ravi_fast_2019}, \citealt{macquart_census_2020}).

Irrespective of whether a given line of sight is probing halo gas, or the intra-group/cluster medium, or the clumpy mass distributions that characterize the filamentary IGM, an insight into the ionization states, densities and chemical compositions in the absorbing clouds is possible only when information on many ions diagnostic of the different gas phases is available. In this paper, we present the analysis of five distinct intervening metal line systems in the redshift interval $z = 0.6 - 1.3$ towards the background quasar PG~$1522+101$. Combining archival $HST$/COS, $HST$/STIS and Keck/HIRES observations, we have information on near-UV (NUV), far-UV (FUV) and extreme-UV (EUV) metal lines from nearly twenty different ionic species that represent a wide range of ionization energies from $13$~eV - $350$~eV, including the successive ionizations of oxygen from {\OII} to {\OVI}, and also {\CII} to {\CIV}, {\NeV} to {\NeVIII} and {\MgX}. Three of the absorbers also cover the EUV lines of {\HeI} at $584.334$~{\AA}, $537.029$~{\AA}, in one of which it is a $> 3\sigma$ detection. 

Information on many ions spanning a continuous series of ionization energies is valuable for probing small scale density-temperature structures within the absorbing material that are otherwise difficult to detect. Furthermore, the multiple ionization stages of oxygen in the five systems presents the possibility of constraining [O/H] in {\OVI} bearing gas, a parameter that is crucial in the estimations of the absorber population's contribution to the closure density [i.e., $\Omega_b(\OVI)$, \citealt{danforth_low-z_2005}, \citealt{tripp_high-resolution_2008}, \citealt{savage_properties_2014}). From the perspective of absorber-galaxy associations, the line of sight probes diffuse gas spread over a redshift epoch that immediately follows the peak in global star formation rate. The increased rate of stellar winds and supernova feedback that ensue rapid star formation is expected to significantly enhance the covering fraction of metals around galaxies and alter the distributions of cold, warm and hot gas phases in their haloes at $z \lesssim 1$ (e.g., \citealt{simcoe_observations_2006}, \citealt{wakker_relationship_2009}, \citealt{tumlinson_large_2011}, \citealt{hummels_constraints_2013}, \citealt{barai_galactic_2013}, \citealt{peeples_budget_2014}). The redshift range also encompasses the time period where spectroscopic imprints on the morphological evolution of galaxies towards the Hubble sequence can be seen from the kinematics of absorption lines directly probing gas within galaxy potential wells (\citealt{mshar_kinematic_2007}, \citealt{rodriguez_hidalgo_evolution_2012}). 

The paper is organized as follows. In Section \ref{sec:data}, we provide information on the archival COS, STIS, and HIRES spectra and our data analysis methods. We begin the subsequent section by describing our approach to ionization modeling, the assumptions made and the choice of extragalactic ionizing background radiation. This is followed by subsections where we address individual absorption systems in detail along with results from their modeling. In Section \ref{sec:DSR}, we conclude with a discussion and summary of the results. The line measurements for all the ions and the system plots showing all available important transitions are included as appendix material. 

\section{Spectroscopic Data in the UV \& Optical}
\label{sec:data}

The spectra for the QSO PG~$1522+101$ include archival data from $HST$/COS (Prog ID. 11741, PI. Todd Tripp), $HST$/STIS (Prog ID. 13846, PI. Todd Tripp) and Keck/HIRES (Prog ID. U066Hb, PI. Xavier Prochaska). Table \ref{tab:instrument} lists the parameters of the instruments used in our study. The COS FUV gratings G130M, and G160M and the NUV grating G185M offer a combined wavelength coverage from $1100$~{\AA} to $2100$~{\AA} with an average signal-to-noise ratio of $S/N \sim 10$~per $20$~{\kms} resolution element, after Nyquist sampling. The data were initially reduced with the CalCOS calibration pipeline. The CalCOS wavelength uncertainties of $\sim 15$~{\kms} vary with wavelength and become as large as $40$~{\kms} at the edges of both detector segments. To improve the CalCOS wavelength calibration, a further set of customized velocity alignment and co-addition steps were applied, following the procedures described in detail in the appendix of \citet{wakker_nearby_2015}. We re-calibrate the CalCOS wavelengths in individual detector segment exposures by cross-correlating relatively strong and well observed galactic ISM lines in individual extractions of the same QSO to determine velocity offsets as a function of wavelength between the different segment observations. The different segment observations are then combined. The absolute wavelength scale is then established as a function of wavelength by comparing the velocities of ISM absorption lines with 21 cm H I emission spectra having high velocity accuracy.  The full procedure produces final calibrated spectra with velocity errors estimated to be $\sim 10$~{\kms}. The final spectrum is an improvement over the coadded version available through the $HST$ Legacy Archive\footnote{https://archive.stsci.edu/missions-and-data/hst-spectroscopic-legacy-archive-hsla}. The STIS E230M grating has a spectral resolution of $10$~{\kms} over the wavelength range $1600$~{\AA} to $3100$~{\AA} but the data is of a lower $S/N$ ratio compared to COS. The HIRES spectra were obtained from the Keck Observatory Archive (KOA)\footnote{https://koa.ipac.caltech.edu/cgi-bin/KOA/nph-KOAlogin}. The orders were extracted using the MAKEE package (MAuna Kea Echelle Extraction) written by T. Barlow\footnote{http://www.astro.caltech.edu/~tb/makee/} and inverse variance weighted coaddition was performed to generate the final one-dimensional spectrum. The HIRES spectrum has a $S/N \sim 25 $~per~$7$~{\kms} resolution. 

\begin{table*}
\centering\setlength{\tabcolsep}{5pt}
\renewcommand{\arraystretch}{1.5}
	\begin{tabular}{c|c|c|c|c|c|c}
			\hline
			\centering
			 Instrument  & Wavelength Coverage$^a$ & Spectral Resolution & Velocity Resolution & Exposure Time$^b$ & S/N per resolution element$^c$ & Program ID\\
			 & ({\AA}) & $R = \lambda/\Delta \lambda$ & $\Delta v $~{\kms} & $t$~(ks) & & \\ \hline
			 
			 $HST$/COS$^d$ & $1100 - 2100$ & $12,000 - 20,000$ & $17 - 20$ &   $54.60$    & $5 - 20$ & 11741, 13846 \\
		
			 $HST$/STIS & $1600 - 3100$ & $\sim 30,000$ & $\sim 10$ & $20.88$  & $4 - 11$ & 13846 \\


			 Keck/HIRES & $3047 - 5895$ & $\sim 47,000$ & $\sim 7$ & $00.80$ & $10 - 25$ & U066Hb \\

    \hline
	\end{tabular}
	\vspace{0.5cm}
	\caption{Table describes the instrument parameters of the archival data used in this study. $^a$ The wavelength range consists of different gratings and frames and thus small coverage gaps in between. 
	$^b$ Total exposure time listed is the sum of multiple observations with multiple instrumental settings. $^c$ The S/N in each line profile can be estimated from the noise level plotted at the bottom of each panel illustrated in Figures \ref{fig:0675fit} to \ref{fig:1277fit}. $^d$ The COS LSF has broad non-Gaussian wings which affects the measurements on narrow and closely separated line features. }\label{tab:instrument}
\end{table*}


\section{Analysis of Absorption Systems \& Ionization Modeling}
\label{sec:analysis}

The absorbers identified along this sightline are at $z = 0.67556, z = 0.72885, z = 1.09457, z = 1.16592,$ and $z = 1.27768$, which would place them at $\Delta v \sim  -95185~{\kms}$, $\Delta v = -86663~{\kms}$, $\Delta v = -31560~{\kms}$, $\Delta v = -21598~{\kms}$,  and $\Delta v = -6552~{\kms}$ from the emission redshift of the quasar ($z = 1.32801 \pm 0.00044$) \citep{hewett_improved_2010}. The systems were identified from a search for {\OVI} doublets along this sightline with simultaneous information on {\OII} to {\OV} offering a continuous range of diagnostics for the gas phases in the absorbers. For the two absorbers at z $<$ 1, the {\OV}~629~{\AA} line is below the COS G130M grating coverage. The cumulative $HST$ and $Keck$ spectra offer coverage of a large number of interesting ultraviolet and optical metal line transitions for all the absorbers. For instances where a given line is covered by multiple instruments, the spectrum with the best $S/N$ ratio was used for analysis. Low order polynomials were used to fit the local continuum, avoiding evident absorption features. Equivalent width and profile measurements were carried out on the continuum-normalized spectra. The integrated column densities were measured using the apparent optical depth (AOD) method of \citep{savage_analysis_1991} by converting velocity-resolved flux profiles of unsaturated lines into apparent column densities. For saturated lines, the AOD method provides a lower limit on the column density. Only those lines with a significance of $\gtrsim 3 \sigma$ are considered as detected. For the others, useful upper limits are estimated from the $3 \sigma$ equivalent width limits, obtained by integrating over the same velocity range as other prominent detected lines, and determining the corresponding column density from the linear regime of the curve-of-growth. Voigt profile fitting was also performed on these lines using the VPFIT routine (ver.10.4, \citep{carswell_vpfit:_2014}) after convolving the model profiles with the instrument line spread functions for COS\footnote{http://www.stsci.edu/hst/instrumentation/cos/performance/spectral-resolution} and STIS and with a Gaussian of $v \sim 6.6$~{\kms} for HIRES spectra. Multiple lines from a given ion are fitted together, with the fitting procedure optimizing on the unsaturated lines. This means that in the case of species like {\HI}, the column density and b-value are predominantly constrained by the unsaturated higher order Lyman transitions. In the case of single lines that are saturated, fitting is done simultaneously with lines of similar ionization from other elements for which better constraints or multiple lines are available (e.g., {\CIII} and {\OIII}, {\CIV} and {\OIV}). In the three absorbers at z $>$ 1, the {\CIVdblt} lines in the higher S/N and resolution HIRES data were used to establish the component structure of lines seen in the COS data. Components with velocity centroids that are within 10 \kms (the velocity calibration uncertainty) are treated as coinciding in velocity. We adopt multi-component fitting for saturated lines such as Ly-${\alpha}$ only when there is evidence of kinematic sub-components in the metal lines. The atomic line list and oscillator strength values used for fitting are from \citet{morton_atomic_2003} for lines with $\lambda > 912$~{\AA} and from \citet{verner_atomic_1996} for lower wavelengths. 

Photoionization modelling using Cloudy(17.01) (\citealt{ferland_2017_2017}) was carried out to derive the physical conditions and chemical abundances in the absorbers. The gas clouds are modeled as plane parallel slabs of constant temperature and density, with zero dust content. The ionization in the cloud is taken to be dominated by photoionization by the extragalactic UV background radiation at the absorber redshifts, for which we have used the model given by \citealt{khaire_new_2019} (fiducial Q18 model; hereafter KS19). In Figure \ref{fig:KS19}, we show the specific intensity ($4{\pi}{\nu}{J_\nu}$) of the KS19 background for the mean redshift of $\langle z \rangle \sim 0.99$ of our sample along with ionization energies of key species present in the absorbers. In addition, we also explore collisional ionizations caused by energetic free electrons colliding with atoms and ions using the models of \citealt{gnat_time-dependent_2007} as a potential source of ionization. 

Assuming the solar relative elemental abundances of \citealt{asplund_chemical_2009}, photoionization simulations were run for the observed values of {\HI} column density, and gas densities ranging from $10^{-6}$ to $10^{-1}$ cm$^{-3}$. A suite of ionization models were generated by varying metallicities over a wide range to determine the physical conditions that best explains the observed column densities and the temperatures implied by the line widths. Multiple components in an absorber are modeled separately. Absorption components of different ions that are aligned in velocity (within 10 \kms of each other) are considered together for a single phase model. In such cases, multiphase models are sought only when the observed column densities are significantly different from the predictions from a narrow range of density phase solutions with reasonable relative elemental abundance variations. The abundances we derive based on the ionization models will carry uncertainties from the {\HI} and metal line column densities in addition to the uncertainty in the shape of the EBR. The latter can induce errors of $\sim 0.5$~dex in the derived metallicities (\citealt{howk_strong_2009}). 

In cases where the centroid velocities of the lines and their Doppler $b$ parameters imply a single phase, the best constraints on the physical conditions are provided by the column density ratios between adjacent ionization stages of oxygen. In Figure \ref{fig:ocomp} is shown the observed column densities of the oxygen ions in the five absorbers. General conclusions can be drawn on the ionization based on a comparison of these successive ionic column densities. In four out of five of these absorbers the $N(\OIV) > N(\OV) > N(\OIII) > N(\OVI)$\footnote{{\OV} covered and detected only in the $z = 1.09457$, $z = 1.16592$ and $z = 1.27768$ absorbers.}. This trend is valid under photoionization equilibrium at gas densities in the range of $n_{\H} = (0.2 - 2) \times 10^{-3}$~{\cc}. Here the lower bound is set by the $N(\OIV) > N(\OV) > N(\OVI)$, and the upper bound by $N(\OIV) > N(\OIII)$. In all these absorbers, therefore, the low to high ionization stages of oxygen can exist in photoionized gas with densities that are within an order of magnitude of each other. This is only a general estimate, and does not involve the nuanced phase structure these absorbers can have with the {\OVI} predominantly from gas that is significantly ionized compared with the intermediate ions such as {\OIII} and {\OIV}. 

Since the high ions, especially {\OVI}, can have an origin in warm collisionally ionized plasma, we consider the predictions from such models as well. In Figure \ref{fig:ocomp}, we show the column density predictions of {\OIV}, {\OV}, and {\OVI} in a cloud of solar metallicity and arbitrary {\HI} column density, for temperatures at which equilibrium (CIE) and non-equilibrium collisional ionizations (NECI) are significant.  The computations are based on the models of \citet{gnat_time-dependent_2007}. As seen in Figure \ref{fig:ocomp}, at T $< 5 \times 10^6$~K  the {\OV} and {\OVI} predictions under NICE depart significantly from their CIE values, with the column density predictions differing by $> 1$~dex. The presence of metals enhances the cooling rate of the plasma significantly. The temperature of the gas declines faster with the recombination between ions and electrons lagging behind. This results in enhanced high ion fractions at lower temperatures in comparison with CIE. With decrease in metallicity, the difference between CIE and NECI predictions tend to get less. Furthermore, even at temperatures where collisions tend to dominate the ionization, photoionization from the extragalactic background radiation can raise the ionization levels of gas to some degree (\citealt{oppenheimer_non-equilibrium_2013}). These aspects are also considered in our analysis of the absorbers. In the subsequent sections, we discuss the ionization models for each absorber in detail, the results of which are summarized in Table \ref{tab:modsum}. The full system plots (Appendix Figures from B1 to B5) and table of measurements (Refer Appendix Tables from B1 to B10) are listed in the appendix section.

The QSO PG~$1522+101$ was observed as part of an HST program to study {\NeVIII} bearing warm-hot gas, which excluded lines of sight with strong Lyman Limit Systems. The target selection may have led to a bias towards high ionization absorbers. However, low ionization phases traced by {\CII}, {\OII}, {\SiII} are commonly detected in intervening {\OVI} and {\NeVIII} systems (e.gs., \citealt{narayanan_cosmic_2012}, \citealt{meiring_qso_2013}, \citealt{hussain_hstcos_2015}, \citealt{pachat_detection_2017}, \citealt{rosenwasser_understanding_2018}). The simultaneous non-detections of low ions and {\NeVIII} in the {\OVI} absorber sample presented here can be suggestive of the absorbers probing relatively narrow density-temperature phase structures.


\renewcommand{\thefigure}{1}
\begin{figure} 
\centering\includegraphics[scale=0.35,clip=true,trim=2cm 0cm 0cm 2cm]{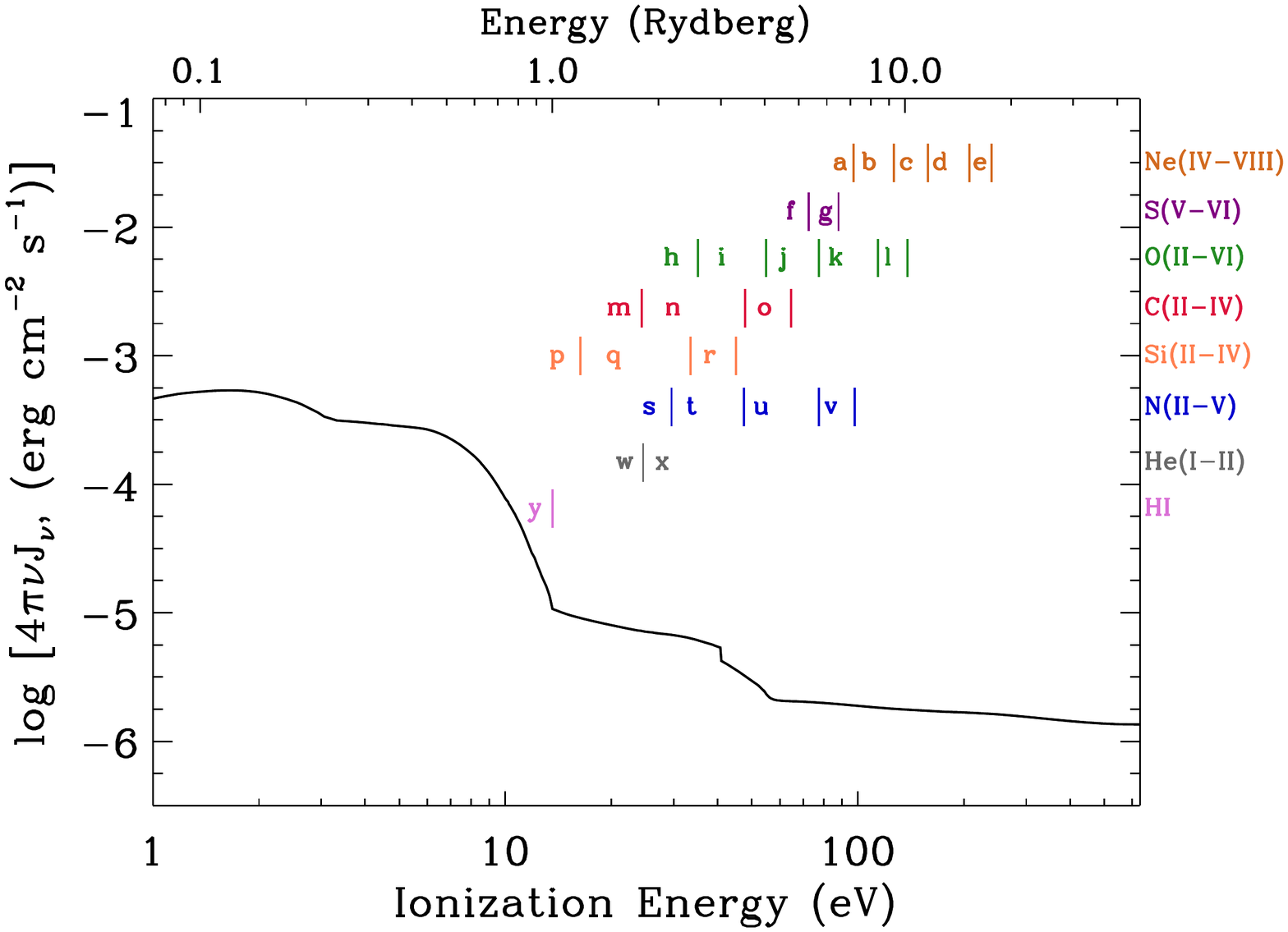}
     \vspace{-0.8cm}\caption{SED of the extragalactic background radiation (KS19), \citet{khaire_new_2019}) which is an improvement over the \citet{haardt_modelling_2001} model as it incorporates the most recent measurements of the quasar luminosity function (\citealt{croom_2df-sdss_2009}, \citealt{palanque-delabrouille_luminosity_2013}) and star formation rate densities (\citealt{khaire_star_2015}). The vertical tick marks signify the creation and destruction energies of key species covered in our sample of absorbers.}
     \label{fig:KS19}
\end{figure}

\renewcommand{\thefigure}{2}
\begin{figure} 
\includegraphics[scale=0.7,clip=true,trim=5.2cm 0cm 0cm 0cm]{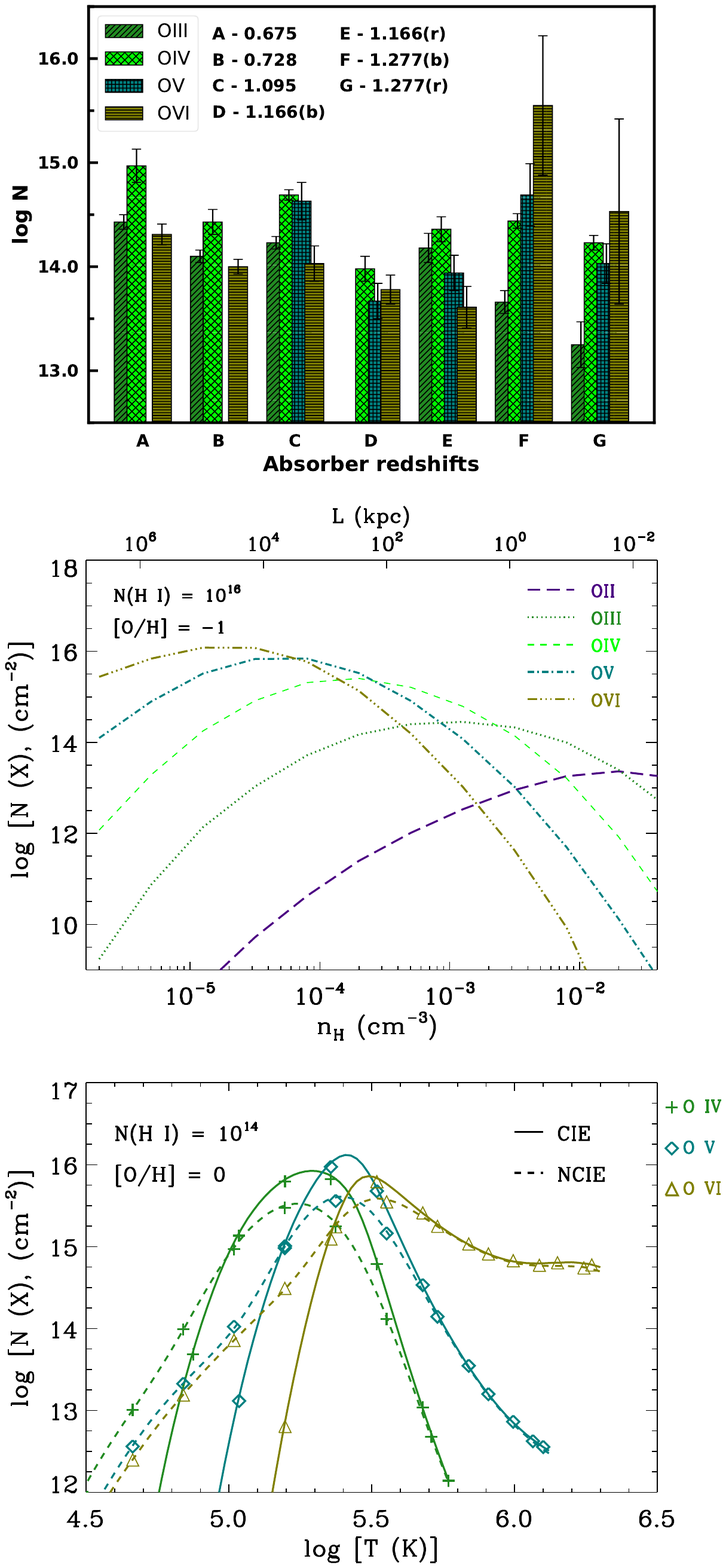}
\vspace{-0.8cm}
     \caption{\textit{Top} panel shows comparison of the column densities of the different oxygen ions in the five absorbers. In the $z = 1.165$ and $z = 1.277$ absorbers, the negative and positive velocity components are separately measured and indicated as "b" and "r". The {\OV} is not covered by the COS spectra for $z < 1.09$. The panel on the \textit{middle} shows a fiducial PIE model computed for an average value of $\log~N(\HI)$ for our sample, an oxygen abundance of one-tenth solar, and an ionizing background corresponding to the mean redshift of our sample. The different curves are the PIE  model predicted column densities for the five successive ionization stages of oxygen. \textit{Bottom} panel shows the column densities of {\OIV}, {\OV}, and {\OVI} in a cloud of solar metallicity and $N(\HI) \sim 10^{14}$~{\cmsq}, for the range of temperatures at which collisional ionization equilibrium (CIE) and non-equilibrium collisional ionizations (NECI) are significant.  The computations are based on the models of \citet{gnat_time-dependent_2007}}.
     \label{fig:ocomp}
\end{figure} 


\section{The z$_{abs}~=~0.67556$ absorber}
\subsection{Characterization of the z$_{abs} = 0.67556$ absorber}
\label{sec:0675D}

This is the lowest redshift {\OVI} absorber along this sightline that we report on. The absorber has detections of {\OIII}, {\OIV}, and {\OVI}. The redshifted {\OV}~$629$ line is below the far-UV coverage of COS. The system plot for this absorber is shown in Figure \ref{fig:0675fit} (the full version in Appendix Figure B1) and the line measurements are listed in Appendix Tables B1 and B2. In addition to the oxygen lines, the COS spectrum also shows {\CIII}. Key species including {\CII}, {\NII}, {\OII}, {\AlII}, {\SiII}, {\NIII}, {\AlIII}, {\SiIII}, and {\NeVIII} are non-detections. The {\CIV}~$1548$ line is affected by contamination from Galactic {\MnII}~$2594$. The {\CIV}~$1551$ line is detected as a weak feature. The high $S/N$ HIRES data shows the {\MgIIdblt} lines as non-detections, consistent with the non-detections of {\CII}, {\OII}, and {\SiII} lines in the COS data, since these low ions have similar ionization potentials. The AOD integrated column densities of the various oxygen lines are comparable to their Voigt profile fitted values to within $1\sigma$ uncertainty, indicating that unresolved saturation is either absent or not significant. The absorption in {\OIII} and {\CIII} coincides with {\HI}, but is centered $25$~{\kms} from {\OIV} and {\OVI} (see Figure \ref{fig:0675nav}). Such velocity offsets between high and low ions is common to many {\OVI} absorbers (e.g., \citealt{savage_multiphase_2011}, \citealt{fox_high-ion_2013}) and are thought to occur when the line of sight is probing a flow of warm interface gas between a cold cloud and a hot surrounding medium (\citealt{boehringer_steady_1987}, \citealt{savage_properties_2006}). 

Neutral hydrogen is detected in a number of Lyman series lines. The unsaturated {\HI} lines from $937$ to $919$~{\AA} and the non-detections of higher order transitions constrains the {\HI} column density to $\log~[N(\HI),~\cmsq] = 15.87~{\pm}~0.04$, and $b$-parameter to $b(\H) = 37~{\pm}~2$~{\kms}.  The {\HI} and metal line absorption profiles in this systems are modeled with a single component since there are no adequate constraints in the data to adopt a rigid multi-component model. Kinematic sub-components that are unresolved by COS or STIS, if present, will be an additional source of  uncertainty in the temperature and ionization model parameters we infer. The wings of the {\Lya} between $75 <$ v $< 175$~{\kms} is contaminated by the {\Lyg} feature associated with the $z = 1.09457$ {\OVI} absorber (refer Appendix Figure B3). The system plot of Figure \ref{fig:0675fit} shows the extent of this contamination. As mentioned earlier, the {\HI}, {\CIII}, and {\OIII} lines have an offset of $\Delta v \sim 25$~{\kms} relative to the {\OIV} and {\OVI}. The offset is evident in the apparent column density comparison plots of Figure \ref{fig:0675nav}. The {\OIV}~$787$ line, covered by the G130M grating of COS, occurs in between the {\OIII}~$702$ and {\OIII}~$832$ lines. The {\OIII} lines are well aligned relative to each other, but offset from the {\OIV}. Furthermore, the {\OIV}~$787$ observed by the G130M grating is well aligned with the {\OVI} doublet lines that appear at the edge of the G160M spectrum. The velocity difference is therefore unlikely to be an artifact of COS wavelength calibration. Such offsets between {\OVI} and ions of lower ionization is usually an indication of multiphased gas, a possibility we consider during the ionziation modeling of the absorber. 
The {\HI} lines are only as broad as the well-aligned {\CIII} and {\OIII} lines indicating a predominance of non-thermal broadening. The $b(\HI) = 36~{\pm}~2$~{\kms}, and $b(\CIII) = 33~{\pm}~6$~{\kms} imply a $T < 4.7 \times 10^4$~K based on the $1\sigma$ uncertainties in $b$. The $b = 39~{\pm}~11$~{\kms} for the kinematically offset {\OVI} places a much higher limit of $T < 2.5 \times 10^6$~K for the {\OVI} bearing gas. 

\renewcommand{\thefigure}{3A}
\begin{figure} 
\centering
  \includegraphics[scale=0.65,clip=true,trim=4cm 0.2cm 4cm 0.6cm]{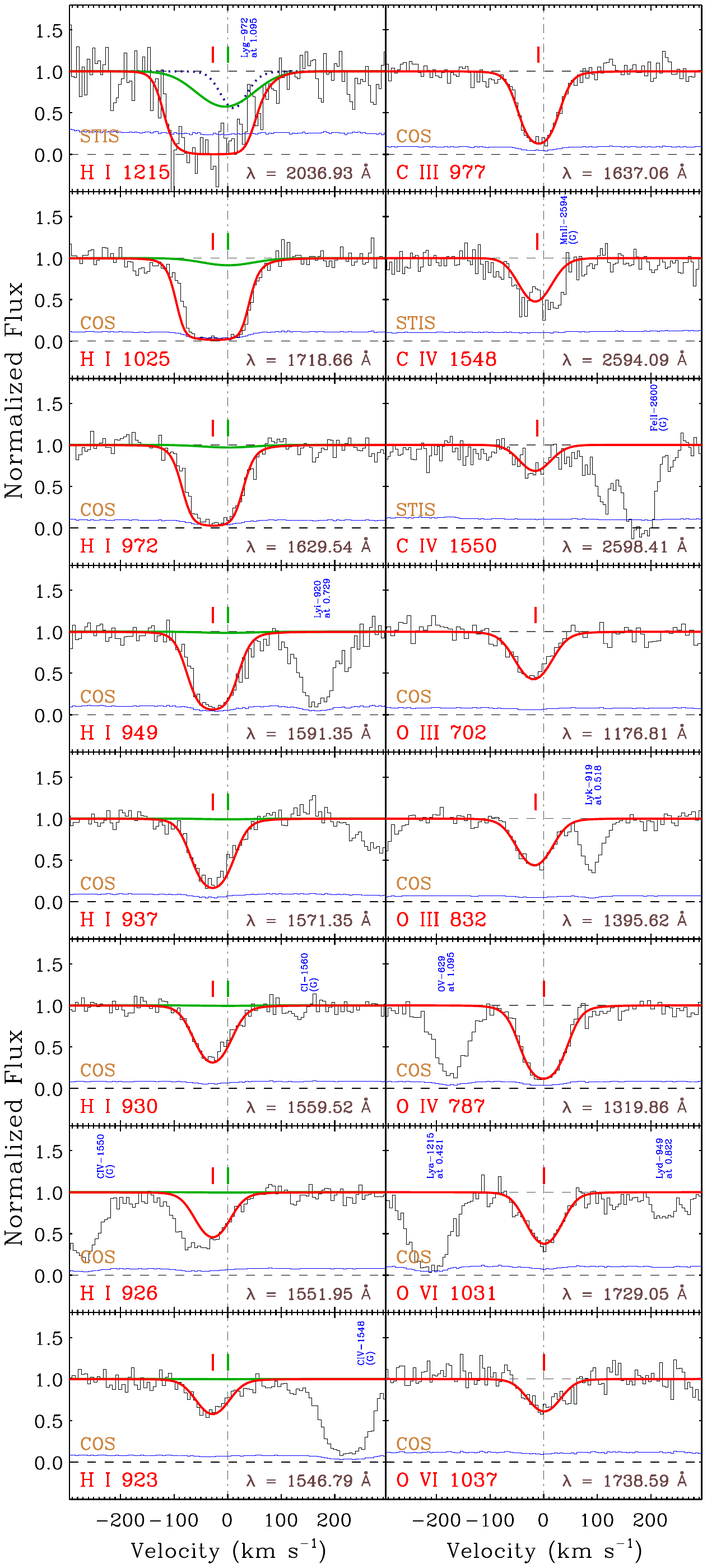}
	\caption{System plot of the z$_{abs} = 0.67556$ absorber, with continuum-normalized flux along the Y-axis and the velocity scale relative to the redshift of the absorber along the X-axis. The $v = 0$~{\kms}, marked by the \textit{dashed-dotted} vertical line, indicates the absorber redshift. The $1\sigma$ uncertainty in flux is represented by the \textit{blue} curve at the bottom of each panel. The \textit{red} curves are the best-fit Voigt profiles. The observed wavelength of each transition is also indicated in the respective panels. Interloping features unrelated to the absorber are labeled. In the Lyman Lines panels, the synthetic profile of a BLA with $\log [N(\HI)] = 13.7$ and $b(\H) = 80$~{\kms} associated with {\OVI} is also shown (green solid curve). The blue dashed curve in the Lyman $\alpha$ panel shows the contaminating feature of Lyman $\gamma$ of the absorber at 1.09457. \label{fig:0675fit}}
\end{figure}

\renewcommand{\thefigure}{3B}
\begin{figure*}\vspace{-2cm}
      \centerline{\vbox{\centerline{\hbox{\hspace{1.5cm}\includegraphics[clip=true, trim=0cm 0cm 3cm 2cm,scale=0.4]{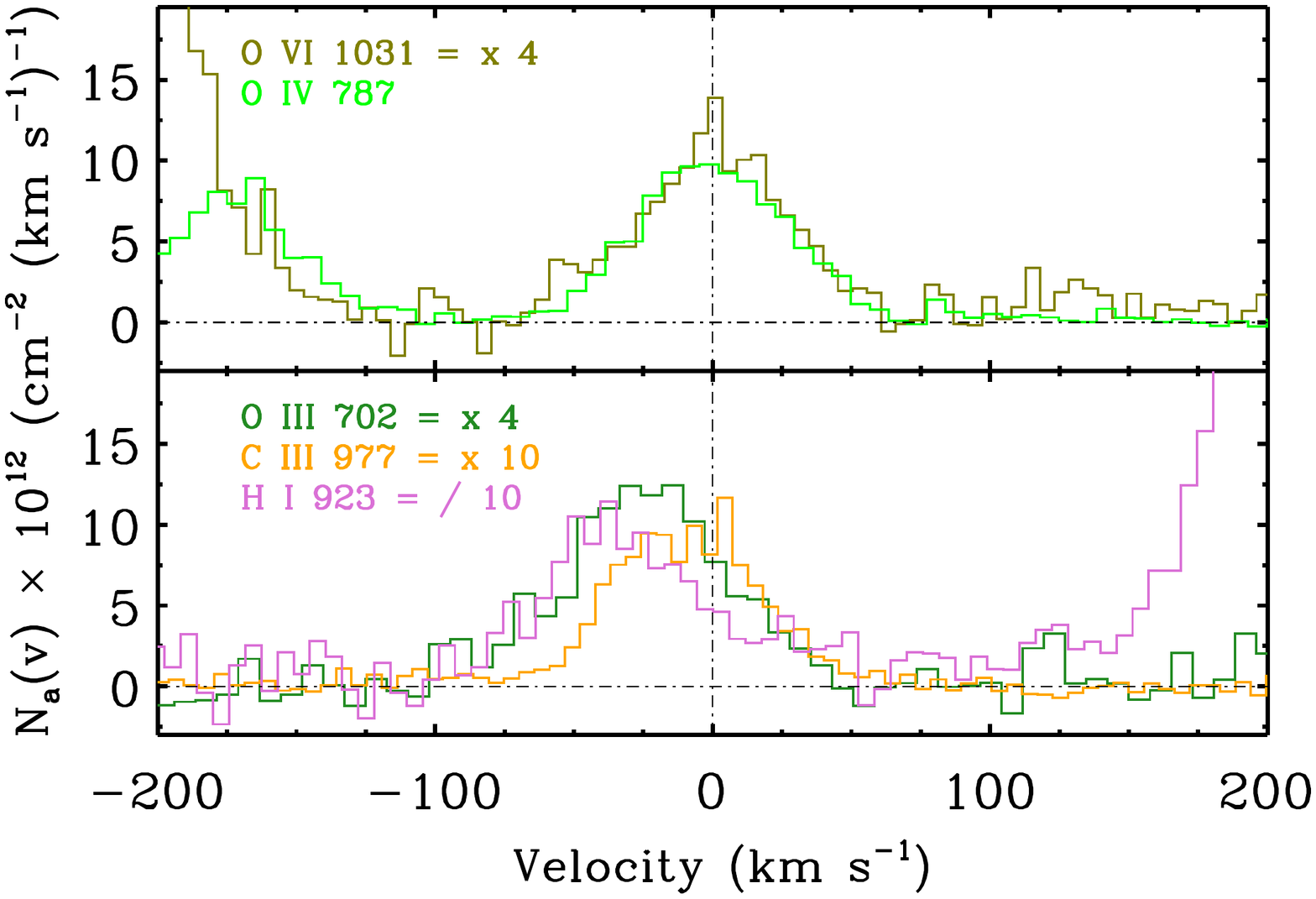}
\includegraphics[clip=true, trim=8cm 0cm 3cm 0cm,scale=0.46]{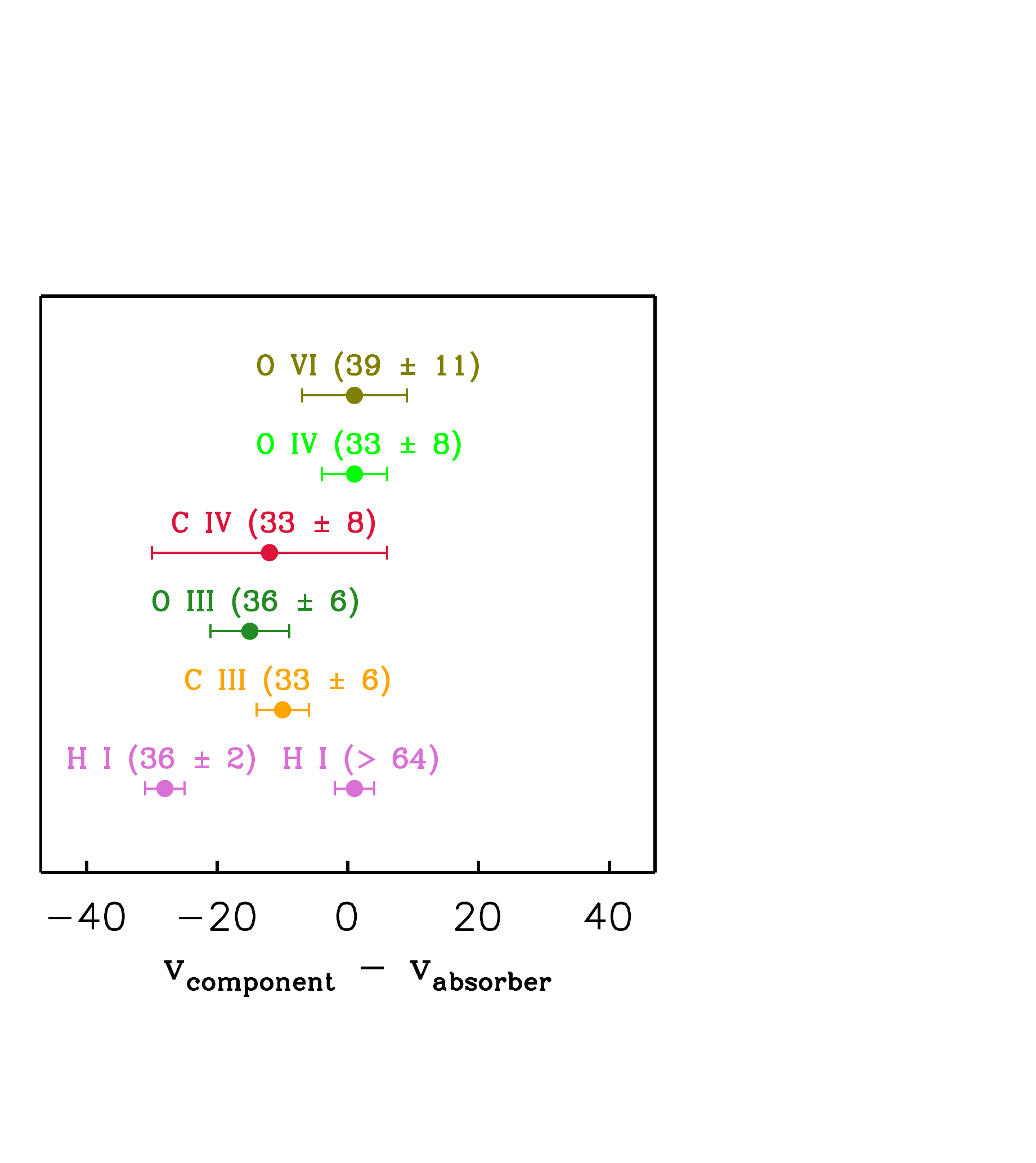}
}}
}} \vspace{-1cm}
     \caption{Apparent column density comparison between the different detected species of the z$_{abs} = 0.67556$ absorber. The high ions are plotted on the top panel and the intermediate ions on the bottom panel. An offset of $25 \kms$ is seen between the intermediate and high ions. The $b$ parameter of {\OVI} also appears to be a bit higher than the other species. The figure on the right shows the velocity centroids of each line with their uncertainty relative to the absorber redshift, as derived from profile fitting. The $b$-parameter measured for the various ions are indicated within the parentheses.}
     \label{fig:0675nav}
\end{figure*} 

\subsection{Ionization Models for the z$_{abs} = 0.67556$ absorber}
\label{sec:0675M}

The photoionization model results are shown in Figure \ref{fig:0675model}. The observed $N(\CIII)/N(\CIV) = 0.07~{\pm}~0.27$ is used to constrain the density at $n_{\H} \sim (0.2 - 0.7) \times 10^{-3}$~{\cc}. The column density predictions from the PIE models yield an [O/H] $= -0.9 \pm 0.1$ for the {\OIII} to be produced at the mean of that density range, with the uncertainty in the metallicity a cumulative of the measurement errors in {\HI} and {\OIII}. The predicted column density ratio of {\CIII} to {\OIII} is a constant over two orders of magnitude in density ($n_{\H} \sim 10^{-4} - 10^{-2}$~{\cc}), indicating that these two species would trace the same phase of the absorber. The observed column densities of these ions thus offer a direct insight into the relative elemental abundance between C and O. For the model predicted {\CIII} and {\OIII} column densities to match their observed values at the mean density of $n_{\H} \sim 0.5 \times 10^{-3}$~{\cc}, the PIE models require [C/O] $=-0.5 \pm 0.2$~dex. For this phase, the PIE models also predict the observed {\CIV} and {\OIV} to within their $1\sigma$ uncertainty. The {\OVI} however is under-produced by $\sim 0.75$~dex. The column density ratio plot of Figure \ref{fig:0675model} also reflect this where the observed {\OIV} to {\OVI} column density ratio matches with its model predicted value at a density of $n_{\H} = 0.2~\times~10^{-3}$~{\cc}  which is about five times lower. The line of sight could be tracing a column of gas with a narrow spread of densities from $n_{\H} = (0.2 - 0.7)~\times~10^{-3}$~{\cc} where all the observed metal ions are produced (\textit{yellow} zone in Figure \ref{fig:0675model}). This density range also constraints the total hydrogen column density to within $\log N(\H) = 19.93 - 19.20$~{\cmsq}, gas temperatures to $T = (2.9 - 2.2) \times 10^4$~K, pressure of $p/K = (4.6 - 15.9)$~{\cc}~K, and thickness along the line of sight of $L = (172.4 - 7.2)$~kpc. Such a medium is also consistent with the non-detections of {\CII}, {\SiII}, and {\MgII}, that are generally tracers of denser and more compact gas.

However, the $\Delta v \sim 25$~{\kms} velocity offset of {\OVI} and {\OIV} with {\OIII}, {\CIII} and {\HI} implies at least two distinct gas phases. The bulk of the {\OVI} could be from gas at $T > 10^5$~K where collisions dominate the ionization. The $b(\OVI) = 39~{\pm}~11$~{\kms} suggests that the temperature in this {\OVI} phase can be as high as $T < 2.5 \times 10^6$~K ($1\sigma$ upper limit), assuming negligible non-thermal broadening. Such a warm/hot phase can contribute significant amounts of {\OIV} as well. It is difficult to disentangle the separate contributions to the observed {\OIV} from the photoionized and collisionally ionized phases. The {\OIV} absorption coinciding in velocity with {\OVI} points to a significant portion of the {\OIV} coming from the {\OVI} phase. Nonetheless, we consider the {\OIV} to {\OVI} observed column density ratio as an upper limit while constraining the warm {\OVI} phase. In gas that is subjected to CIE, the observed $\log [N(\OIV)/N(\OVI)] \leq 0.66$ is valid for $T > 0.3 \times 10^6$~K as shown by the collisional ionization models of Figure \ref{fig:0675model}. Given the low neutral fraction at such temperatures ($f_{\HI} = 1.95 \times 10^{-6}$, \citealt{gnat_time-dependent_2007}), the {\HI} from this phase is expected to be a shallow feature, with a $b(\HI) \geq 64$~{\kms} corresponding to a $T > 0.2 \times 10^6$~K from the {\HI} - {\OVI} combined $b$-parameters. Figure \ref{fig:0675fit} shows the possible presence of such a broad-{\Lya} (BLA). The BLA column density has to be $\log~N(\HI) \leq 13.7$ for the cumulative model profile to not exceed the absorption seen in {\Lya} and the higher Lyman transitions. At this limiting temperature, the observed $N(\OVI)$ is reproduced by the CIE models for a [O/H] = $-0.9$, identical to the abundance derived for the cooler photoionized phase using {\OIII} as the constraint. The abundance is not adequately constrained due to lack of information on the exact temperature of the {\OVI} gas, and the {\HI} associated with the {\OVI}. The BLA properties are comparable to the characteristic value predicted for absorption tracing hot circumgalactic gas around galaxies with virial halo masses of $\log~(M/M_{\odot}) \sim 10^{11}$ (\citealt{richter_hot_2020}). 

An alternative is for the {\OVI} to be produced via photoionization. Assuming a $\log~[N(\HI)/\cmsq] = 13.7$, the PIE models require oxygen abundance to be at least [O/H] $\geq -0.5$, with a corresponding density lower limit of $\log~[n(\H),{\cc}] \geq -5.0$, with higher densities requiring higher metallicites. The [O/H] thus estimated is assuming $\log~[N(\HI)/\cmsq] = 13.7$ corresponding to the upper limit in the column density of the BLA. Lower values of N({\HI}) for the BLA would require higher [O/H] for the {\OVI} to be recovered. The PIE model for the lower limit on density yields a temperature of $T = 4.9 \times 10^4$~K, which predicts a thermal line width of $b \sim 7$~{\kms} that is a factor of $\sim 4$ less than the observed $b(\OVI)$. The metallicity will have to be super-solar for the {\OIV} to be produced at detectable levels from this photoionized {\OVI} phase. This would also mean that the photoionized low and high phases would have a metallicity gradient of more than an order of magnitude. Metallicity differences of as much as $\sim 1$~dex between aligned components have been reported before (\citealt{prochter_keck_2010}, \citealt{tripp_hidden_2011}, \citealt{crighton_metal-poor_2013}, \citealt{rosenwasser_understanding_2018}, \citealt{lehner_cos_2019}, \citealt{zahedy_characterizing_2019}). The differences often stem from the low and high ions tracing gas of different enrichment history, particularly when the line of sight is probing material proximate to galaxies where superwinds and AGN outflows can create patchy metal distributions (e.g., \citealt{zonak_absorption_2004}, \citealt{veilleux_galactic_2005}). An alternative explanation involves the intermediate ions tracing photoionized gas, and the {\OVI} (and some substantial fraction of {\OIV}) coming from a warm ($T \gtrsim 3 \times 10^{5}$~K) higher ionization phase with a metallicity similar to the photoionized gas.

\renewcommand{\thefigure}{3C}
\begin{figure} 
     \vspace{-1cm}
     \includegraphics[scale=0.74, clip=true, trim=5cm 2cm 0cm 0cm ]{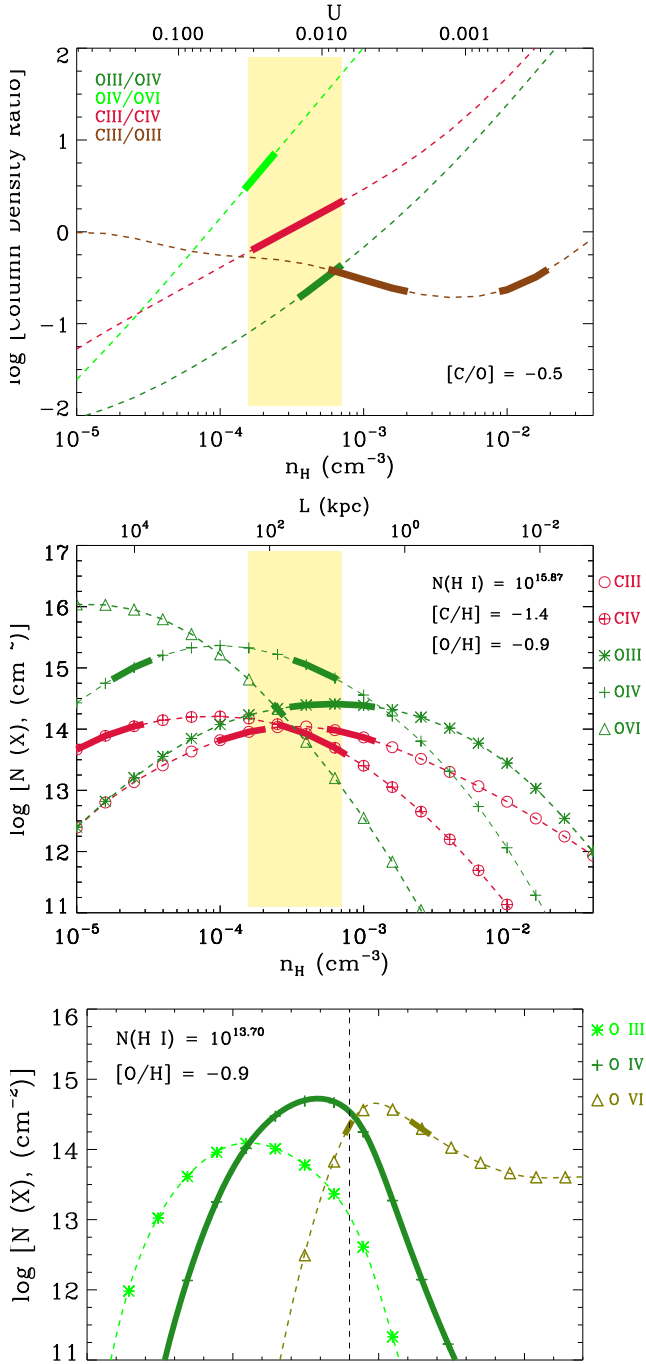}
	\caption{Photoionization models for the z$_{abs}$ $=$ 0.67556 absorber. On the {\textit{top}} panel is shown the model predictions (\textit{thin} lines) and the observed (\textit{thick} lines) column density ratios between ions plotted against gas density. These ratios confine the gas density to the narrow range indicated by the {\textit{yellow} strip}. On the {\textit{middle}} panel are predictions of column densities (thin lines) along with their $1\sigma$ observed values (thick lines). The density range allowed by the column density ratios is represented by the yellow region. The elemental abundances in the models are varied to reproduce the observed column densities from a single or nearly similar gas phases. The ionization parameter (u) and line of sight thickness for a given density are given by the \textit{top} X-axis of the panels. The {\textit{bottom}} panel shows column density predictions from CIE models for different equilibrium temperatures. The vertical dashed line in the bottom panel marks the temperature at which the observed column density ratio between {\OIV} and {\OVI} is recovered by the CIE models.} 
     \label{fig:0675model}
\end{figure}

\section{The z$_{abs}~=~0.72885$ absorber}
\subsection{Characterization of the z$_{abs} = 0.72885$ absorber}
\label{sec:0728D}

The system plot displaying the key transitions in this absorber is shown in Figure \ref{fig:0728fit} (the full version is available in Appendix Figure B2) and line measurements are given in Appendix Tables B3 and B4. The absorber has {\CIII}, {\OIII}, {\OIV}, and {\OVI} ions detected at $\geq 3\sigma$. Apparent column density comparison between the detected lines is shown in the \textit{top} panel of Figure \ref{fig:0728nav}. The {\OVI}~$1037$~{\AA} line is severely contaminated by {\OIV}~$787$ from $z = 1.277$ and {\Lya} from $z = 0.475$. We consider the absorption at $\lambda = 1784.05$~{\AA} as {\OVI}~$1031$~{\AA}. The feature is weaker and more spread out compared to any of the other detected ions (see $N_a(v)$ comparison of Figure \ref{fig:0728nav}). Equivalent width estimation over the velocity range in which absorption is seen shows the feature to be of $\gtrsim 3 \sigma$ significance. 

A Voigt profile model recovers the same column density as the integrated apparent column density with a $b(\OVI) = 61~{\pm}~11$~{\kms}. In the absence of unidentified line contamination, or unresolved component structure, which we cannot rule out, the broader kinematic profile of {\OVI} implies an origin in a  separate phase with $T < 5.0 \times 10^6$~K compared with other metal ions and {\HI} which are considerably narrower and hence tracing cooler gas.

The {\MgIIdblt} lines covered by HIRES and the {\CIVdblt} from STIS are non-detections. The absence of {\MgII} in the higher $S/N$ HIRES spectrum, is consistent with the non-detections of {\CII}, {\NII}, {\OII} and {\SiII} in the COS and STIS data, as all these ions possess similar ionization potentials. It further alludes to the absence of a prominent high density ($n_{\H} > 10^{-3}$~{\cc}) low ionization phase in the absorber. Similarly, the non-detections of {\CIV} and {\NIV}, despite the presence of a comparatively strong {\OIV},  hints at non-solar [C/O] and [N/O] in the absorbing gas. The narrow weak feature coinciding with the expected location of {\NeVIII}~$770$ is identified as {\NV}~$1238$ from $z=0.075$, confirmed by the presence of {\HI}~$1215$, and {\CIVdblt} lines from the same redshift. 

The full range of Lyman series lines are covered by the combined COS and STIS data. A simultaneous fit to these lines with a single component yields $\log~[N(\HI)] = 16.50~{\pm}~0.02~{\cmsq}$ and $b(\HI) = 26~{\pm}~2$~{\kms}. The {\Lya} and {\Lyb} features span a broad kinematic range of $-100$ to $200$~\kms with the metals coinciding in velocity with the stronger component. The weaker {\HI} absorption at v $\sim +150$~{\kms} is modeled using a single component profile as shown in Figure \ref{fig:0728fit}. While constraining the metallicity of the absorbing gas, we consider the column density of {\HI} that is coincident with the metal lines. The obtained value of $N(\HI)$ is slightly lower than the value quoted by \cite{lehner_bimodal_2013} and \cite{wotta_low-metallicity_2016} $(16.66 \pm 0.05)$. We proceed with our measurement which is constrained by a large number of Lyman transitions. The $b(\HI) = 26~{\pm}~3$~{\kms} and $b(\OIII) \sim b(\CIII) = 17~{\pm}~3$~{\kms} yield a temperature of $T = (0.8 - 4.3) \times 10^4$~K in this intermediate ion phase, which is two orders of magnitude lower than the temperature upper limit implied by the broad {\OVI}.

\renewcommand{\thefigure}{4A}
\begin{figure*} 
	\centering
	\includegraphics[scale=0.8,clip=true, trim = 0cm 0cm 0cm 1cm]{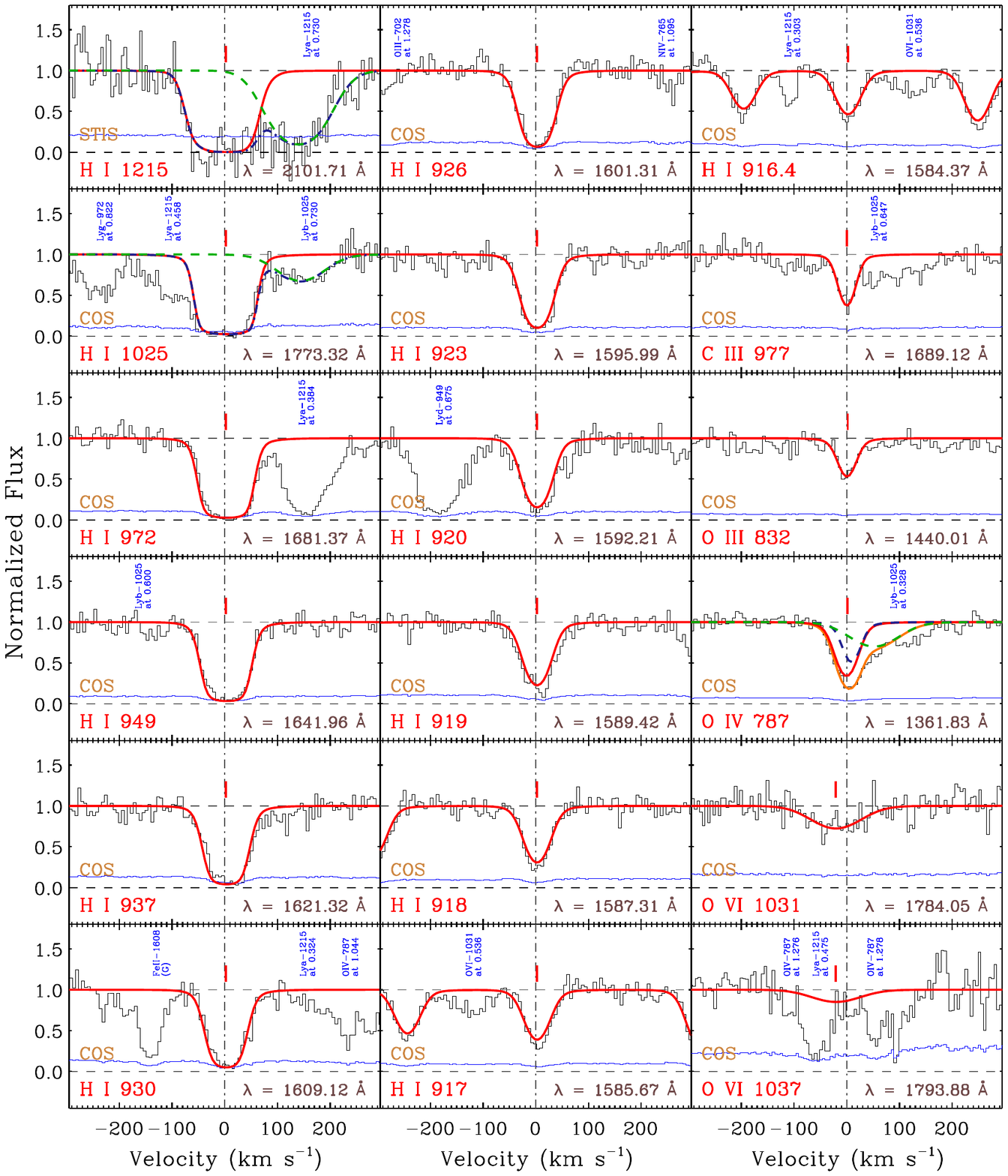}
	\vspace{-6.5cm}
	\caption{System plot of the z$_{abs}$=0.72885 absorber, with continuum-normalized flux along the Y-axis and the velocity scale relative to the redshift of the absorber along the X-axis. The $v = 0$~{\kms}, marked by the \textit{dashed-dotted} vertical line, indicates the absorber redshift. The $1\sigma$ uncertainty in flux is indicated by the \textit{blue} curve at the bottom of each panel. The \textit{red} curves are the best-fit Voigt profiles. The observed wavelength of each transition is also indicated in the respective panels. Interloping features unrelated to the absorber are also labeled. The dashed green curves in the {\HI}~1215 {\AA} and {\HI}~1025 {\AA} panels show a fit of the corresponding {\HI} lines at z $\sim 0.730$ and the blue curve shows the composite fit. The {\OIV}~787 {\AA} is severely contaminated by {\HI}~1025 {\AA} at z $\sim 0.327$ and $0.328$. The contamination was accounted for in the fit and is displayed as dashed blue and green curve in the panel.} \label{fig:0728fit}
\end{figure*}

\renewcommand{\thefigure}{4B}
\begin{figure*}\vspace{-2.5cm}
\centerline{\vbox{\centerline{\hbox{\hspace{1.5cm}\includegraphics[clip=true, trim=0cm 0cm 3cm 2cm,scale=0.4]{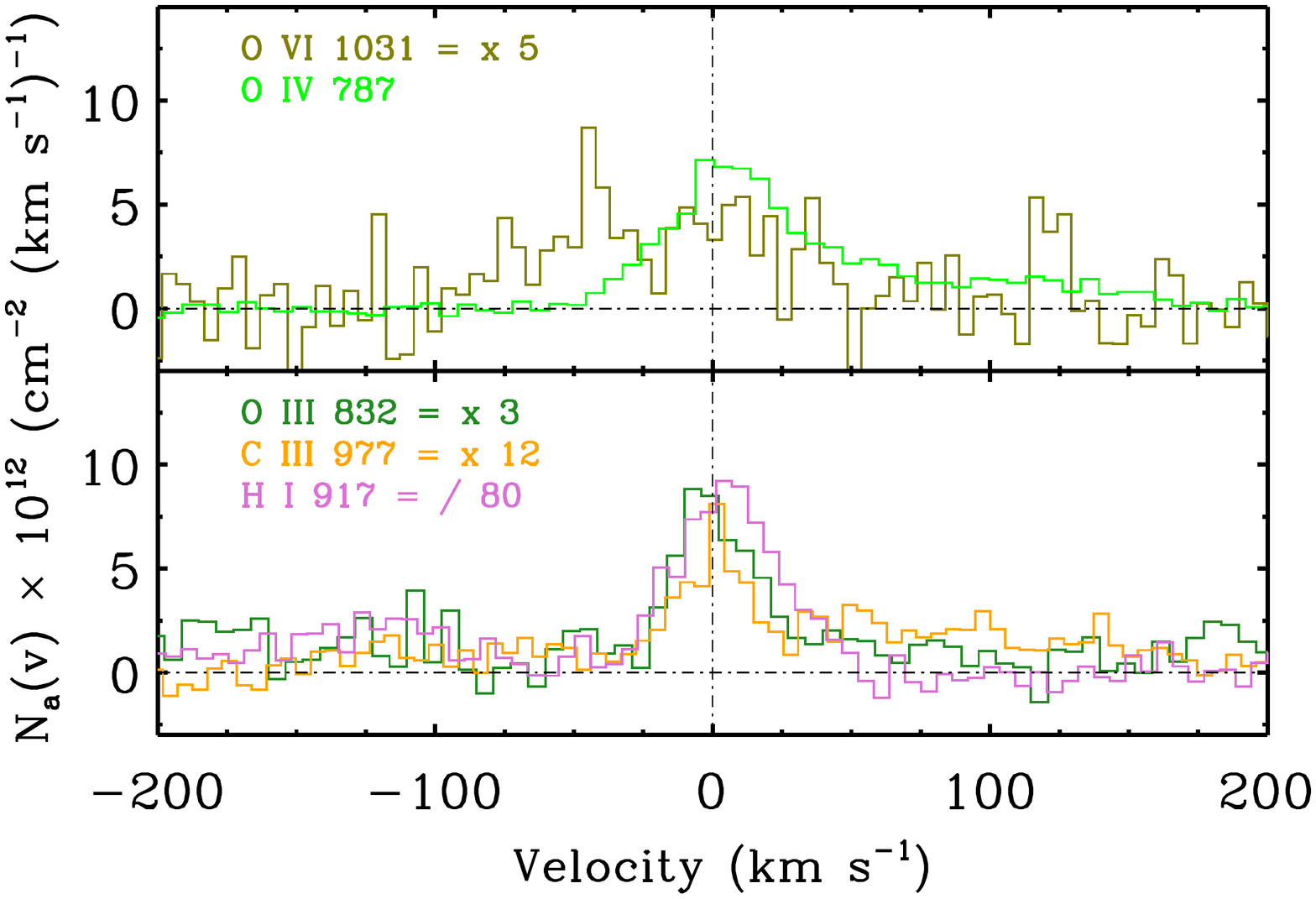} 
\includegraphics[clip=true, trim=8cm 0cm 3cm 0cm,scale=0.46]{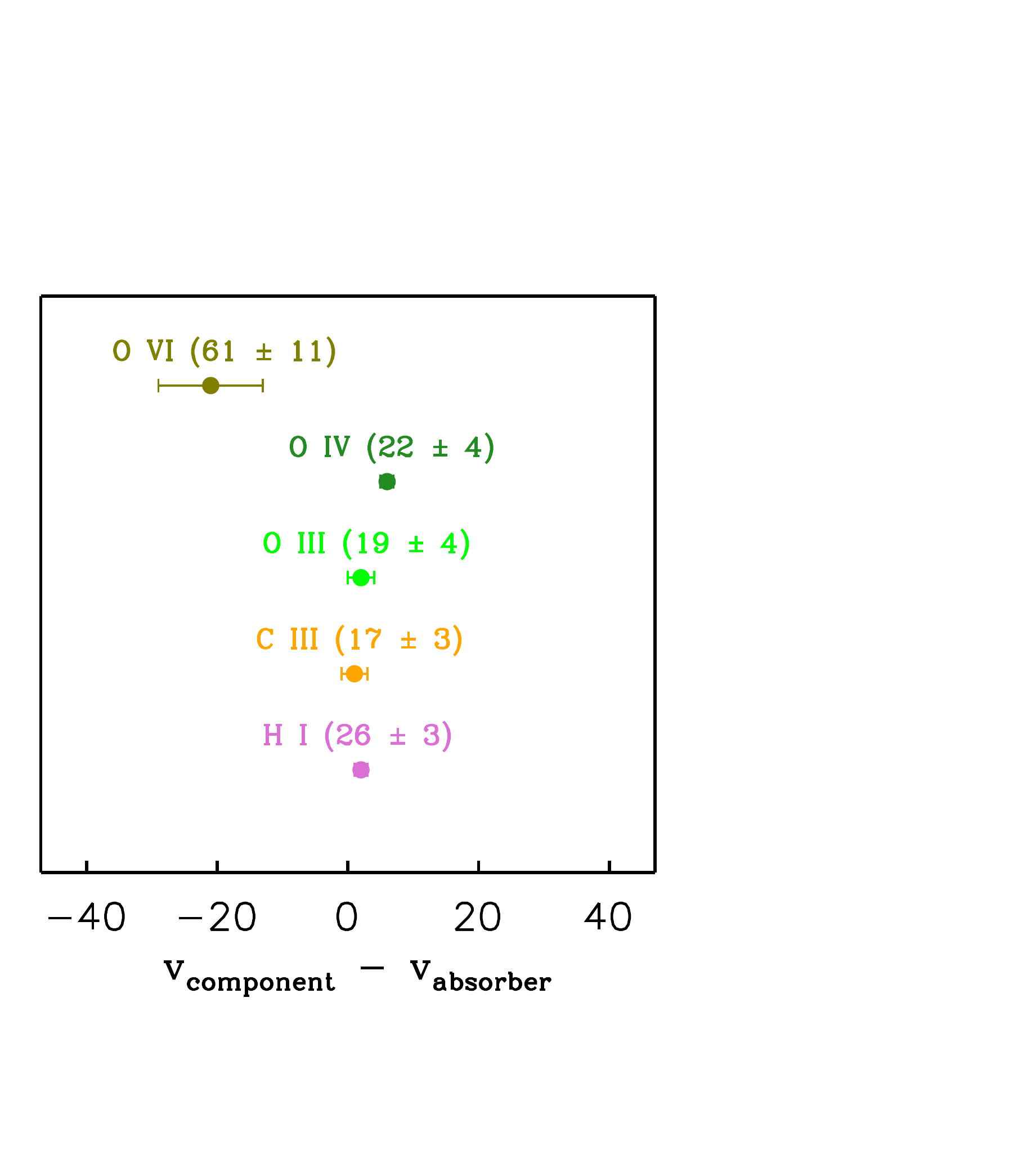}
}}
}} \vspace{-1cm}
     \caption{Apparent column density comparison of different detected species of the absorber at z$_{abs} = 0.72885$. The high ions are plotted on the top panel and the intermediate ions on the bottom panel. The figure on the right shows the velocity centroid of each line derived from profile fitting relative to the redshift of the system. The $b$-parameter measured for the various ions are indicated within the parentheses.}
     \label{fig:0728nav}
\end{figure*} 

\subsection{Ionization Models for the z$_{abs} = 0.72885$ absorber}
\label{sec:0728M}

The PIE model predictions for this absorber are given in Figure \ref{fig:0728model}. The constraints imposed by the column density ratio between {\OIII}, and {\OIV} suggest the gas density to be $n_{\H} = 7 \times 10^{-4}$~{\cc}, if the two ions are predominantly from the same gas phase. The observed ratio between {\CIII} and {\OIII}, both of which are measurements from unsaturated lines, is also valid at this density when the relative abundance is [C/O] $= -0.8 \pm 0.1$~dex. The column density predictions from the PIE models yield an [O/H] $= -2.0 \pm 0.1$ with the uncertainty in metallicity being a cumulative of the measurement errors in {\HI} and {\OIII}. At lower abundances, {\CIII} and {\OIII} will be under-produced at all densities, whereas higher abundances will not yield a model solution that explains the intermediate ions. This phase with $n_{\H} = 7 \times 10^{-4}$~{\cc} results in $\log N(\H) = 20.03$~{\cmsq}, $T = 2.7~\times~10^4$~K, $p/K = 18.3$~{\cc}~K and $L = 59.14$~kpc. \citealt{lehner_bimodal_2013} estimate the absorber to be a complex multi-phase system with the lower ionization phase arising from $n_{\H} < 1.26~\times~10^{-3}$~{\cc} and [X/H] $< -2.0$. These metallicity and density upper limits are consistent with the values we infer. However, the solar [C/$\alpha$] which they report is inconsistent with our findings. They used the non-detection of {\MgII} to estimate [C/$\alpha$] while we use the detected {\CIII} to {\OIII} column density ratio, which is nearly uniform in photoionized gas over five orders of magnitude in density (see \textit{top} panel of Figure \ref{fig:0728model}), and hence a better tracer of relative elemental abundance. 

The column density of {\OVI} predicted from the PIE model described above is an order of magnitude lower than its observed value. The significantly broader {\OVI} profile compared to the narrow metal ions and {\HI} suggest the presence of an additional gaseous component at $T < 5 \times 10^6$~K. The {\HI} in this highly ionized phase should produce a BLA absorption with $b(\HI) < 280$~{\kms}. The presence of such a broad {\HI} is not evident in the {\Lya}, suggesting that the feature could be too weak to be detected against the strength of the absorption from cooler gas. Specific constraints on this gas phase, such as its [O/H] or exact temperature cannot be gleaned from the {\OVI} profile alone. 

From the analysis, one can only conclude that there is a dominant cool photoionized medium in this absorber with [C/O] $\sim -0.8$~dex, and a warm/hot phase traced by the {\OVI}. The broad {\OVI} is similar to the $T = (0.6 - 1.6) \times 10^6$~K {\OVI} reported in \citet{savage_o_2010} where the broad {\Lya} absorption with a predicted $b(\HI) \sim 150$~{\kms} was too weak to be detected. The absorber, in that instance, was hypothesized as tracing either a hot intergalactic gas filament or the circumgalactic medium of a pair of luminous galaxies at impact parameters of $\rho \lesssim 120$~{kpc}.  

\renewcommand{\thefigure}{4C}
\begin{figure} 
     \includegraphics[scale=0.53,clip=true, trim=2.6cm 0cm 0cm 2.6cm]{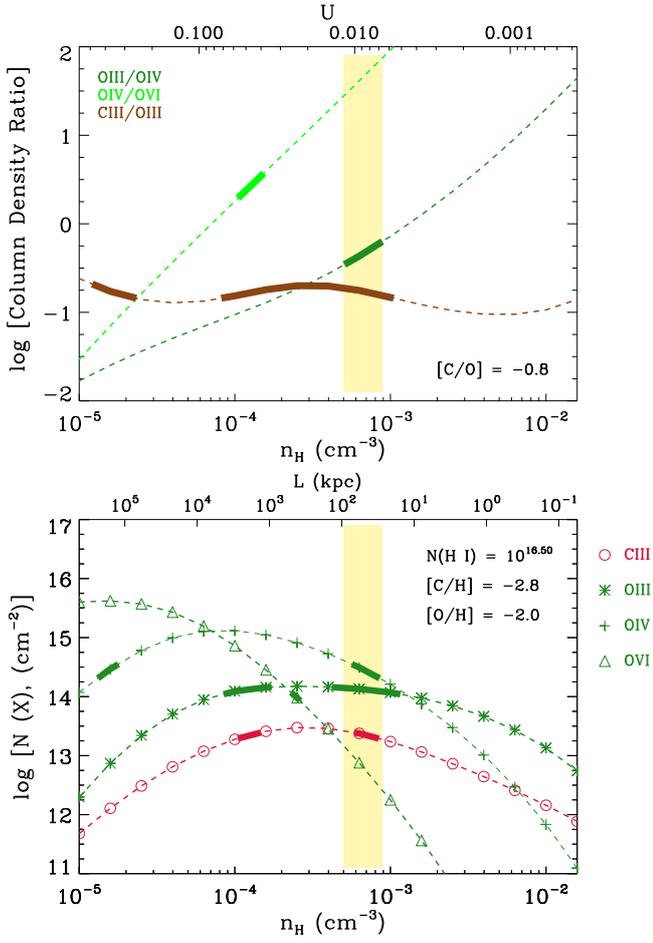}
	\vspace{-1.2cm}
     \caption{Photoionization equilibrium models for the z$_{abs}$ $=$ 0.728850 absorber. The model predicted ratio between column densities of various ionic species at different densities is shown in the \textit{left} panel, and the column density predictions are in the \textit{right} panel. The \textit{thick} portions of each curve represent the observed column density ratios with its $1\sigma$ uncertainty (left-panel) and the measured column density (right-panel). The \textit{yellow} region marks the narrow density range that simultaneously satisfies the column density ratios between the ions of oxygen. The abundances that are required for the models to reproduce the observations are indicated in the panels. The ionization parameter (u) and line of sight thickness for a given density are given by the \textit{top} X-axis of the panels.}
     \label{fig:0728model}
\end{figure}

\section{The z$_{abs} = 1.09457$ Absorption System}
\subsection{Characterization of the z$_{abs} = 1.09457$ absorber}
\label{sec:1094D}

The system plot for the key transitions in the absorber is shown in Figure \ref{fig:1094fit} (the full version in Appendix Figure B3), and the AOD and Voigt profile fit measurements are listed in Appendix Tables B5 and B6. The combined COS and STIS spectra cover five successive ionization stages of oxygen, viz., {\OII}, {\OIII}, {\OIV}, {\OV} and {\OVI} in their multiple lines. Among these, barring {\OII}, all the other ions are detected at $\geq 3\sigma$, with single component line profiles. The non-detection of {\OII} is consistent with the non-detections of {\CII}, {\MgII}, and {\SiII}, all of which possess similar ionization potentials. The {\CIVdblt} lines are well detected in the higher $S/N$ HIRES data. The single component absorption for {\CIV} at $7$~{\kms} resolution justifies the single component profile adopted for the metal lines and {\HI} in the $HST$ data. The apparent column density profiles and their integrated values for the four {\OIV} lines are all within $0.1$~dex of each other suggesting little unresolved saturation in each of them (See Figure \ref{fig:1094nav} and Table B5). A simultaneous profile fit to these four lines offer a robust estimate on the column density and $b$-parameter for {\OIV}. Adopting the $b$-value of {\CIV} also yields good fit to the four {\OIV} lines, with column density that is within $1\sigma$ of the free-fit value. The {\OV}~$629$ line is strong and saturated. The {\OV} ion has previously been detected in only two other intervening absorbers at $z \lesssim 1$ (\citealt{howk_strong_2009}, \citealt{narayanan_cosmic_2011}). The apparent column density integrated measurement for {\OV} is $0.54$~dex lower than the column density obtained from a free fit, suggesting line saturation that is unresolved. Though line profile fitting corrects for saturation to some degree, we found that the column density of {\OV} can be varied by about $\sim 0.24$~dex by changing the $b$-parameter over a permissible range yielding satisfactory fits to the {\OV}~$629$ line feature. The column density we therefore adopt for {\OV} is $\log[N(\OV)] = 14.63 \pm 0.24~{\cmsq}$. The {\NIII}~$684$ and $685$ lines are detected, but both are severely contaminated. A column density upper limit for {\NIII} is obtained by integrating the {\NIII}~$685$ profile over the same velocity range as {\OIII}. The {\NIV}~$765$ is a marginal $3\sigma$ feature, whereas the {\NVdblt} lines covered by STIS are non-detections.

All other metal ions are non-detections or information on them is severely obscured by line contamination. The region corresponding to {\NeVIII}~$770$~line is contaminated by {\Lya} from an absorber at $z = 0.32773$. The {\NeVIII}~$780$ line is a non-detection. The other non-detections include {\CII}, {\OII},  {\MgII}, {\AlII}, {\SiII}, {\FeII}, {\AlIII}, {\SiIII}, {\SIV}, {\SV}, {\SVI} and {\MgX}. The non-detections of {\NeVIII}, {\SVI} and {\MgX} are an indication of the absence of hot collisionally ionized gas with $T > 10^{5.5}$~K, as these ions are diagnostic of such thermal conditions. The {\CIII}~$977$ line appears strong, and is covered by the G185M NUV grating of COS and the E230M grating of STIS. The STIS spectrum however is at a very low $S/N$ of $\sim 3$ per $10$~{\kms} resolution. A single component free fit to the {\CIII}~$977$ COS feature by adopting the $b$-value of {\CIV} yields $\log~[N(\CIII),~{\cmsq}] = 14.16 \pm 0.20$~{\cmsq}. However, for the saturated line, this solution is not unique. We therefore adopt a conservative lower limit of $\log~N(\CIII) >  13.6$~{\cmsq} obtained from the AOD method. The detections of {\CIII} and {\OIII}, along with non-detection of {\SiIII} point to non-solar [C/Si] and [O/Si] abundances. The other significant non-detection is of {\HeI}~$537$ and $584$ lines, which puts an upper limit of $\log~[N(\HeI),~{\cmsq}] < 13.0$~{\cmsq}. 

The STIS and COS observations combined offer coverage of a number of Lyman series lines. Though the {\Lya} is saturated, the non-detection of higher order Lyman lines yield a measurement of $\log[N(\HI),~{\cmsq}] = 14.68~{\pm}~0.15$ with $b(\HI) = 25 \pm 3~\kms$ obtained from a simultaneous Voigt profile fit to all the uncontaminated Lyman lines. The profile fit was done assuming a single component, as there is no explicit evidence for additional components in the metal lines or the {\HI} that is kinematically coincident with the metals. We also fit the additional three components seen in {\Lya} from $+125$ to $+350$~{\kms} as shown in Figure \ref{fig:1094fit}. The non-detection of these components in {\Lyb} serves as a good constraint for fitting these lines. No metals lines are detected at this velocity. The first among these offset components could potentially be tracing warm gas as hinted by its broad b-parameter of $\sim 46$~{\kms}. The absence of metal absorption at these offset velocities limits us from exploring such a possibility further. The well detected oxygen lines all have profiles similar to that of {\HI} indicating that they could all be tracing a kinematically homogeneous medium. The centroid of metal lines and {\HI} obtained from fitting are also within $|\Delta v| \lesssim 15$~{\kms} of each other, with the differences primarily coming from the lower S/N STIS spectra compared to HIRES and COS (see Figure \ref{fig:1094nav}). The {\HI} column density is comparable to the {\OIV} column density, pointing at high [O/H] abundance, as the photoionization models later reveal. The $b$-values of {\HI} and metal ions including their $1\sigma$ uncertainty suggest the temperature in the absorber to be in the range $T = (0.3 - 3.6) \times 10^4$~K.

\renewcommand{\thefigure}{5A}
\begin{figure}
	\includegraphics[clip=true, trim=4cm 0cm 1cm 2cm, scale=0.64]{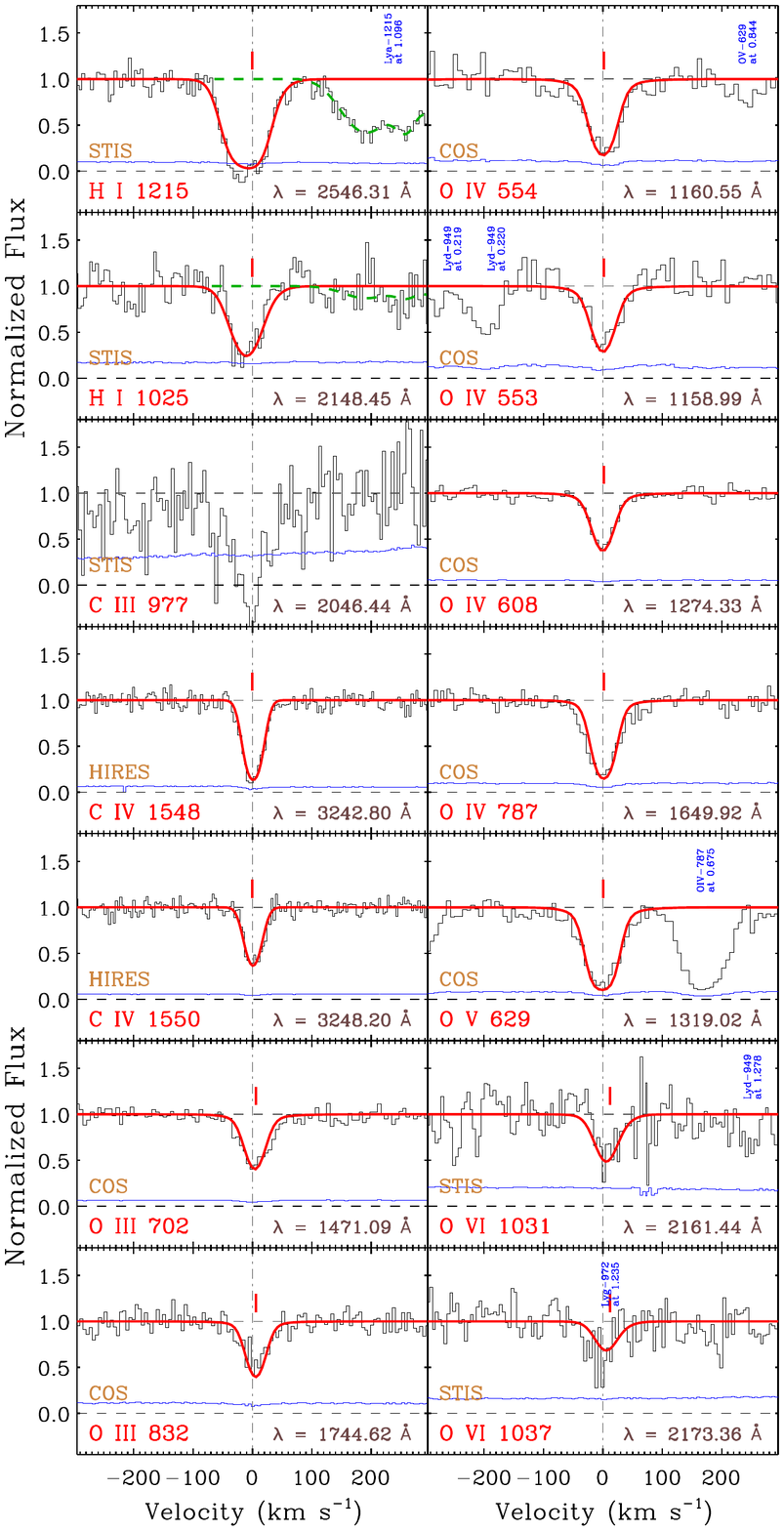}
	\vspace{-1.4cm}
	\caption{System plot of the z$_{abs}$=1.09457 absorber, with continuum-normalized flux along the Y-axis and the velocity scale relative to the redshift of the absorber along the X-axis. The $v = 0$~{\kms}, marked by the \textit{dashed-dotted} vertical line, indicates the absorber redshift. The $1\sigma$ uncertainty in flux is indicated by the \textit{blue} curve at the bottom of each panel. The \textit{red} curves are the best-fit Voigt profiles. The observed wavelength of each transition is also indicated in the respective panels. Interloping features unrelated to the absorber are also labeled. Due to the low $S/N$, a satisfactory Voigt profile model is not obtained for {\CIII}~977. The dashed green curves in the {\HI}~1215 {\AA} and {\HI}~1025 {\AA} panels show a three component fit of the corresponding {\HI} lines at z $\sim 1.096$.} \label{fig:1094fit}
\end{figure}

\renewcommand{\thefigure}{5B}
\begin{figure*} 
     \centerline{\vbox{\centerline{\hbox{\hspace{2cm}\includegraphics[clip=true, trim=0cm 0cm 0cm 1cm,scale=0.4]{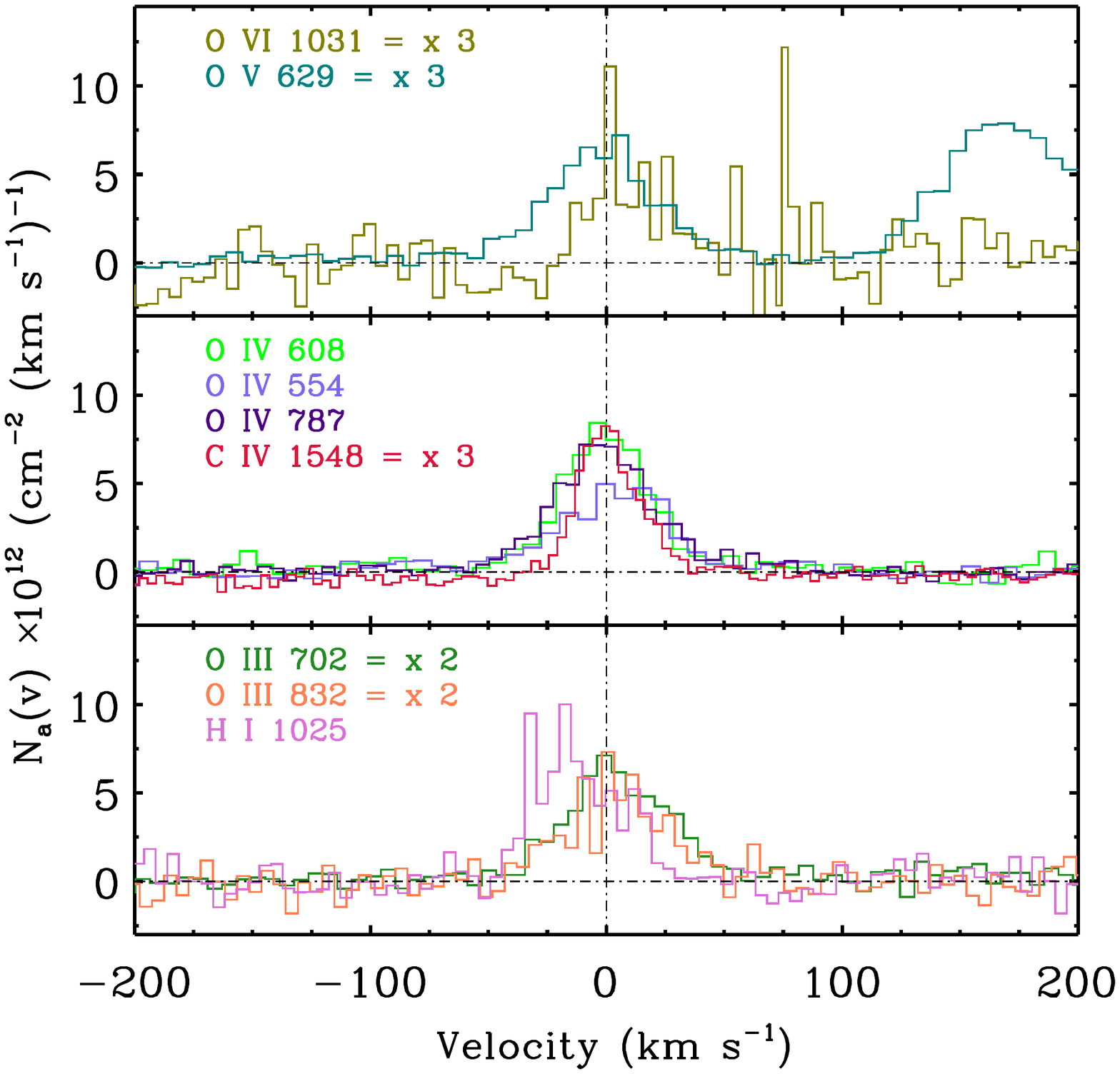} \includegraphics[clip=true, trim=7.5cm 0cm 2cm 0cm,scale=0.5]{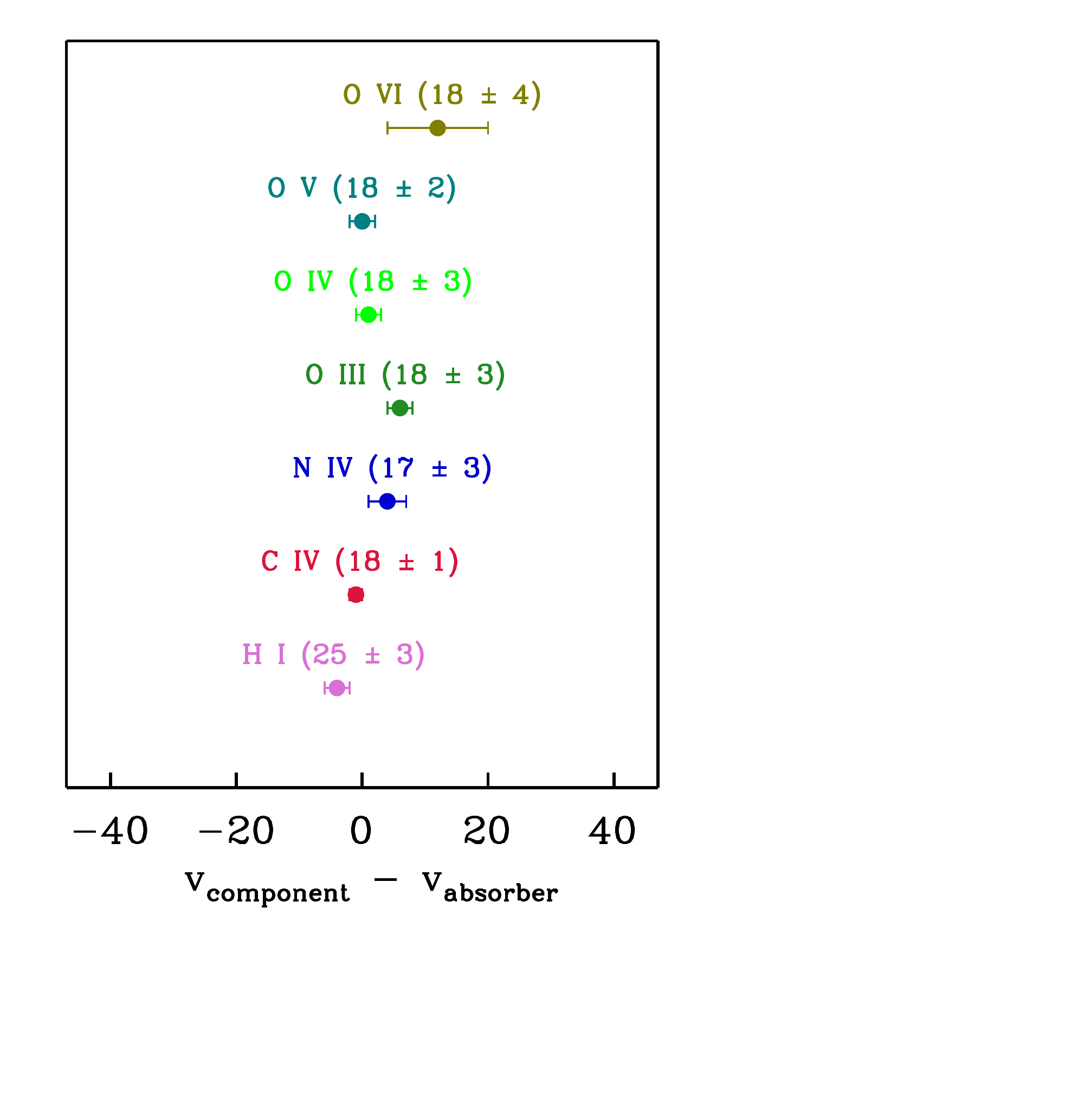}
}}
}} \vspace{-2cm}
     \caption{Apparent column density comparison of different detected species of the absorber at z$_{abs} = 1.09457$. The high ions are plotted on the top panel and the intermediate ions on the bottom panel. The figure on the right shows the velocity centroid of each line derived from profile fitting relative to the redshift of the system. The $b$-parameter measured for the various ions are indicated within the parentheses.}
     \label{fig:1094nav}
\end{figure*} 

\subsection{Ionization Models for the z$_{abs} = 1.09457$ absorber}
\label{sec:1094M}

 Information on five successive ionization stages of oxygen in this absorber offers robust constraints on the density, independent of metallicity as well as the {\HI} column density. In Figure \ref{fig:1094model}, the observed column density ratios between the different ionization stages of oxygen are compared with the model predictions. The $N(\OIII)/N(\OIV) = -0.46~{\pm}~0.08$, the $N(\OIV)/N(\OV) = 0.06~{\pm}~0.19$, and the $N(\OV)/N(\OVI) = 0.6~{\pm}~0.25$ are valid for densities within the range of $n_{\H} = (0.3 - 1.1)~\times~10^{-3}$~{\cc}. It is unrealistic to expect a single uniformly photoionized medium to simultaneously explain five ionization stages of oxygen. Assuming that the kinematically coincident components are also co-spatial, the range may be suggestive of the narrow density-temperature differences that could be present in the absorber at scales unresolved. At the spectral resolutions of the data, such small-scale inhomogeneity will get smoothed-out. The measurements only reveal integrated column densities along the line of sight. Unlike the previous two absorbers, there is no explicit evidence for the {\OVI} to be tracing a separate warm phase. 
 
 The oxygen abundance is constrained primarily by $N(\OIII)$. The observed $N(\OIII)$ cannot be justified for [O/H] $\leq +0.2$ at any density. For [O/H] $= +0.2 \pm 0.2$, the PIE models predict column densities of {\OIII}, {\OIV}, {\OV} and {\OVI} that agree with their respective observed values within the density range given by the models based on the observed column density ratios. The uncertainty in metallicity is a cumulative of the measurement errors in {\HI} and {\OIII}. The narrow density range also predicts the observed {\CIII} to {\CIV} and {\NIII} to {\NIV} column density ratios. The {\CIV} column density, which is very well measured from the doublet lines in the HIRES spectrum, require [C/O] = $-0.2 \pm 0.1$ for an origin from the same phase as the oxygen. For the same reason, the observed $N(\NIV)$ require a [N/O] = $-0.4 \pm 0.4$. The non-detections of Ne, Mg, Si and S are also consistent with the predictions from the ionization models. For the density range of $n_{\H} = (0.3 - 1.1)~\times~10^{-3}$~{\cc}, the PIE models predict a total hydrogen column density in the range of $\log[N(\H),~{\cmsq}] = 18.61 - 17.89$, a gas temperature and pressure of $T = (1.1 - 1.7) \times 10^4$~K and $p/K = (4.8 - 12.3)$~{\cc}~K, over a path length of $L = (4.7 - 0.2)$~kpc. The temperature is within the $(0.3 - 3.6) \times 10^4$~K range predicted by the separate $b$-values of {\HI} and metal lines. 

The non-detection of the {\HeI}~584 line can be used to establish an upper limit on the He abundance in the absorber. At the mean density of $n_{\H} = 0.7  \times 10^{-3}$~{\cc}, the neutral fraction of H and He are $f_{\HI} = 3.7 \times 10^{-4}$, and $f_{\HeI} = 8.6 \times 10^{-5}$ respectively. Applying these ionization corrections to $N(\HeI) < 10^{13.0}$~{\cmsq}, and $N(\HI) = 10^{14.68}$~{\cmsq}, we estimate a He abundance of $y_p = $ N(He)/N(H) $< 0.0892 $, and a He mass-fraction of $Y_p = 4y_p/(1 + 4y_p) < 0.26304$. The mass-fraction thus obtained is consistent with the $Y_p = 0.24672~{\pm}~0.00017$ given by primordial nucleosynthesis models (\citealt{pitrou_precision_2018}), and CMB anisotropy measurements (\citealt{planck_collaboration_planck_2016}). Quasar absorption line measurements of gas with near-primordial chemical composition estimate the [He/H] mass fraction as $0.25^{+0.033}_{-0.025}$ \cite{cooke_measurement_2018}. 

\renewcommand{\thefigure}{5C}
\begin{figure} 
     \includegraphics[scale=0.53,clip=true, trim=2.6cm 0cm 0cm 2.6cm]{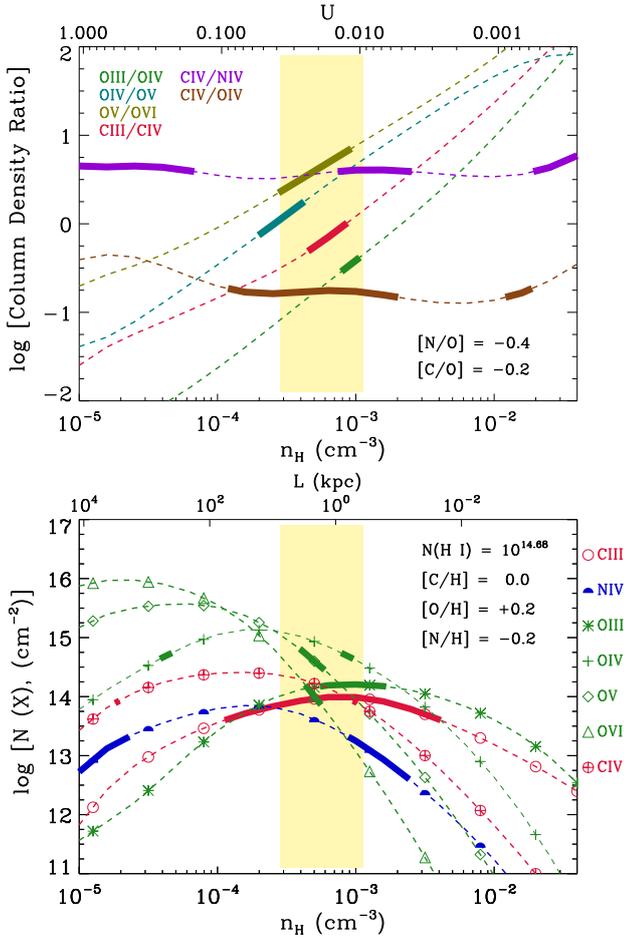}\vspace{-1.2cm}
     \caption{Photoionization equilibrium model for the z$_{abs}$ $=$ 1.09457 absorber. On the \textit{left} panel is the prediction for the ionic column density ratios (\textit{thin} lines) and their observed values (\textit{thick} lines) plotted for various gas number densities. The ratio between successive ions of oxygen limits the plausible gas density to a narrow range indicated by the \textit{yellow} strip. On the \textit{right} panel is the model-predicted column densities of various species (thin lines) and their observed values (thick lines). The density range allowed by the column density ratios is represented by the yellow region. The elemental abundances in the models are varied to reproduce the observed column densities from similar gas phase. The ionization parameter (u) and line of sight thickness for a given density are given by the \textit{top} X-axis of the panels.}
     \label{fig:1094model}
\end{figure}

\section{The z$_{abs} = 1.16592$ Absorption System}
\subsection{Characterization of the z$_{abs} = 1.16592$ absorber}
\label{sec:1165D}

The system plot for the key transitions in the absorber is shown in Figure \ref{fig:1165fit} (the full version is Appendix Figure B4), and the AOD and Voigt profile fit measurements are listed in Appendix Tables B7 and B8. Information on five different ionization stages of oxygen are available for this absorber. The {\OIII}, {\OIV}, {\OV} and {\OVI}~$1031$ are $> 3\sigma$ detections, whereas {\OII} is a non-detection. The {\OVI}~$1037$ line predicted from a fit to the $1031$ line is weak and consistent with its formal non-detection in the STIS spectrum with $S/N \sim 7$ per $0.05$~{\AA} resolution. The {\OIV} ion is detected in four transitions. The {\OIV}~$554$ line is contaminated by the interstellar {\NI}~$1200$. The integrated apparent column densities of the other three uncontaminated {\OIV} lines are within $0.1$~dex of each other indicating that unresolved saturation is nearly absent. The {\OV}~$629$ also has only mild levels of unresolved saturation at the line core as revealed by its $0.07$~dex lower integrated apparent column density compared to the profile fit. The non-detections of high ionization state ions such as {\NeVIII}, and {\MgX} indicate the absence of warm-hot temperatures ($T > 10^{5.5}$~K) and high degree of ionization characteristic of collisionally ionized gas.

The {\OIV} lines show asymmetry in the negative velocity portion of their profiles, which is more evident in {\OV}~$629$ (see $N_{a}(v)$ comparison in Figure \ref{fig:1165nav}). The excess absorption is also seen at a lesser significance in the {\CIVdblt} lines detected by HIRES. A two component model was therefore adopted to model the metal lines. The mild excess absorption at the negative velocity edge is not distinctly evident in the {\OVIdblt}, {\CIII}~$977$ and {\SiIII}~$1206$ lines covered by the noisier STIS data. For the {\OVI} a two component model still yields a satisfactory fit, but for {\CIII} and {\SiIII} a single component was adequate. 

The Lyman series absorption lines from {\Lya} to {\HI}~$930$ covered by STIS are strong saturated lines. The COS coverage of Lyman transitions continue from {\HI}~$918$, which are weak and the higher order transitions are non-detections. The $S/N$ of STIS spectra is inadequate to know the exact kinematic substructure in {\HI}. Taking a cue from the metal lines detected by COS, we fit the {\HI} lines  simultaneously with a two component model. The contribution from the two components is separately shown in the system plot of Figure \ref{fig:1165fit}. The free-fit positions the two {\HI} components at the same velocity as the components seen in the unsaturated metal lines. The bulk of the absorption is dominated by the central component. The $v \sim -25$~{\kms} component has an {\HI} column density that is two orders of magnitude lower. 

The metal lines all have similar $b$-values in each component of the absorption (see \textit{right} panel Figure \ref{fig:1165nav}). For the negative velocity component, $b(\HI) = 29~{\pm}~3$~{\kms} and $b(\OV) = 22~{\pm}~7$~{\kms} imply metal line broadening that is significantly non-thermal with a temperature upper limit of $T < 5.2 \times 10^4$~K (considering the $1\sigma$ limits). For the stronger positive velocity component, the $b(\HI) = 21~{\pm}~1$~{\kms} and $b(\OV) = 10~{\pm}~1$~{\kms} result in $T = (1.8 - 2.6) \times 10^4$~K. 

The COS G130M spectrum covers {\HeI}~$584$ and $537$ lines, both of which are detections. The {\HeI}~$584$ line is contaminated from $-70 \lesssim v \lesssim -30$~{\kms}, and hence these particular pixels were de-weighted while profile fitting. The apparent column density comparison between these two lines shown in \textit{bottom} row of Figure \ref{fig:1165nav} shows unresolved saturation in the line core for the stronger {\HeI}~$584$ line. By adopting the method given by \cite{savage_analysis_1991}, the AOD measured column density of {\HeI} will be $\log~N_a(\HeI) = \log~N_a(\HeI)~537~+~\Delta \log~N_a = 14.36 + 0.22 = 14.58$, which is consistent with the $\log N(\HeI) = 14.48~{\pm}~0.17$ obtained from simultaneous Voigt profile fitting of the two lines. The detection of {\HeI} constrains the density in the absorber as discussed in the next section \ref{sec:1165M}.  

\renewcommand{\thefigure}{6A}
\begin{figure*} 
	\centering
	\includegraphics[scale=0.8,clip=true, trim= 0cm 0cm 0cm 1cm]{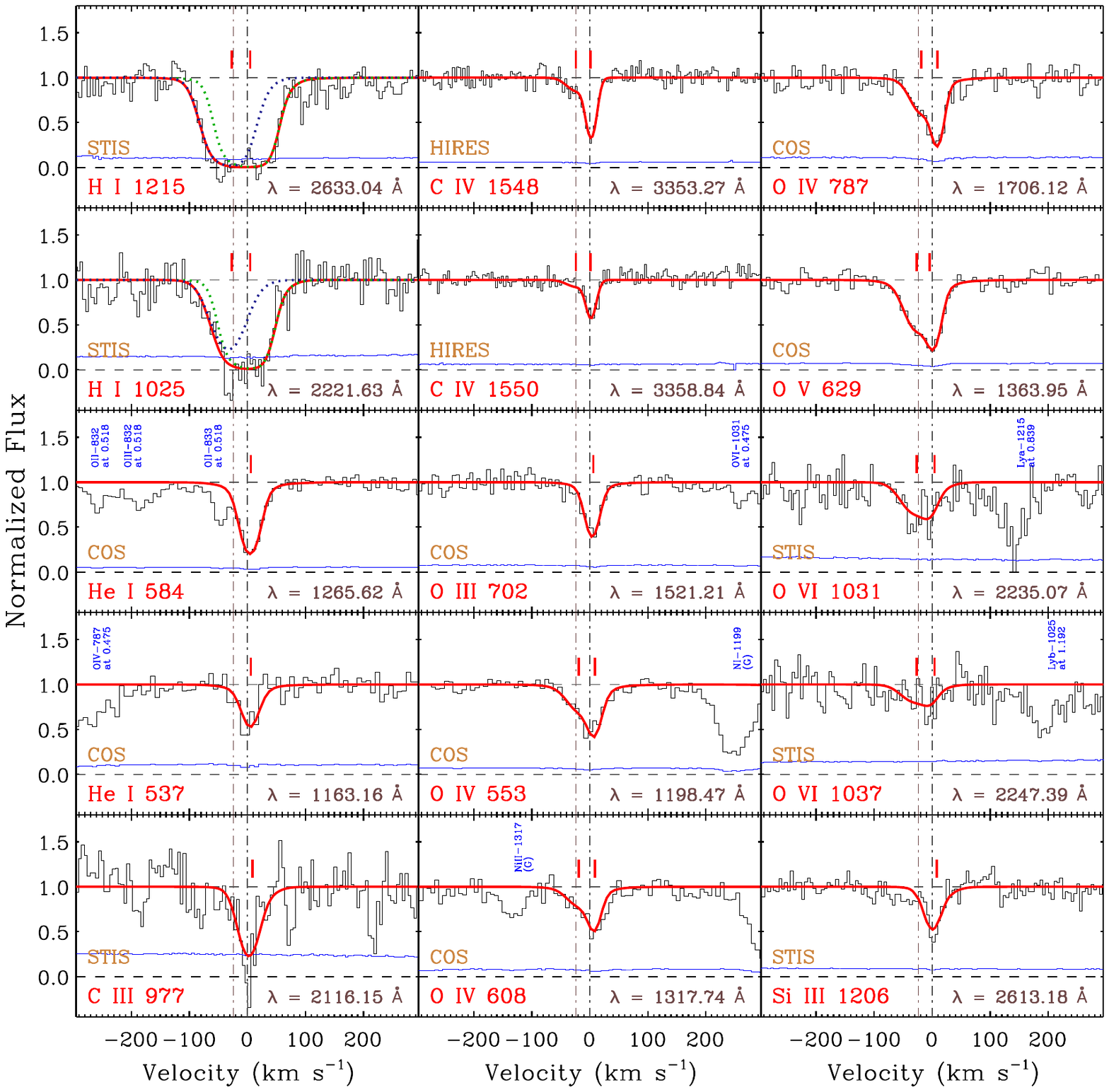}
	\vspace{-9cm}
	\caption{System plot of the z$_{abs}$=1.16592 absorber, with continuum-normalized flux along the Y-axis and the velocity scale relative to the redshift of the absorber along the X-axis. The $v = 0$~{\kms}, marked by the \textit{dashed-dotted} vertical line, indicates the absorber redshift. The $1\sigma$ uncertainty in flux is indicated by the \textit{blue} curve at the bottom of each panel. The \textit{red} curves are the best-fit Voigt profiles. The observed wavelength of each transition is also indicated in the respective panels. Interloping features unrelated to the absorber are also labeled. In the Lyman Panels, marked in green and blue are the two components of {\HI}. \label{fig:1165fit}}
\end{figure*}

\renewcommand{\thefigure}{6B}
\begin{figure*} \vspace{-0.6cm}
\centerline{\vbox{\centerline{\hbox{\hspace{2.5cm}\includegraphics[scale=0.4]{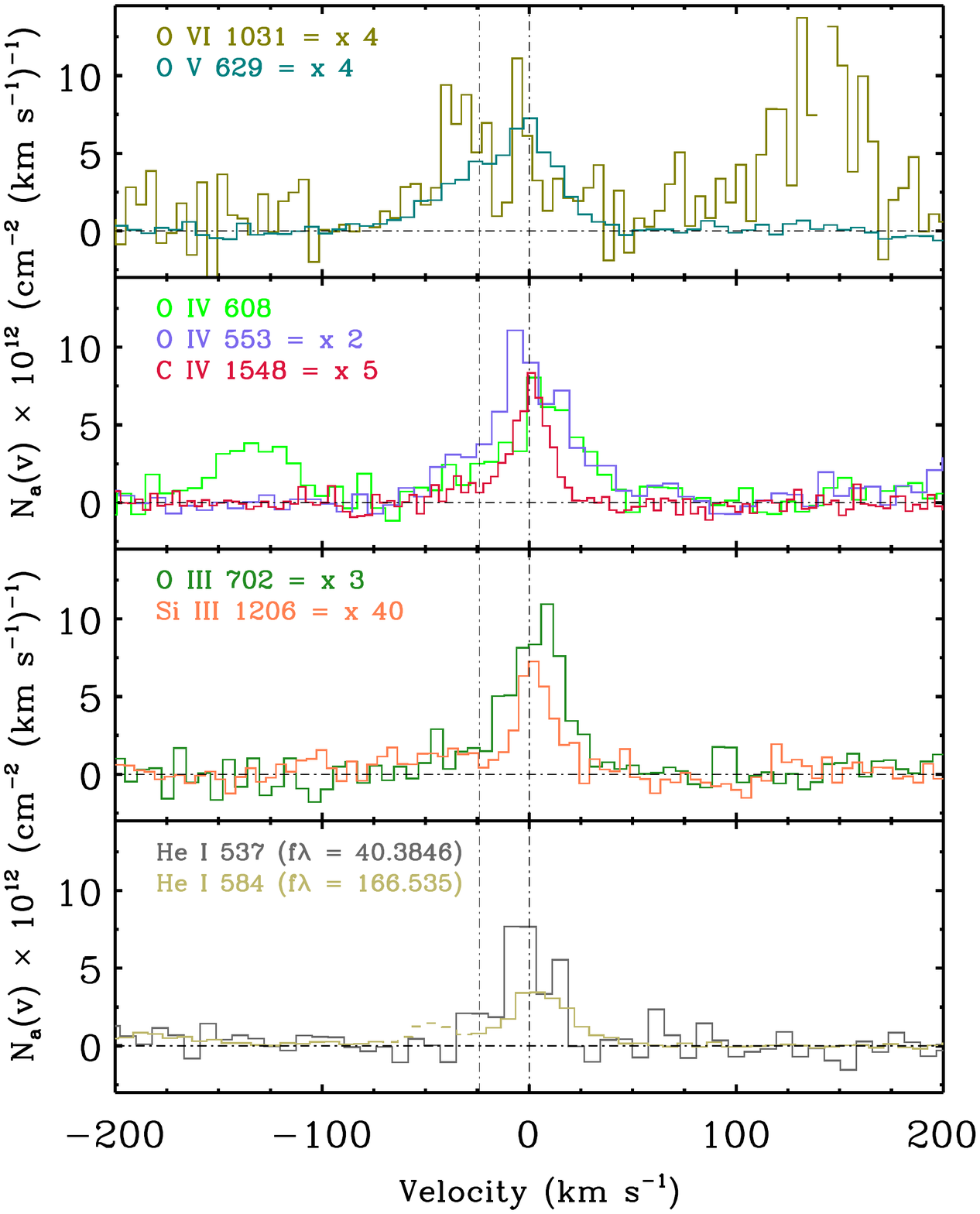} 
\includegraphics[clip=true,trim=8cm 0cm 4cm 0cm,scale=0.52]{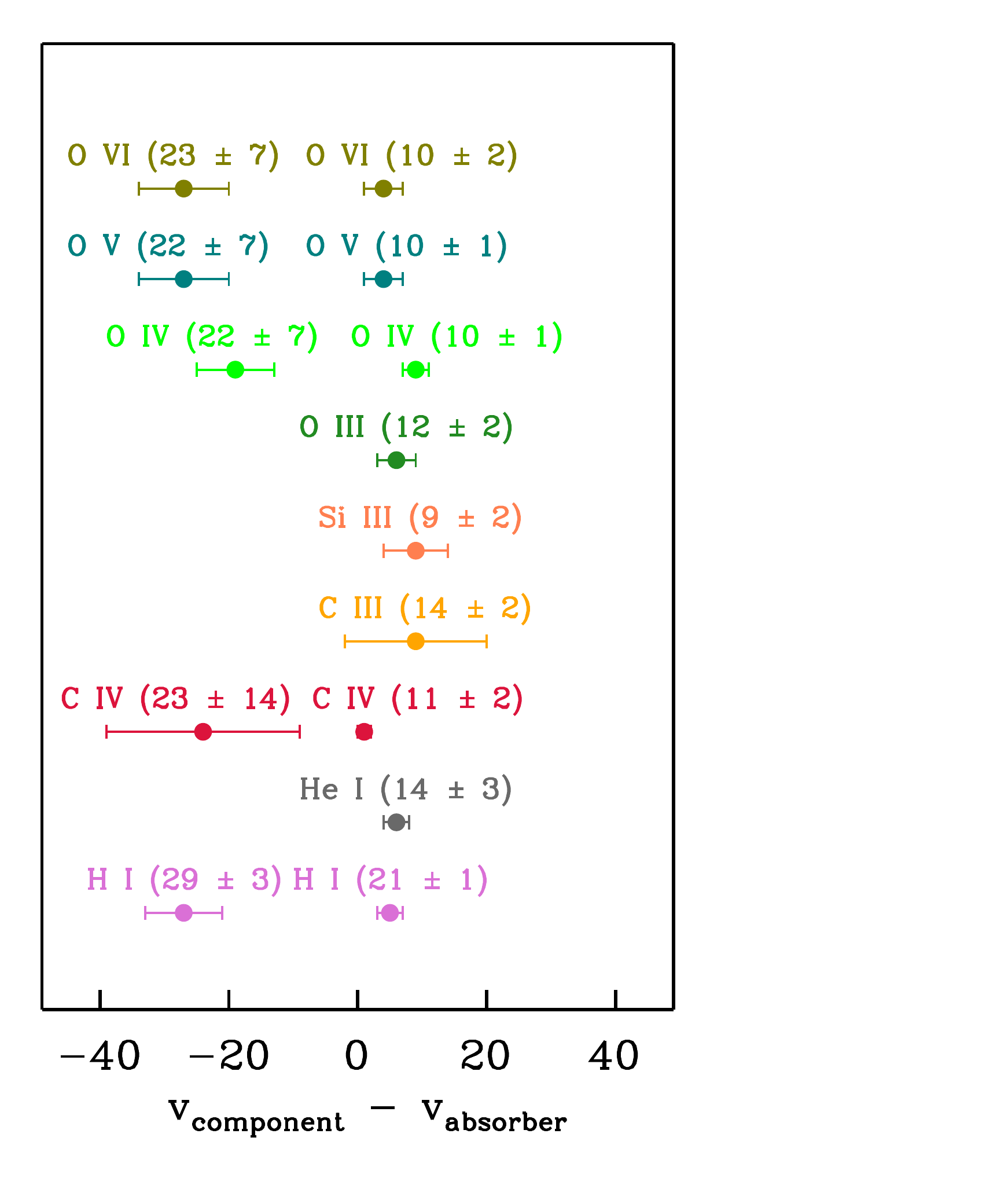}
}}
}} \vspace{-0.6cm}
\caption{Apparent column density comparison of different detected species of the absorber at z$_{abs} = 1.16592$. The high ions are plotted on the top panel and the intermediate ions on the bottom panel. The figure on the right shows the velocity centroid of each line derived from profile fitting relative to the redshift of the system. The $b$-parameter measured for the various ions are indicated within the parentheses.}
     \label{fig:1165nav}
\end{figure*} 

\subsection{Ionization Models for the z$_{abs} = 1.16592$ absorber}
\label{sec:1165M}

The two components of the absorption are modeled separately. The photoionization curves are shown in Figure \ref{fig:1165model}. The observed {\OIII} to {\OIV}, and {\OIV} to {\OV} column density ratios for the central component are recovered by the PIE models over the density range $n_{\H} = (0.4 - 2.5) \times 10^{-3}$~{\cc}. The {\OV} to {\OVI} ratio require slightly lower densities. Similar to the previous absorber, the range in density can be suggestive of mildly different gas phases that are kinematically adjacent contributing to the absorption. The constraints are however inadequate for modeling such a scenario with multiple phases. 

An [O/H] $= -2.1 \pm 0.2$ is the minimum oxygen abundance required to produce {\OIII} at the density of $n_{\H} = 0.95 \times 10^{-3}$~{\cc} corresponding to its peak ionization fraction. The uncertainty in metallicity is a cumulative of the measurement errors in {\HI} and {\OIII}. The same oxygen abundance also produces the observed {\OIV}, {\OV} and {\OVI} within a density range of $n_{\H} = (0.4 - 2.5) \times 10^{-3}$~{\cc} as shown in Figure \ref{fig:1165model}.  The observed column density of {\SiIII} is also recovered within this density range for [Si/H] $= -2.0 \pm 0.25$. The observed {\CIV} however requires [C/H] $= -2.4 \pm 0.15$ for the {\CIV} to be originating from the same environment as the oxygen ions. For the density range of $n_{\H} = (0.4 - 2.5)~\times~10^{-3}$~{\cc}, the models predict a total hydrogen column density in the range $\log N(\H) = 20.60 - 19.70$~{\cmsq}, a gas temperature and pressure of $T = (2.3 - 3.4)~\times~10^4$~K and $p/K = (14.7 - 57.5)$~{\cc}~K, and path length thickness of $L = (296.2 - 6.5)$~kpc. The temperature given by the model agrees with the $T = (1.8 - 2.6) \times 10^4$~K from the $b$-values of {\HI} and metal lines. 

A more binding constraint on the density can be derived by equating the (He/H) abundance in the absorber to its primordial value. From the photoionization model we find that at $n_{\H} = 0.4~\times~10^{-3}$~{\cc} ($\log~[n_{\H},~{\cc}] = -3.36 \pm 0.25$), the He and H ionization corrections of $f_{\HeI} = 8.95~\times~10^{-6}$ and $f_{\HI} = 8.27~\times~10^{-5}$ result in (He/H) $= 0.0843$ which is close to the primordial (He/H) $= 0.085^{+0.015}_{-0.011}$. The density thus obtained, which is within the range predicted by the photoionization models of the other ions, yield $\log N(\H) = 20.6$~{\cmsq}, $T = 3.4~\times~10^4$~K, $p/K \sim 14.7$~{\cc}~K, and $L = 296.2$~kpc. 

The inferred length scale points at the absorption coming from a photoionized phase occupying bulk of the CGM of a massive galaxy (\citealt{muzahid_probing_2014}, \citealt{prochaska_probing_2011}). A direct example of galaxies possessing such a widely extended envelope of diffuse gas has come from the pair QSO sightline observations by \citet{muzahid_probing_2014} in which they arrived at a size of $\sim 330$~kpc for the photoionized {\OVI} absorbing halo around a $1.2$ L$^{*}$ galaxy. The absorption along either sightlines with a transverse separation of $280$~kpc suggested a nearly coherent medium of high ionization gas with a photoionization model predicted line of sight length scale of $\sim 700$~kpc. An alternative to such a large-scale CGM origin is for the absorption to be tracing a high column density node in the network of intergalactic filaments and sheets, which typically possess a thickness of several hundred kpc (\citealt{zhang_physical_1998}). The estimated low metallicity of $\sim 1/100$-th of solar is consistent with such an origin as well.

The blueward offset component with log [N(\HI)/\cmsq] = $14.69 \pm 0.16$ requires a higher metallicity of [X/H] $= -1.2 \pm 0.4$ for the observed {\CIV}, {\OIV}, {\OV}, and {\OVI} to be produced in the density range of  $n_{\H} = (1 - 6.3)~\times~10^{-4}$~{\cc}. The models predict a total hydrogen column density in the range $\log N(\H) = 19.42 - 18.50$~{\cmsq}, a gas temperature and pressure of $T = (4.1 - 2.9)~\times~10^4$~K and $p/K = (4.1 - 18.2)$~{\cc}~K, and path length thickness of $L = (85.8 - 1.6)$~kpc. The temperature given by the model agrees with the upper limit of $T < 5.2 \times 10^4$~K obtained from the $b$-values of {\HI} and metal lines.

\renewcommand{\thefigure}{6C}
\begin{figure*} 
     \centering
     \vspace{-1cm}
     \includegraphics[scale=0.85,clip=true,trim=0.5cm 11.5cm 0cm 0cm]{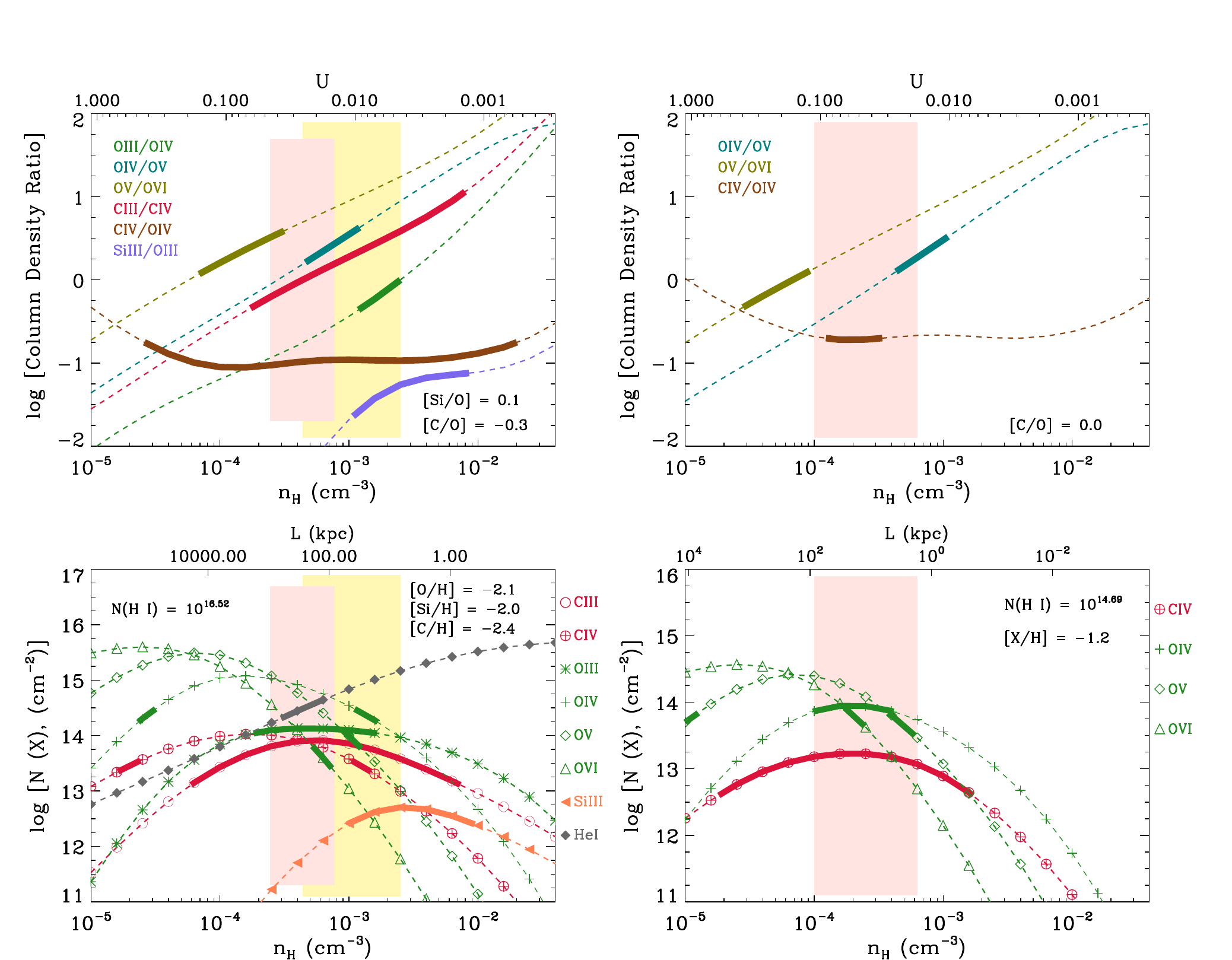}
    \caption{Photoionization equilibrium model for the two components of z$_{abs}$ $=$ 1.16592 absorber towards PG1522+101. The left panels display the PIE model for the red component and the right panels display the PIE model for the blue component. The model of the red component assumes an abundance of $-1.8$ solar for all elements and the model of the right component assumes an abundance of $-1.2$ solar for all elements. On the \textit{top} is shown the model predicted (\textit{thin} lines) and the observed (\textit{thick} lines) column density ratios of successive ionization stages of the same element. This ratio, independent of metallicity, constrains the gas density in the absorber to a narrow range indicated by the \textit{yellow} strip. On the \textit{bottom} panels are the model-predicted column densities of various species (thin lines), along with their observed values (thick lines), plotted against gas density. The density range allowed by the column density ratios is represented by the yellow region. The elemental abundances in the models are varied to reproduce the observed column densities (as many as possible) from a single phase. The ionization parameter (u) and line of sight thickness for a given density are given by the \textit{top} X-axis of the panels. On the \textit{left bottom} panel are the range of abundances permitted by the photoionization models for the various elements. The \textit{pink} zone shows the density range for which the primordial He/H can be recovered.}
    \label{fig:1165model}
\end{figure*}

\section{The z$_{abs} = 1.27768$ Absorption System}
\subsection{Characterization of the z$_{abs} = 1.27768$ absorber}
\label{sec:1277D}

The system plot covering important transitions is shown in Figure \ref{fig:1277fit} (the full version is Appendix Figure B5) and the line measurements are listed in Appendix Tables B9 and B10. The absorber is displaced by $\Delta v = -6540$~{\kms} from the emission redshift of the quasar, which is only marginally higher than the statistical cut-off of $|\Delta v| \leq 5000$~{\kms} frequently adopted for associated absorbers (\citealt{richards_determining_1999}, \citealt{misawa_census_2007}). In the ionization modelling section ahead (Sec \ref{sec:1277M}), we consider the enhancement of the ionization radiation due to the absorber’s possible proximity to the background quasar.

The STIS spectrum covers {\HI} lines from {\Lya} to {\HI}~$930$. Higher order Lyman lines are covered by COS. The lines higher than {\HI}~$949$ are non-detections. The {\Lyg} is strongly contaminated by {\Lya} from  $z = 0.822$. The unsaturated {\Lyb} shows two velocity components which when fitted simultaneously with {\Lya} gives a total {\HI} column density of $\log~N(\HI) = 14.49~{\pm}~0.19$, which makes it the {\OVI} absorber with the lowest {\HI} column density along this sightline. The two component absorption is distinctly visible in the narrower metal lines. The weak feature coincident with {\HI}~$949$ is most likely contamination as the simultaneous fitting reveals. 

Oxygen is seen in {\OIII}, {\OIV}, {\OV}, and {\OVI} ionization stages. The weak {\OIII} and the multiple {\OIV} lines also indicate a two component structure identical to {\HI}. The $N$ and $b$ of these components in {\OIV} are constrained from a simultaneous fit to the {\OIV}~554, {\OIV}~608, and {\OIV}~787 lines which are unsaturated. The {\CIVdblt} lines in the high resolution HIRES data show the two component structure to be narrow with $b \sim 10$~{\kms} (refer to the $N_a(v)$ comparison in Figure \ref{fig:1277nav}). The intermediate and high ionization metal lines such as {\OIV}, {\OV}, {\OVI}, {\NeV}, {\NeVI}, and {\SVI} seen at lower $S/N$ and/or resolution by COS and STIS are also likely to possess narrow $b$-values if they are tracing the same gas phase. Indeed, the free-fit done for {\OIV} does yield such a result. 

The positive velocity side of the {\OV}~$629$ is contaminated by {\NIII} from the $z = 1.09457$ absorber (refer Appendix Figure B3). The contaminated region was excluded during profile fitting. The {\OV} line is saturated at the line core as shown by the $0.5$~dex lower column density from the AOD integration compared to the profile fitting. We applied a two component profile for the saturated {\OV} by simultaneously fitting it with the well measured {\OIV}. Even so, the uncertainty in {\OV} column density is likely to be larger than what the profile fitting routine yields. The {\OVI}~$1031$ line is strong and saturated. A two component model was used to fit the {\OVI}~$1031$ line by adopting the same $b$ value as {\OIV} for the stronger and saturated component. The effect of line saturation is reflected in the uncertainty in the column density which comes out as $\sim 0.8$~dex. The $1037$~{\AA} line of the {\OVI} doublet is heavily blended with {\Lya} at $z = 0.945$. The extent of contamination is evident in the system plot (refer Appendix Figure B5) where the expected $1037$~{\AA} profile is shown based on the Voigt model for the {\OVI}~$1031$~{\AA} line. 

The $b$ parameters of the various metal lines are similar to each other to within their $1\sigma$ uncertainty. The profile fit shows the {\HI} lines to be broader in comparison to metal lines. The $b(\H) \sim 25$~{\kms}, $b(O) \sim b(C) \sim 9$~{\kms} imply the temperature to be $T \sim 3.5 \times 10^4$~K in both components. 

Additionally, there is detection of {\NeV}, and {\NeVI} where the same velocity structure is evident. The COS spectra of these lines show an asymmetry that is consistent with the presence of a second component, albeit at a lower significance compared to the oxygen ions. This is most likely due to the Ne absorption being intrinsically weaker due to the low cosmic abundance of Ne, and also the decline in spectral resolution towards shorter wavelengths. There is also detection of {\SVI~933} in the STIS data, but a sub-component structure is not conspicuous at the lower $S/N$. The important non-detections in the absorber include {\CII}, {\CIII}, {\SiII}, {\NeVIII}, {\MgX} and {\HeI}.

\renewcommand{\thefigure}{7A}
\begin{figure*} 
	\centering
	\includegraphics[scale=0.8,clip=true, trim= 0cm 0cm 0cm 1cm]{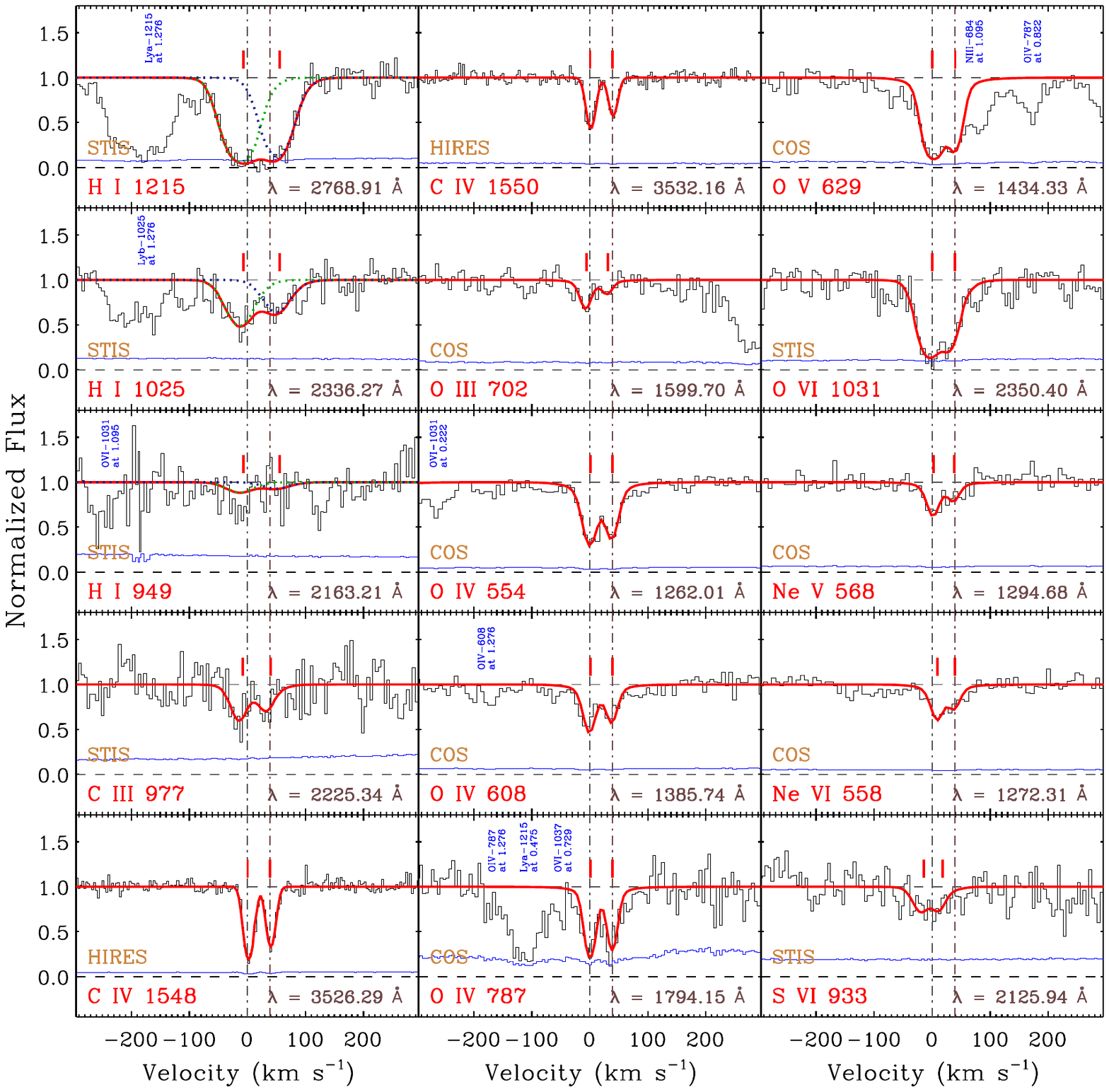}
	\vspace{-9cm}
	\caption{System plot of the z$_{abs}$=1.27768 absorber, with continuum-normalized flux along the Y-axis and the velocity scale relative to the redshift of the absorber along the X-axis. The $v = 0$~{\kms}, marked by the \textit{dashed-dotted} vertical line, indicates the absorber redshift. The $1\sigma$ uncertainty in flux is indicated by the \textit{blue} curve at the bottom of each panel. The \textit{red} curves are the best-fit Voigt profiles. The observed wavelength of each transition is also indicated in the respective panels. Interloping features unrelated to the absorber are also labeled. \label{fig:1277fit}}
\end{figure*}

\renewcommand{\thefigure}{7B}
\begin{figure*}\vspace{-1cm}
     \centerline{\vbox{\centerline{\hbox{\hspace{1.5cm}\includegraphics[clip=true, trim=0cm 0cm 0cm 1cm,scale=0.41]{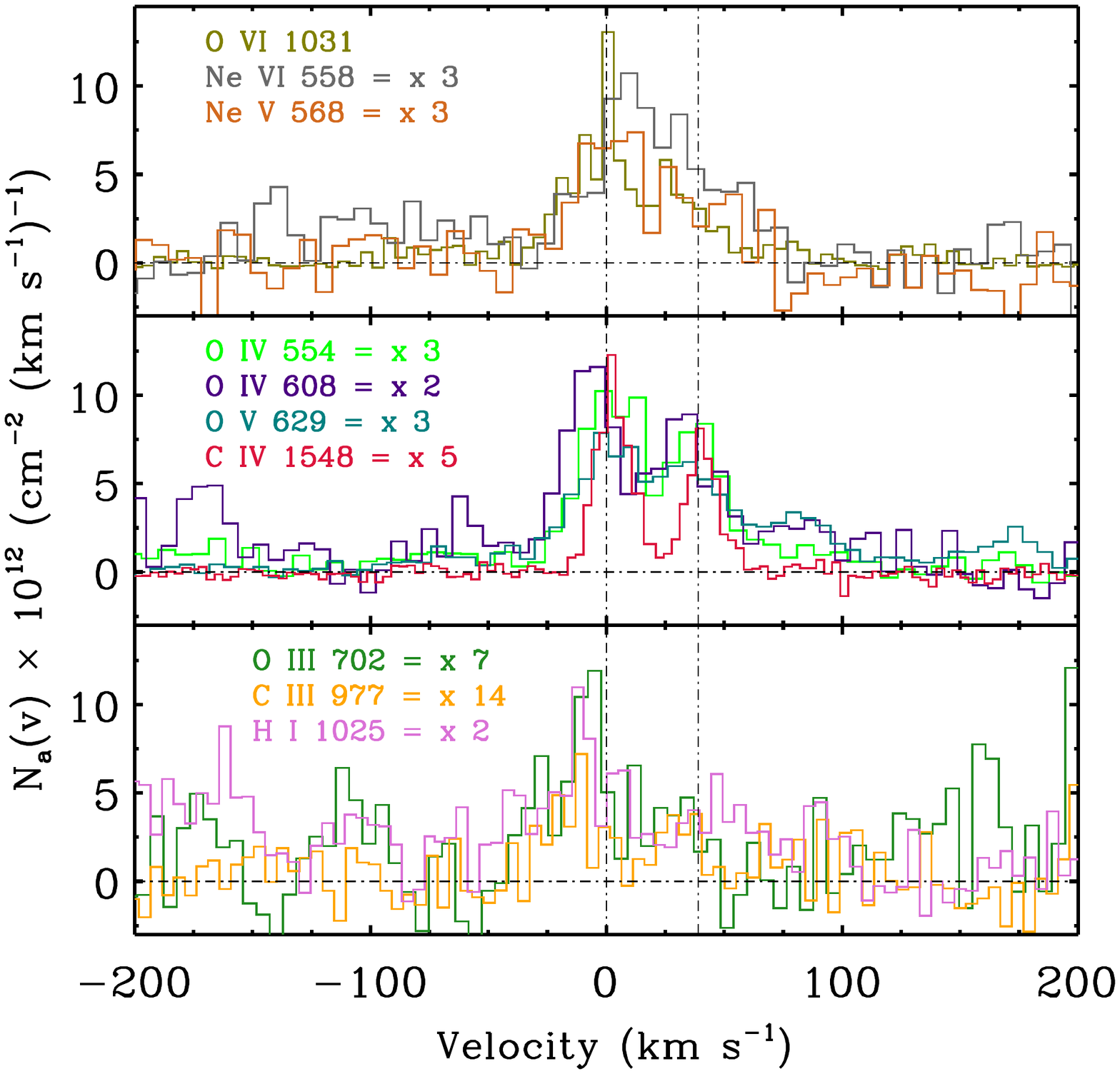} 
\includegraphics[clip=true, trim=6.5cm 0cm 3cm 0cm,scale=0.5]{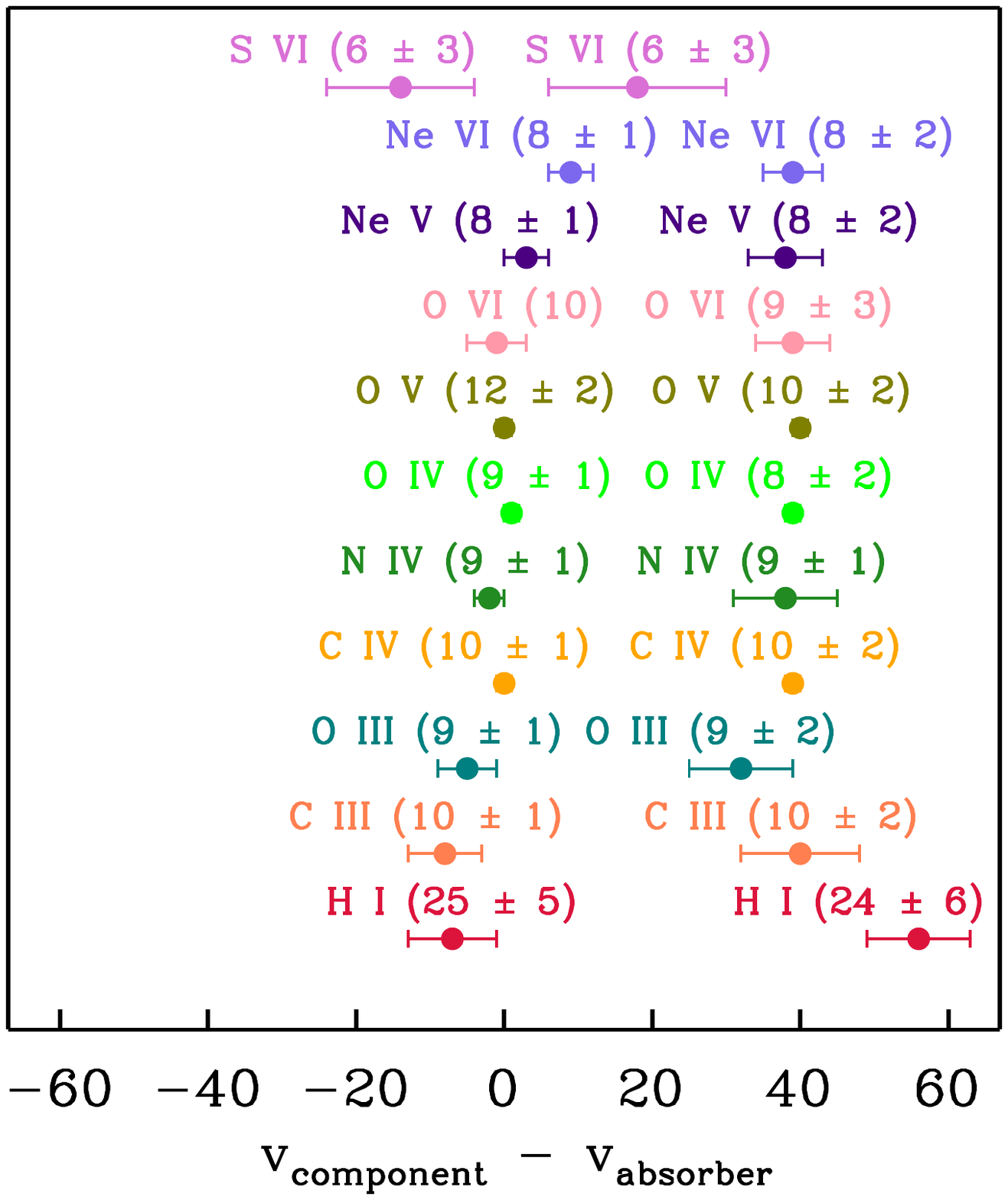}
}}
}} \vspace{-2cm}
     \caption{Apparent column density comparison of different detected species of the absorber at z$_{abs} = 1.27768$. The high ions are plotted on the top panel and the intermediate ions on the bottom panel. The figure on the right shows the velocity centroid of each line derived from profile fitting relative to the redshift of the system. The $b$-parameter measured for the various ions are indicated within the parentheses. The {\SVI} data is very noisy although it appears to have a $15 \kms$ offset. The velocity separation between the two {\HI} components is strongly influenced by the noisy 1025 line.}
     \label{fig:1277nav}
\end{figure*} 

\subsection{Ionization Models for the z$_{abs} = 1.27768$ absorber}
\label{sec:1277M}

The absorber is separated from the background quasar by $\Delta v = -6540$~{\kms}. The difference in redshift between the absorber and the quasar corresponds to a comoving separation of $105$~Mpc.\footnote{estimated using Nick Gnedin’s cosmology calculator for the flat universe  https://home.fnal.gov/~gnedin/cc} We assess whether the radiation from the background quasar plays any role in the ionization by estimating its intensity at $\sim 910$~{\AA} near the absorber redshift. The intrinsic emission from the quasar is modeled by fitting a power-law continuum to the flux calibrated $HST$ spectrum shifted to the quasar rest-frame. Using a flat cosmology model with $\Omega_m =0.3$ and H$_0$ = 70 km s$^{-1}$ Mpc$^{-1}$, we obtain the intrinsic intensity near the Lyman limit to be $j_{\nu} = (5.2 \times 10 ^{-26})~e^{\tau_{\mathrm{eff}}} $~erg~s~$^{-1}$~{\cmsq}~Hz$^{-1}$~sr$^{-1}$ where $\tau_{\mathrm{{eff}}}$ is the effective optical depth between the quasar and the absorber (see Appendix section A for a more detailed calculation). In comparison, the KS19 ionizing background has a $910$~{\AA} intensity of $j_{\nu}^{UVB} = 3.7 \times 10^{-22}$~erg~s~$^{-1}$~{\cmsq}~Hz$^{-1}$~sr$^{-1}$. The quasar ionizing radiation at the location of the absorber is thus several orders of magnitude weaker than the extragalactic UV background and is therefore unlikely to have regulated the ionization in the absorber in any significant way. We therefore proceed with the KS19 model for the background radiation field. 

In the PIE models, the two components of this absorber, separated by $|\Delta v| \sim 40$~{\kms}, are considered separately. The observed $N(\CIII)/N(\CIV)$, $N(\OIII)/N(\OIV)$, $N(\OIV)/N(\OV)$, $N(\OV)/N(\OVI)$ and $N(\NeV)/N(\NeVI)$ are the principal measurements constraining the density. The column density in either component is well determined for all the ions except {\OV}, and {\OVI} which have unresolved saturation and also {\SVI} in which the component structure is less evident because of the absorption being weak and at lower $S/N$. It can be noticed from the PIE column density ratio curves of Figure \ref{fig:1277model} that there is no common density where these five column density ratios exactly overlap, indicating that there exists a multiphase structure. In both components, the {\OIII} to {\OIV}, {\CIII} to {\CIV} and {\OIV} to {\OV} ratios roughly coincide at a density of $n_{\H} \sim 4.5 \times 10^{-4}$~{\cc}, whereas the {\NeV} with {\NeVI} is recovered at $n_{\H} \sim 1.0 \times 10^{-4}$~{\cc} for solar abundance pattern. The {\OV} to {\OVI} ratio is less constraining because of the large uncertainty in both column densities. Yet, the ratio with its $1\sigma$ cumulative error suggests lower densities of $n_{\H} \sim (1 - 10) \times 10^{-5}$~{\cc} if they are tracing photoionized gas. 

The predictions from the PIE models for both components are shown in the \textit{bottom} panels of Figure \ref{fig:1277model}. Rather than a single density, the column densities of all ions are recovered to within their $1\sigma$ uncertainty if we consider the density range of $n_{\H} = (1 - 7) \times 10^{-4}$~{\cc}. It is possible that the absorbing medium may not have uniform ionization throughout. Assuming that the kinematically coincident components are also co-spatial, the line of sight is most likely probing a region with differences in ionization parameter at small physical scales that are unresolved, resulting in absorption from ions of different ionization levels roughly coinciding in velocity space.

Such a photoionized medium requires [O/H] and [Ne/H] to be solar and the [C/H] and [N/H] to be sub-solar. Oxygen and Neon enrichment happens through Type II SN explosions involving $\geq 10$~M$_{\odot}$ stars (\citealt{iwamoto_nucleosynthesis_1999}, \citealt{nomoto_nucleosynthesis_2006}), whereas C is returned into the ISM primarily through mass loss during the AGB phase of intermediate mass stars ($\sim 1 - 8$~M$_{\odot}$) and from Type Ia SNe (\citealt{thielemann_operation_1993}, \citealt{chiappini_oxygen_2003}). The difference in stellar evolution time-scales between massive and intermediate mass stars can give rise to sub-solar [C/O], especially in a medium that is recently enriched. The absorbing gas could be tracing such a medium of high metallicity primarily enriched via contemporaneous Type II SN events. 

Interestingly the {\OVI} is barely explained by the above PIE phase over the wide density range considered, indicating the presence of a distinct higher ionization phase. Due to line saturation, the {\OVI} $b$-parameter is not independently constrained. In deriving the Voigt profile model, we have assumed $b(\OVI) \sim b(\OIV) \sim b(\CIV) \sim 10$~{\kms}, which implies a maximum temperature of $T \sim 9.8 \times 10^4$~K. If the {\OVI} is tracing a phase separate from {\OIV} and {\CIV}, the temperature of that phase will have to be $T < 4 \times 10^5$~K for the {\NeVIII} to remain a non-detection. Such an upper bound on the temperature leads to $b(\OVI) < 20$~{\kms} in either component. 

To summarize, the absorption line information suggests the presence of multiphase photoionized gas having $n_{\H} = (1 - 7) \times 10^{-4}$~{\cc} and [O/H] $\sim 0.0 \pm 0.2$, with relative elemental abundances favouring Type II SNe enrichment history. The uncertainty in metallicity is a cumulative of the measurement errors in {\HI} and {\OIII}. No firm conclusions can be drawn on the origin of {\OVI} or the properties of gas it traces. The ion could be from a low density ($n_{\H} \lesssim 10^{-5}$~{\cc}) highly photoionized medium, or gas at $T \sim (2 - 4) \times 10^5$~K where collisionial ionizations are dominant.

The multiphase nature and the symmetry between the profiles of the two components raises the possibility that the line of sight is passing through a biconical outflow such as superwind or superbubble shell created by multiple supernovae. Such expanding shells can give rise to warm/hot interface layers between the hot (T $\sim 10^{6} - 10^{7}$~K)  interior of the bubble and the outer shell of cooler (T $\sim 10^{4}$~K) interstellar gas swept-up by the supernovae (e.gs., \citealt{heckman_absorption-line_2000}, \citealt{savage_stis_2001}, \citealt{bond_evidence_2001}), or between hot gas flowing alongside fragments within the shell. The latter scenario was proposed by \citet{heckman_fuse_2001} to explain the absorption created by superwind driven material in the starburst dwarf galaxy NGC 1705. The {\OVI} was hypothesized as due to mixing of the hot outflowing material and the cooler fragments of swept-up material resulting in gas that is initially heated to T $> 3 \times 10^{5}$~K, and subsequently radiatively cooling. The {\OV}, {\OVI}, {\NeV} and {\NeVI} in the z = 1.277 absorber can be from such cooling interface gas with the lower ions tracing less ionized shells or fragments. At T $\lesssim 10^{5}$~K, such gas would produce little {\NeVIII}, consistent with what we observe.

\renewcommand{\thefigure}{7C}
\begin{figure*} 
\vspace{-1cm}
     \centering
      \includegraphics[scale=0.85,clip=true,trim=0.5cm 0cm 0cm 1cm]{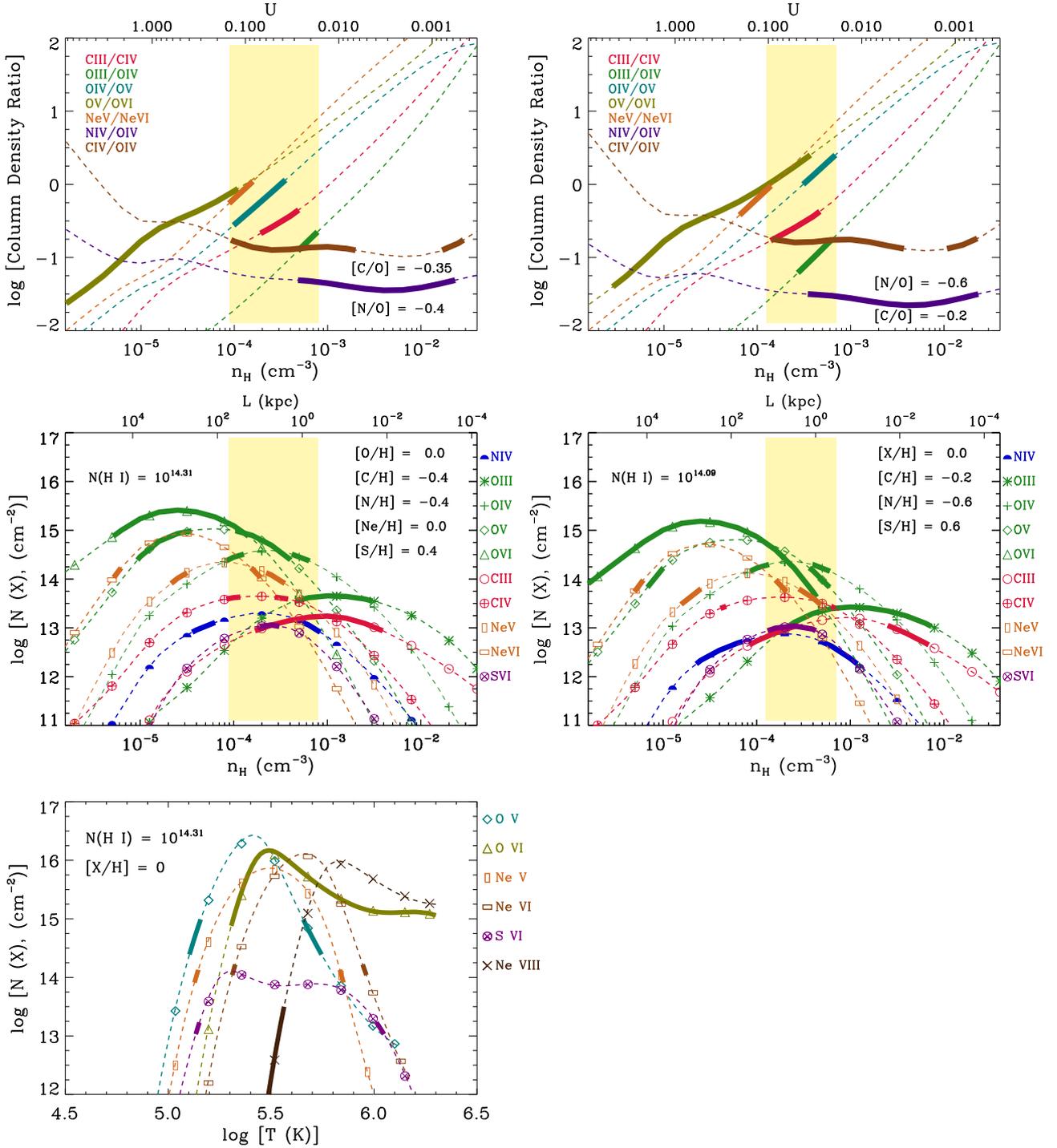}
     \vspace{-2.5cm}
     \caption{Photoionization equilibrium model for both components of the z$_{abs}$ $=$ 1.27768 absorber towards PG1522+101. The left panels display the PIE model for the blue component and the right panels display the PIE model for the red component. On the \textit{top} is shown the model predicted (\textit{thin} lines) and the observed (\textit{thick} lines) column density ratios of successive ionization stages of the same element. This ratio, independent of metallicity, constrains the gas density in the absorber to a narrow range indicated by the \textit{yellow} strip. On the \textit{bottom} panels are the model-predicted column densities of various species (thin lines), along with their observed values (thick lines), plotted against gas density. The density range allowed by the column density ratios is represented by the yellow region. The elemental abundances in the models are varied to reproduce the observed column densities (as many as possible) from a single phase. The line of sight thickness for a given density is given by the \textit{top} X-axis. On the \textit{left bottom} panel are the range of abundances permitted by the photoionization models for the various elements.}
     \label{fig:1277model}
     
\end{figure*}

\section{Discussion \& Summary of Results}
\label{sec:DSR}

We report on the physical and chemical properties of five multiphase absorbers from $0.67 - 1.28$ towards the quasar PG~$1522+101$. The $HST$/COS and STIS spectra of this sightline cover several important redshifted FUV and EUV metal absorption lines in addition to Lyman series lines, all of which serve as diagnostic of the gas phases and abundances. The Keck/HIRES spectra provide additional robust constraints through coverage of {\CIV} and {\MgII}. The results of photoionization modelling of the systems are summarized in Table \ref{tab:modsum}.

Surveys of {\OVI} absorbers have previously reported an anti-correlation between $\log~[N(\OVI)/N(\HI)]$ and $\log~N(\HI)$. \citet{savage_properties_2014} explained this trend as due to the {\OVI} in absorber samples spanning a significantly narrow range in column density when compared to {\HI}. A negative correlation is expected from plotting $1/N(\HI)$ against $N(\HI)$. However, it is also seen from Figure \ref{fig:s14l14} that the trend remains largely unaffected even when the column density ratio of {\OVI} to {\HI} is plotted. This happens due to the {\OVI} column density remaining nearly a constant in absorbers spanning seven orders of magnitude in {\HI}. \citet{danforth_low-z_2005} reason that this would be the case if {\OVI} and {\HI} are tracing multiple gas phases. The {\OVI}, in their empirical model, comes from a shell-like ionized structure surrounding a low ionization cloud possessing bulk of the {\HI}. The arbitrary size of the neutral cloud would imply that the {\HI} column densities will be different between absorbers while ionizations by energetic photons and electrons would result in a highly ionized boundary layer of similar {\OVI} column densities across absorbers. 

As shown in Figure \ref{fig:s14l14}, the {\OVI} to {\HI} column density ratio in our sample follows the anti-correlation trend with {\HI} similar to previous studies. The consistently well measured $\log~N({\OIII})$ and $\log~N({\OIV})$ sample a combined narrow range of $13.2 - 15.0$ compared to $\log~N(\HI) \sim 14.4 - 16.5$. The {\OVI} in the absorber at z $= 1.27768$ is not well constrained and thus the upper end of the range of $\log~N(\OVI)$ in our sample is not accurately established. We find that both N({\OIII})/N({\HI}) and N({\OIV})/N({\HI}) obey the anti-correlation with N({\HI}) (Spearman's rank correlation $\rho$ = -0.85, p = 0.00023). The non-detections of {\CII}, {\OII}, and {\SiII} indicate that the gas is at intermediate to moderately high ionization levels. The observed anti-correlation trend is therefore more likely due to inefficient mixing of metals with gas. As \citet{schaye_large_2007} argue, metals displaced from galaxies could be confined to patchy zones within larger {\HI} clouds of the CGM or IGM. Observations sampling such regions will yield a wide range for the column densities of {\HI} depending on the absorber environment and the arbitrary line of sight probing it. From ionization models we derive an overabundance of oxygen relative to carbon in all the absorbers with [C/O] in the range of $0.0 \pm 0.4$ to $-0.8 \pm 0.1$ relative to solar suggesting that the gas has preferentially been enriched by feedback from Type II SNe (\citealt{telfer_extreme-ultraviolet_2002}). Despite this, the metallicity being low for the $z = 0.67556$, $z = 0.72885$ and $z = 1.16592$ absorbers can be due to poor mixing of gas and metals. The lowest metallicity absorbers in our sample are indeed partial Lyman limit systems (see Figure \ref{fig:nhirel}). Their low metallicities are consistent with one of conclusions of the COS CGM Compendium (CCC) survey (\cite{wotta_low-metallicity_2016}) that more than half the pLLS systems at $z \lesssim 1$ have [X/H] $< -1.0$. Depending on the amount of neutral hydrogen present along a line of sight, inferred metal abundances can turn out as high or low, for a clumpy distribution of metals. This would also mean that metallicities of CGM gas inferred from moderate resolution absorption line studies cannot be used as a definitive marker to distinguish inflows of gas into galaxies, which are necessarily metal poor, from feedback driven metal-rich outflows. Additional information from the kinematics of the absorbing gas, comparison with the ISM metallicities of galaxies in close proximity, and their orientation with respect to the line of sight are needed to draw conclusions on the physical origin of the absorbing gas (e.g., \citealt{nielsen_magiicat_2015}, \citealt{lan_circumgalactic_2018}, \citealt{kacprzak_relationship_2019}).

Based on the observed trend of declining gas column densities with increasing galaxy impact parameters, \citealt{prochaska_probing_2011} had proposed that strong {\Lya} ($W_r > 300$~m{\AA}) and {\OVI} ($W_r > 70$~m{\AA}) preferentially trace gas linked to the CGM of $L > 0.1L^*$ galaxies, than the intergalactic clouds of the cosmic web. The {\HI} and metal absorption in intervening systems are also often associated with the multiphase CGM of dwarf galaxies, though the halo covering fractions of gas of various ionization are found to be lower in dwarfs compared to more massive systems. \citet{johnson_possible_2015} estimated that as much as 20\% of the {\OVI} absorption at low redshift (z $\sim 0.2$) can potentially be from the CGM of star-forming field dwarf galaxies ($10^{7.7} <$ (M$^{*}/M_{\odot}) < 10^{9.2}$). Similarly, the COS-Dwarfs study found that $\sim 60\%$ of {\CIV} absorbers with W$_{r} > 0.1$~{\AA} are statistically consistent with tracing the CGM of dwarfs (M$^{*} < 10^{10}$ M$_{\sun}$, \citealt{bordoloi_cos-dwarfs_2014}). Such photoionized CGM gas with {\HI} column densities in the sub-Lyman limit ranges ($15.0~< \log~[N/\cmsq] < 18.0$) possess a wide range of metallicities from near-pristine to super-solar, due to the diverse processes that are at work at the interface regions of galaxies with the IGM (\citealt{werk_cos-halos_2014}, \citealt{keeney_characterizing_2017}, \citealt{lehner_cos_2019}, \citealt{wotta_cos_2019}). The absorbers in our sample have metal and {\HI} equivalent widths and column densities that support close association ($\rho <$ R$_{vir}$) with galaxies than the IGM. Among them, the metal-poor absorbers ($z = 0.67556, 0.72885, 1.16592$) could potentially be tracing clouds of large {\HI} column resulting in diluted inferred metal concentrations along the line of sight. The farthest absorber ($z = 1.27768$) has strong and saturated {\OVI} with $\log N(\OVI) \gtrsim 14.6$, and strong {\OV} as well. Such strong {\OVI} absorbers are closely associated with the extended environments of star forming galaxies of wide mass ranges (\citealt{chen_probing_2009}, \citealt{wakker_relationship_2009}, \citealt{tumlinson_large_2011}, \citealt{johnson_possible_2015}, \citealt{bielby_quasar_2019}, \citealt{rudie_column_2019}). 

Straightforward evidence for a warm thermal phase is only present in two out of five absorbers in our sample. The first is the $z = 0.67556$ absorber where the {\OIV} and {\OVI} are displaced in velocity from the {\CIII}, {\OIII}, and {\HI} with the simplest explanation for {\OVI} in this case being collisional ioniziation at $T \sim 2.5 \times 10^5$~K, at which its  ionization fraction peaks. The other is the $z = 0.72885$ absorber where the broader line width of {\OVI} compared to {\OIV}, {\OIII}, and {\HI} clearly indicates a higher phase with a T < $5~\times~10^{6}$~K associated with a very broad BLA that is undetected. In rest of the absorbers, photoionization by the UV background is a feasible mechanism for the production of {\OVI} and lower ionization metals. Here we note that \citet{zahedy_characterizing_2019} had found similar differences in centroid velocities and line widths between {\OVI} and low ions in some of the absorbers identified with the halos of Luminous Red Galaxies (LRGs). Their estimates show that these massive ellipticals (M$_{*} > 10^{11}$ M$_{\sun}$) retain CGM mass comparable to that of star-forming massive galaxies. The density prediction from PIE models based on intermediate and high ions as constraints comes out in the range of $n_{\H} \sim (1 - 25) \times 10^{-4}$~{\cc}. The range indicates that the line of sight is probing a multiphase medium, though the exact density-temperature structure is blurred at the resolution of COS. \citet{roca-fabrega_cgm_2019} show that the ionization of the CGM (and the origin of {\OVI}) is redshift dependant, with photoionization playing a dominant role in the epochs between $0.5 \lesssim z \lesssim 2$ due to the enhanced intensity of the extragalactic UV background following the peak in cosmic AGN activity and star-formation rates. On the other hand, collisional ionization is predominant at the very high ($z \gtrsim 2$) and very low ($z < 0.5$) redshift epochs. Their simulations further show that photoionization tends to dominate the production of {\OVI} in low mass halos (halo virial mass of $M_h \lesssim 10^{10}~M_{\odot}$), and even the outskirts of high-mass galaxies. Thus, it is possible for the photoionized absorbers in our sample to have a circumgalactic origin.   

Finally, we comment on the non-detection of {\NeVIII} in all five absorbers presented here. The presence of {\NeVIII} in {\OVI} absorbers have been decisive in revealing the presence of gas with T~$\sim (0.5 - 1.5) \times 10^6$~K associated with the IGM and the gaseous halos of  luminous galaxies (\citealt{savage_detection_2005}, \citealt{savage_multiphase_2011}, \citealt{narayanan_cosmic_2011}, \citealt{tripp_hidden_2011}, \citealt{meiring_qso_2013}, \citealt{hussain_hstcos_2015}, \citealt{qu_hot_2016}, \citealt{pachat_detection_2017}, \citealt{bordoloi_formation_2017}, \citealt{rosenwasser_understanding_2018}, \citealt{burchett_cos_2019}). Based on a small sample of {\NeVIII} detections in {\OVI} absorbers, \citet{narayanan_detection_2009} had predicted the redshift number density of {\NeVIII} systems to be $\sim 1/7~\times$~dN({\OVI})/dz $\sim 2.1$ at z~$< 0.5$. Through agnostic stacking \footnote{Presuming every absorption feature to be a {\NeVIII}~$770$~{\AA} line, stacking them in the rest frame of the absorber, and searching for {\NeVIII}~$780$~{\AA} feature in the stacked spectrum.} of absorption lines along 26 high S/N COS spectra, \citet{frank_agnostic_2018} arrive at a similar dN/dz~$\sim 1.38~(+0.97, -0.82)$ for the interval $0.47 \lesssim z \lesssim 1.34$. On the other hand, \citet{burchett_cos_2019}, from a search for {\NeVIII} in the CGM of 29 galaxies over a similar redshift path, found 9 {\NeVIII} detections (with all of them also having {\OVI}), arriving at a higher value dN/dz $\sim 5$. 

The absorbers in the PG~$1522+101$ sightline presented here cover a redshift path length of $\Delta z = 0.6021$\footnote{$\Delta z$ is taken as the difference between the lowest and highest redshift absorber in our sample, i.e., ${\delta}_{z} = 1.2776 - 0.6755$}. If we adopt the dN/dz $\sim 5$ from \citet{burchett_cos_2019}, we expect three {\NeVIII} detections with log [N(\NeVIII)/\cmsq] $> 13.0$ along this sightline. On the other hand, the $dN/dz = 1.38$ from \citet{frank_agnostic_2018} predict no {\NeVIII} detection, consistent with what we find for the pathlength probed. The column density upper limit of log N $\lesssim 13.7$ from the non-detection of {\NeVIII} in the five absorbers presented here implies T $< 10^{6}$~K, since for temperatures larger than that, we expect N({\NeVIII}) $>$ N({\OVI}), for solar [Ne/O]. The absence of {\NeVIII} could mean that the line of sight towards the PG~$1522+101$ may not be passing through the hot components of the WHIM or the diffuse hot coronal halos of galaxies at the absorber redshifts, though some of these could still be probing the warm (T $\sim 10^{5}$~K) and the photoionized phases of the CGM and IGM.

A summary of the key results are as follows :
\begin{enumerate}

   \item We present the absorption line properties and ionization analysis of five {\OVI} absorbers in the combined $HST$/COS, $HST$/STIS, and Keck/HIRES spectra of PG~$1522+101$. A range of metal lines including successive ionizations of oxygen ({\OII} to {\OVI}), {\CII}, {\MgII}, {\SiII}, {\CIII}, {\SiIII}, {\CIV}, {\SiIV}, {\NeV}, {\NeVI}, {\NeVIII}, {\MgX} etc are covered. Three of the absorbers also cover the redshifted EUV lines of {\HeI}~$584.334$, $537.029$~{\AA}, in one of which it is a $> 3\sigma$ detection.
   
    \item A prominent high density ($n_{\H} \gtrsim 10^{-3}$~{\cc}), low ionization phase is absent in all the five absorbers as revealed by the non-detections of {\CII}, {\OII}, {\NII}, {\MgII} and {\SiII}. In four out of the five absorbers, the column densities of oxygen ions follow a trend of $N(\OIV) > N(\OIII) > N(\OVI)$. 
    
    \item The metal ions and {\HI} show simple kinematic structures of only one or two sub-components. The presence of a collisionally ionized warm ($T \sim 10^5$~K) phase can be deduced directly from the data in the $z = 0.675$ and $z = 0.728$ absorbers in which the {\OVI} is broader and has a significant velocity offset with {\OIII} and {\HI}. In the $z = 1.095$ absorber, all observed ions, including {\OVI}, are consistent with photionization in a medium with a narrow range of densities. In the remaining two absorbers, the ionization origin of {\OVI} is ambiguous.  
    
    \item Photoionization equilibrium models require densities spread over the range $n{\H} \sim 10^{-4} - 10^{-3}$~{\cc} to explain the observed {\CIII}, {\OIII}, {\CIV}, and {\OIV} column densities suggesting that the absorbing medium does not have uniform ionization throughout. 

    \item The observed column densities of {\OIII}, {\OIV}, and {\OVI} in all the absorbers are in the relatively narrow ranges of $\log [N/\cmsq] =$~$[13.2, 14.4]$, $[13.9, 15.0]$, and $[13.9, 15.6]$ respectively, whereas the {\HI} column density spans a much wider range of $[14.1 - 16.5]$, resulting in [O/H] values as varied as $1/100$-th of solar to super-solar. Ionization models also predict an overabundance of oxygen relative to carbon in all the absorbers with [C/O] in the range of $0.0 \pm 0.4$ to $-0.8 \pm 0.1$ suggesting that the gas has preferentially been enriched by feedback from Type II SNe. The $z = 0.67556$, $z = 0.72885$ and $z = 1.16592$ absorbers with the lowest [O/H] also have the highest {\HI} column densities, indicating that the inferred low metallicities could be due to poor mixing of gas and metals on small scales. 
    
    \item The {\HeI}~$537, 584$~{\AA} lines are detected in the $z = 1.16592$ absorber with $N(\HeI) = 14.47~{\pm}~0.16$. Intergalactic {\HeI} lines have been detected in only two other instances (\citealt{reimers_hei_1993}, \citealt{cooke_measurement_2018}). This absorber is the lowest redshift {\HeI} detection known till date. The observed {\HeI} to {\HI} column density ratio along with the primordial (He/H) abundance places a strong constraint on the gas density in this absorber. This absorber is found to be tracing metal poor gas with [O/H] $= -1.9 \pm 0.2$. The {\HeI} lines are undetected in the $z = 1.09457$ and $z = 1.27768$ absorbers, for which the suggested upper limits on He-mass fraction are consistent with the primoridal nucleosynthesis value. 
    
    \item The {\NeVIII} and {\MgX} ions are non-detections, pointing to the absence of hot gas with $T > 10^{5.5}$~K. The absence of {\NeVIII} could mean that the line of sight towards the PG $1522+101$ may not be passing through the hot components of the WHIM or the diffuse hot coronal halos of galaxies at the absorber redshifts, though some of these could still be probing the warm (T $\sim 10^5$~K) and the photoionized phases of the CGM and IGM.
 
\end{enumerate}

\begin{table*}
\hspace{-0.1cm}\setlength{\tabcolsep}{3.5pt}
\renewcommand{\arraystretch}{1.6}
	\begin{tabular}{c|c|c|c|c|c|c|c|c|c}
			\hline
			\centering
			 z$_{abs}$  & log N(HI) & n$_{\H}$ & log N$_{\H}$ & $p/k$ & T & L & Origin of & [O/H] & [C/O]\\
			 &  (cm$^{-2}$) & (cm$^{-3}$) & (cm$^{-2}$) & (K cm$^{-3}$)  & (K) & (kpc) & {\OVI} & & \\ \hline
			 
			 0.67556 & $15.87 \pm 0.04$ & ${(1.6 - 7.1)}~\times~10^{-4}$ & $19.93 - 19.20$ & $4.6 - 15.9$  &  ${(2.9 - 2.2)}~\times~10^{4}$ &  $172.4 - 7.2$ & CI & $-0.9 \pm 0.1$ & $-0.5 \pm 0.2$ \\
			 
			 0.72885 & $16.50 \pm 0.02$ & ${(5.0 - 8.9)}~\times~10^{-4}$ & $20.15 - 19.86$ & $14.3 - 22.3 $ &  ${(2.8 - 2.5)}~\times~10^{4}$ &  $92.2 - 26.1$ & CI & $-2.0 \pm 0.1$ & $-0.8 \pm 0.1$ \\ 
			 
			1.09457 & $14.68 \pm 0.15$ & ${(2.8 - 11.2)} \times 10^{-4}$ & $18.61 - 17.89$ & $4.8 - 12.3$  &  ${(1.7 - 1.1)}~\times~10^{4}$ &  $4.7 - 0.2$ & PI & $+0.2 \pm 0.2$ & $-0.2 \pm 0.1$ \\
		    
		   1.16592(-25) &  $14.69 \pm 0.16$ & ${(1 - 6.3)}~\times~10^{-4}$ & $19.42 - 18.50$ & $4.1 - 18.2$  &  ${(4.1 - 2.9)}~\times~10^{4}$ &  $85.8 - 1.6$ & PI/CI & $-1.2 \pm 0.2$ & $0.0 \pm 0.4$ \\
		   
		   1.16592(1) &  $16.52 \pm 0.11$ & ${(4.4 - 25.1)}~\times~10^{-4}$ & $20.60 - 19.70$ & $14.7 - 57.5$  &  ${(3.4 - 2.3)}~\times~10^{4}$ &  $296.2 - 6.5$ & PI & $-2.1 \pm 0.2$ & $-0.3 \pm 0.2$  \\
		   
		  1.27768(0) &  $14.31 \pm 0.09$ & ${1.4 - 7.9}~\times~10^{-4}$ & $18.73 - 17.81$ & $3.3 - 10.8$  &  ${(2.3 - 1.4)}~\times~10^{4}$ &  $12.3 - 0.3$ & PI/CI & $0.0 \pm 0.2$ & $-0.4 \pm 0.2$ \\
		  
		  1.27768(39) &  $14.09 \pm 0.11$ & ${1.3 - 7.1}~\times~10^{-4}$ & $18.56 - 17.63$ & $3.0 - 9.5$  &  ${(2.4 - 1.3)}~\times~10^{4}$ &  $9.4 - 0.5$ & PI/CI & $0.0 \pm 0.3$ & $-0.2 \pm 0.3$ \\

    \hline
	\end{tabular}
	\vspace{0.5cm}
	\caption{Summary of phase solution results from Photoionization Modelling of absorbers. The first column indicates the absorber redshift (z$_{abs}$). Successive columns correspond to the logarithm of neutral hydrogen column density ($\log N(\HI)$) measured from Lyman series lines, the logarithm of total hydrogen column density ($\log N_{\H}$), the solution phase gas density (n$_{\H}$), gas pressure ($p$) normalized by $k$, photoionization equilibrium temperature (T), the path length (L) of the absorber along the sightline indicating the size of the absorber in kpc, and the possible origin scenario that could give rise to the observed {\OVI}. The final two columns indicate the obtained relative abundance of Carbon to Oxygen, [C/O] and the abundance of Oxygen [O/H] in each absorber respectively. We note that the values quoted are extracted from the photoionization models of each indvidual absorber. The [O/H] has an additional systematic uncertainty of $\sim 0.5$~dex coming from the generic EBR model (\citealt{howk_strong_2009}).}\label{tab:modsum}

\end{table*}


\renewcommand{\thefigure}{8}
\begin{figure} 
    \includegraphics[clip=true,trim=2.6cm 1.8cm 2.5cm 2.7cm,scale=0.4]{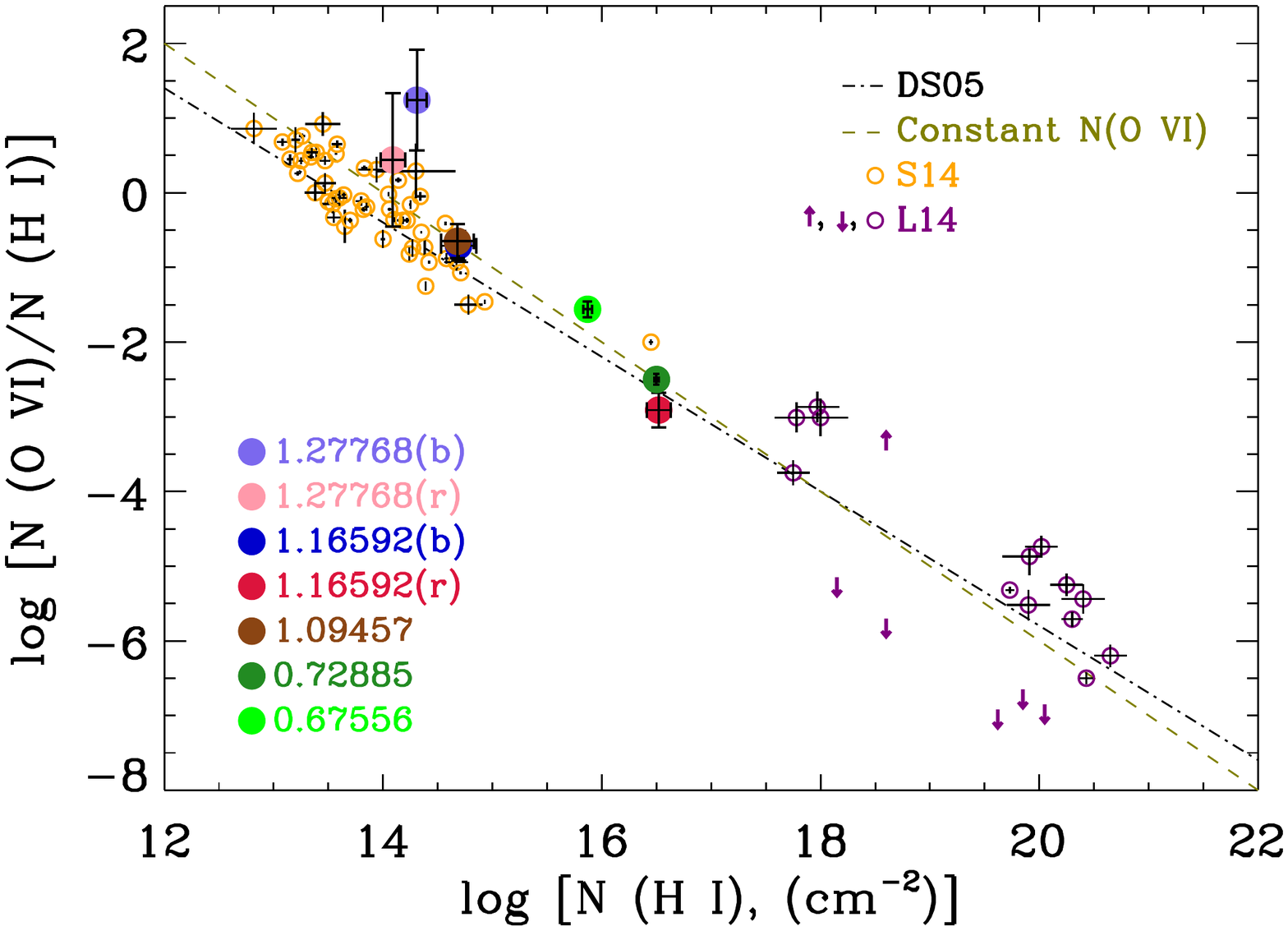}  \includegraphics[clip=true,trim=2.6cm 0cm 2.5cm 2.7cm,scale=0.4]{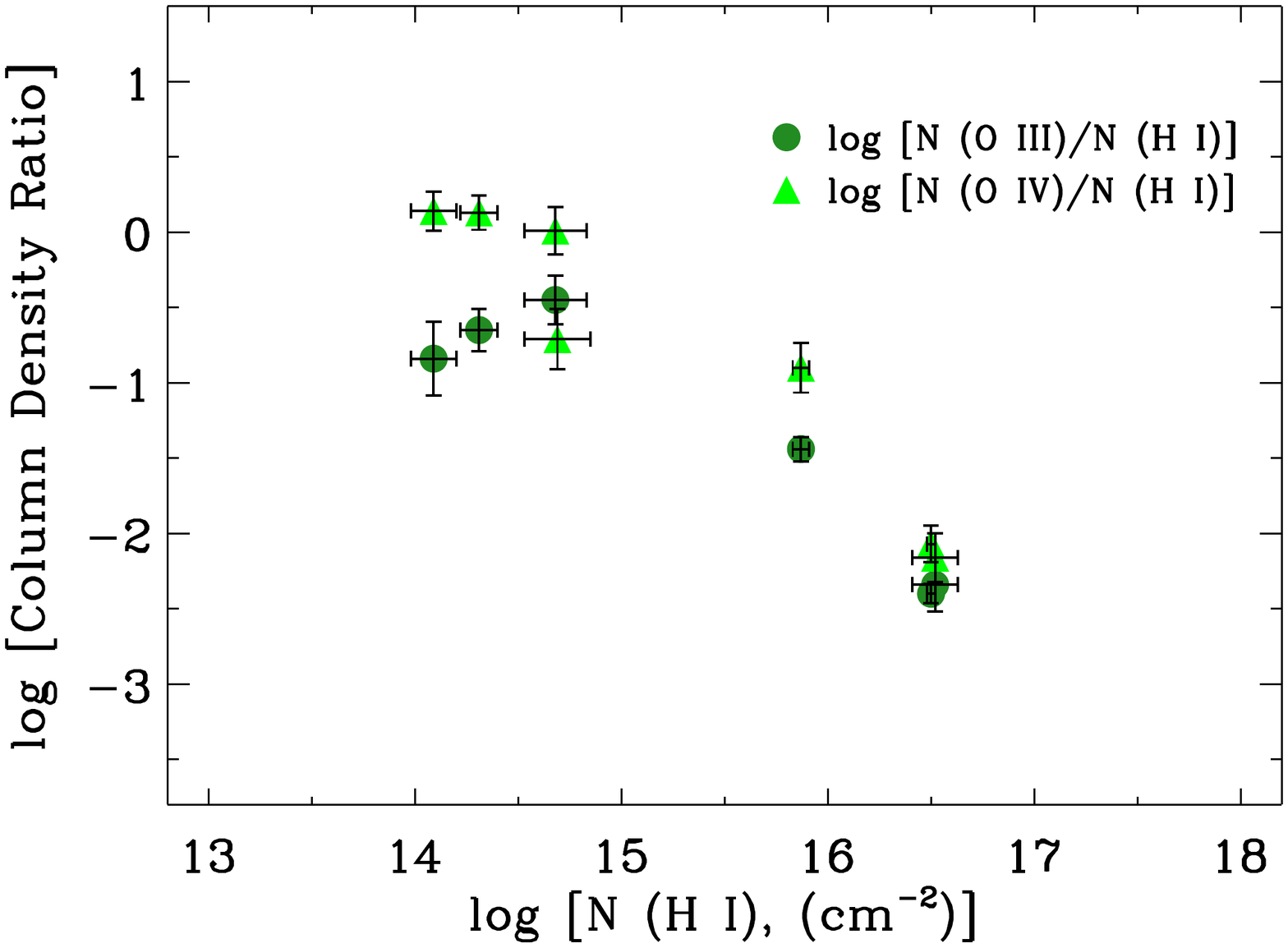}
     \vspace{-1cm}
     \caption{The \textit{top} panel shows the observed {\OVI} to {\HI} column density ratio against the {\HI} column density. Data points are taken from \citealt{savage_properties_2014} and \citealt{lehner_galactic_2014}. The \textit{dash-dot} line is a power law fit to the data, identical to the one given by \citealt{danforth_low-z_2005} formalizing the trend seen between $\log~$[$N (\OVI)/N (\HI)$] and $\log~$[$N(\HI)$]. Measurements on the five absorbers in our sample closely follows the trend. The \textit{dash} line represents the negative correlation from plotting 1/N({\HI}) against N({\HI}) for a constant N({\OVI}) = $10^{14}$. The DS05 curve and the data points closely follow this trend as {\OVI} is roughly constant in 7 orders of magnitude of N({\HI}). The \textit{bottom} panel shows the observed anti-correlation between {\OIII} to {\HI} column density ratio and N({\HI}), and {\OIV} to {\HI} column density ratio against the {\HI} column density ($\rho$ = -0.85, p = 0.0023). The anti-correlation comes about as a result of Oxygen ionization column density in samples of absorbers spanning a much narrower range compared to {\HI}, which can be interpreted as evidence of poor mixing of metals with ambient gas assuming that the kinematically coincident components are also co-spatial.}
     \label{fig:s14l14}
\end{figure}

\renewcommand{\thefigure}{9}
\begin{figure*} 
     \centering
     \includegraphics[scale=0.8, clip=true, trim={1cm 16cm 0cm 1cm}]{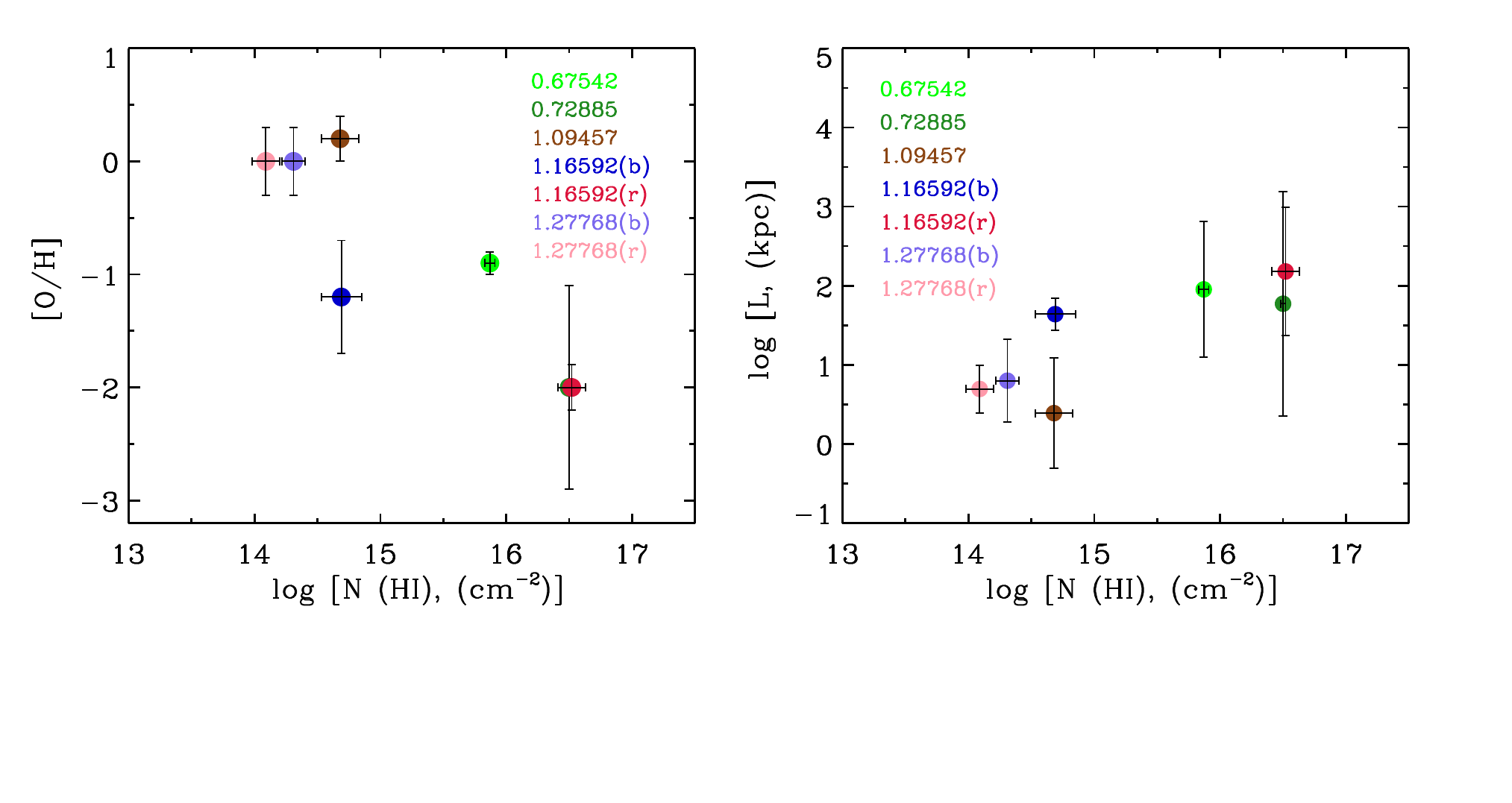}
 	\vspace{-2cm}
     \caption{The panel on the \textit{left} shows the trend between the inferred abundance of oxygen from PIE models and the {\HI} column density for each absorber in our sample. Metals confining to small physical scales results in low values of inferred metallicity in higher {\HI} column density systems, which also leads to large values for inferred path length sizes for the absorbers from the PIE models (\textit{right} panel).However we note that components that are close-by in velocity need not be co-spatial.}
     \label{fig:nhirel}
\end{figure*}

\section*{Acknowledgements}

We thank the referee, Dr. Sean Johnson for the valuable feedback and suggestions on the manuscript. SS thanks Jayadev Pradeep for discussions on line measurements at the initial stages of this work. SS and AN thank Raghunathan Srianand for several useful inputs on modeling in the early stages of this work. We thank the people responsible for the tools and packages used in this work (Cloudy, Makee, VPFIT, Astropy). We thank the people behind the creation and maintenance of COS, STIS, and HIRES instruments and their respective public archives. Support for this work was provided by SERB through grant number EMR/2017/002531 from the Department of Science \& Technology, Government of India. Based on observations made with the NASA/ESA Hubble Space Telescope, support for which was given by NASA through grant HST GO-14655 from the Space Telescope Science Institute. STScI is operated by the Association of Universities for Research in Astronomy, Inc. under NASA contract NAS 5-26555. This research has made use of the Keck Observatory Archive (KOA), which is operated by the W. M. Keck Observatory and the NASA Exoplanet Science Institute (NExScI), under contract with the National Aeronautics and Space Administration. 

\section*{Data availability}

The data underlying this article are from Archival HST and Keck databases and are publicly available from the Mikulski Archive for Space Telescopes (MAST, https://archive.stsci.edu/hst/) and Keck Observatory Archive (KOA, https://koa.ipac.caltech.edu/cgi-bin/KOA/nph-KOAlogin) respectively.




\bibliographystyle{mnras}
\bibliography{PG1522+101} 



\clearpage
\newpage
\appendix
\label{sec:appendix}

\section{Estimating the Ionization Due to Quasar Flux at $z = 1.277$} \label{sec:procalc}

Since the $z = 1.27768$ absorber is close to QSO emission redshift of $z_{\rm em} = 1.328$, it is important to estimate whether the ionizing radiation ($\lambda < 912$~{\AA}) from the QSO is going to have a significant impact on the absorber compared to the ionization brought about by the diffuse extragalactic UV background. The details of such a calculation is provided here.

The angle averaged specific intensity ($j_{\nu_0}$  in units ${\rm  erg\,\, s^{-1}\,\, cm^{-2}\,\, Hz^{-1}\,\, sr^{-1}}$) of the radiation received by the absorber at $z_{\rm abs}$ from QSO at $z_{\rm em}$ is given by, 
\begin{equation}\label{abs_rad}
    j_{\nu_0} = \frac{L^{\rm QSO}_{\nu}}{(4 \pi \Delta D_{L})^2} e^{-\tau _{\rm eff}(\nu_0, \, z_{\rm em}, \, z_{\rm abs} )} 
\end{equation}
where $L_{\nu_0}$ is the specific intensity of the QSO radiation at frequency $\nu_0$ (in units of $\rm  erg\,\, s^{-1}\,\, cm^{-2}\,\, Hz^{-1}$),  $\Delta D_{L} = D_{L}(z_{\rm em }) - D_{L}(z_{\rm abs }) $ where 
$D_{L}$ is 
the luminosity distance, and $\tau_{\rm eff}(\nu_0, \, z_{\rm em}, \, z_{\rm abs})$ is the effective optical depth suffered by
ionizing photons travelling from  $z_{\rm em}$ to $z_{\rm abs}$ which was emitted 
at frequency $\nu = (1+z_{\rm em}) \nu_0/(1+z_{\rm abs})$. 
%
%
The intensity of QSO radiation  ${L^{\rm QSO}_{\nu}}$ at frequency $\nu$ can be
obtained using flux calibrated COS QSO spectrum $f_{\nu}$,
\begin{equation}\label{qso_rad}
    {L^{\rm QSO}_{\nu}} = f_{\nu} 4 \pi D^2_L(z_{\rm em}) e^{\tau _{\rm eff}(\nu' , \, z_{\rm em}, \, 0)}
\end{equation}
where $\tau_{\rm eff}(\nu', \, z_{\rm em}, \, 0)$ is the effective optical depth suffered by ionizing photons travelling from  $z_{\rm em}$ to $z=0$ which was emitted at frequency $\nu = \nu' (1+z_{\rm em}) $. Therefore, combining Eq.~\ref{abs_rad} and \ref{qso_rad}, 
\begin{equation}\label{comb_rad}
    j_{\nu_0} = \frac{f_{\nu}  D^2_L(z_{\rm em})}{4 \pi (\Delta D_{L})^2} e^{\tau _{\rm eff}(\nu', \, z_{\rm abs}, \, 0 )}. 
\end{equation}
One can compare this $j_{\nu_0}$ to the one from extra-galactic UV background, however the spectral slopes of both can be different. Therefore, comparing the H~{\sc i} photoionization rates ($\Gamma_{\rm HI}$) is more intuitive. The $\Gamma_{\rm HI }$ is defined as 
%
\begin{equation}\label{eq.gama}
\Gamma_{\rm HI}=\int_{\nu_{\rm th}}^{\infty}d\nu_0\,
\frac{4\pi\,j_{\nu_0}}{h\nu_0}\,\sigma_{\rm HI}(\nu_0)\,\, ,
\end{equation}
%
where, $\nu_{\rm th}$ is a threshold frequency for ionization of H~{\sc i}, $h$ is the Planck's constant, and $\sigma_{\rm HI}$ is the photoionization cross-section of hydrogen. 

To crudely estimate an order of magnitude contribution from the QSO, we ignore the frequency dependence of the $\tau_{\rm eff}$ in Eq~\ref{comb_rad} and use power-law fits to the flux calibrated rest frame QSO spectrum  $f_{\nu} = f_{\nu_{\rm th}}(\nu/\nu_{\rm th})^{-\alpha}$ and photoionization  cross-section $\sigma_{\rm HI}(\nu_0)= \sigma_{\rm HI}(\nu_{\rm th})  (\nu_0/\nu_{\rm th})^{-\beta}$ which simplifies Eq.~\ref{eq.gama} to 
%
\begin{equation}\label{final_gama}
\Gamma_{\rm HI} \approx \frac{ f_{\nu_{\rm th}} \sigma_{\rm HI}(\nu_{\rm th})  D^2_L(z_{\rm em})}
{h(\alpha + \beta) (\Delta D_{L})^2 } e^{\tau _{\rm eff}(\nu', \, z_{\rm abs}, \, 0 )},
\end{equation}

We fit a power-law to the rest frame QSO spectra ($z_{\rm em } = 1.328$) at $\lambda < 890$~\AA~which is equivalent of fitting power law QSO spectra at $\lambda_{\rm abs} < 912 $~\AA~at the absorber redshift $z_{\rm abs}= 1.28$. We obtain $\alpha = 3.13$ and $f_{\nu_{\rm th}} = 1.43 \times 10^{-27}$ erg s$^{-1}$ cm$^{-2}$ Hz$^{-1}$. Using the value of $\beta = 3$, $\sigma_{\rm HI}(\nu_{\rm th}) = 6.3 \times 10^{-18}$ cm$^{2}$ and luminosity distances obtained assuming flat cosmology parameters ($\Omega_{m} =0.3$, $\Omega_{\Lambda}=0.7$ and $H_0  = 70 $ km s$^{-1}$ MPc$^{-1}$) we obtain
$\Gamma_{\rm HI} \approx 10^{-16} \, e^{\tau _{\rm eff}(\nu', \, z_{\rm abs}, \, 0)}$ s $^{-1}$ which is lower by four orders of magnitude than the fiducial value obtained using the KS19 extragalactic UV background at $z_{\rm abs}$. 

Given the fact that there are no DLAs along the sight-line and the contribution from IGM opacity to $e^{\tau _{\rm eff}}$ is quite small at $z<1$, the extragalactic UVB dominates the ionizing radiation received by abasorber over 
the radiation from QSO by orders of magnitude. Therefore we safely ignore the QSO contribution during the PIE modelling of this absorber.

\section{Absorption Line Measurements and Absorption System Velocity Plots}

The appendix contains full system plots of each absorber (See Figures from \ref{fig:sysplot0675} to \ref{fig:sysplot1277}) and the line measurements of all detected and covered species in each of the systems (See Tables from \ref{0675table1} to \ref{1277table2})


\newcolumntype{Y}{>{\centering\arraybackslash}X}
\begin{table}\centering
\setlength{\tabcolsep}{1.5pt}
\renewcommand{\arraystretch}{1.5}
\caption{Line measurements for the z$_{abs}$ $=$ 0.67556 absorber towards PG~$1522+101$ with the successive columns indicating the equivalent width in the rest-frame of the absorber, the column density measured using the AOD method, or the Voigt profile measured column density and Doppler $b$ parameter. The final column shows the velocity range over which the equivalent width and apparent column densities were integrated, or the velocity centroid for the profile-fitted absorption components. For lines that are not detected with a significance of $\geq 3\sigma$, upper limits are quoted for the equivalent width and the corresponding column density is obtained from the linear part of the curve of growth. The {\HI}~$926$ line suffers from contamination. $^c$ - contaminated lines.} 
\label{0675table1}

\begin{tabularx}{\linewidth}{@{}p{4em}YYYY@{}}

\toprule
\multicolumn{1}{c}{Line} & \multicolumn{1}{c}{$\mathit{W_{r}}(m\mathit{A^{o}})$} &  \multicolumn{1}{c}{log[N ($\mathit{cm^{-2}}$)]} & \multicolumn{1}{c}{b (\kms)} & \multicolumn{1}{c}{v (\kms)}  \\ \midrule

H I 1215    & $>$ 635 & $>$ 14.4 &   & $[-145,95]$ \\
H I 1025    & $>$ 451 & $>$ 15.3 &   & $[-110,85]$ \\
H I 972     & 346 $\pm$ 8 & 15.52 $\pm$ 0.04 &   & $[-110,85]$ \\
H I 949     & 298.4 $\pm$ 9 & 15.73 $\pm$ 0.03 &   & $[-110,85]$ \\
H I 937     & 212 $\pm$ 8 & 15.75 $\pm$ 0.02 &   & $[-110,85]$ \\
H I 930     & 166 $\pm$ 8 & 15.80 $\pm$ 0.01 &   & $[-110,85]$ \\
H I 926     & 213 $\pm$ 7 & 16.11 $\pm$ 0.01 &   & $[-110,85]$ \\
H I 923     & 112 $\pm$ 8 & 15.90 $\pm$ 0.02 &   & $[-110,85]$ \\
H I 920     & 81 $\pm$ 8 & 15.89 $\pm$ 0.02 &   & $[-110,85]$ \\
H I 919     & 84 $\pm$ 9 & 16.02 $\pm$ 0.02 &   & $[-110,85]$ \\
H I 918     & $<$ 25 & $<$ 15.5 &   & $[-110,85]$ \\
H I 917     & $<$ 25 & $<$ 15.6 &   & $[-110,85]$ \\
H I        & &    15.87 $\pm$ 0.04 & 36 $\pm$ 2 & -28 $\pm$ 3 \\
H I        & &    $<$ 13.7  & $<$ 64 & 1 $\pm$ 8 \\

C~II~1334$^c$  & $<$ 197 & $<$ 14.1 &   & $[-80,70]$ \\
C~II~1036   & $<$ 34 & $<$ 13.4 &   & $[-80,70]$ \\
C~III~977   & 236 $\pm$ 7 & 13.82 $\pm$ 0.02 &   & $[-80,70]$ \\
C III        & &   13.90 $\pm$ 0.10 &  33 $\pm$ 6 & -10 $\pm$ 4 \\

C~IV~1548   & 291 $\pm$ 15 & 14.01 $\pm$ 0.03 &   & $[-80,70]$ \\
C~IV~1550   & 137 $\pm$ 15 & 13.89 $\pm$ 0.04 &   & $[-80,70]$ \\
C IV  & &   13.83 $\pm$ 0.25 &  33 $\pm$ 8 & -12 $\pm$ 18 \\

N II 915    & $<$ 22 & $<$ 13.3 &   & $[-80,70]$ \\
N III 989   & $<$ 29 & $<$ 13.5 &   & $[-80,70]$ \\
N~IV~765$^c$   & $<$ 192 & $<$ 14.03 &   & $[-80,70]$ \\
N V 1238    & $<$ 86 & $<$ 13.6 &   & $[-80,70]$ \\
N V 1242    & $<$ 77 & $<$ 13.8 &   & $[-80,70]$ \\
O II 834    & $<$ 18 & $<$ 13.3 &   & $[-80,70]$ \\
O III 702   & 98 $\pm$ 6 & 14.33 $\pm$ 0.01 &   & $[-80,70]$ \\
O III 832   & 130 $\pm$ 5 & 14.42 $\pm$ 0.01 &   & $[-80,60]$ \\
O III        & &   14.43 $\pm$ 0.07  &  36 $\pm$ 6 & -15 $\pm$ 6  \\

O IV 787    & 210 $\pm$ 6 & 14.83 $\pm$ 0.02 &   & $[-80,70]$ \\
O IV        & &   14.97 $\pm$ 0.16 &  33 $\pm$ 8 & 1 $\pm$ 5  \\

O~VI~1031   & 171 $\pm$ 9 & 14.28 $\pm$ 0.02 &   & $[-80,70]$ \\
O~VI~1037   & 93 $\pm$ 11 & 14.25 $\pm$ 0.04 &   & $[-80,70]$ \\
O VI        & &   14.31 $\pm$ 0.10 &  39 $\pm$ 11 & 1 $\pm$ 8 \\

\bottomrule
\end{tabularx}
\end{table}

\newcolumntype{Y}{>{\centering\arraybackslash}X}

\begin{table}\centering
\setlength{\tabcolsep}{1.5pt}
\renewcommand{\arraystretch}{1.5}
\caption{Table of line measurements for the z=0.67556 absorber towards PG~$1522+101$, continued from Table \ref{0675table1}. $^c$ - contaminated lines.}
\label{0675table2}

\begin{tabularx}{\linewidth}{@{}p{4em}YYYY@{}}

\toprule
\multicolumn{1}{c}{Line} & \multicolumn{1}{c}{$\mathit{W_{r}}(m\mathit{A^{o}})$} &  \multicolumn{1}{c}{log[N ($\mathit{cm^{-2}}$)]} & \multicolumn{1}{c}{b (\kms)} & \multicolumn{1}{c}{v (\kms)}  \\ \midrule

Ne~VIII~770 & $<$ 15 & $<$ 13.4 &   & $[-80,70]$ \\
Ne~VIII~780 & $<$ 21 & $<$ 13.8 &   & $[-80,70]$ \\
Mg~II~2796  & $<$ 27 & $<$ 11.8 &   & $[-80,70]$ \\
Mg~II~2803  & $<$ 26 & $<$ 12.0 &   & $[-80,70]$ \\
Al~III~1862 & $<$ 53 & $<$ 12.8 &   & $[-80,70]$ \\
Al~III~1854 & $<$ 56 & $<$ 12.5 &   & $[-80,70]$ \\
Si~II~1526  & $<$ 43 & $<$ 13.2 &   & $[-80,70]$ \\
Si~II~1260$^c$ & $<$ 117 & $<$ 13.0 &   & $[-80,70]$ \\
Si~II~1193$^c$ & $<$ 221  & $<$ 13.7 &   & $[-80,70]$ \\
Si~II~1190$^c$ & $<$ 333 & $<$ 13.8 &   & $[-80,70]$ \\
Si~II~989   & $<$ 29 & $<$ 13.2 &   & $[-80,70]$ \\
Si~III~1206 & $<$ 108 & $<$ 12.7 &   & $[-80,70]$ \\
Si~IV~1393$^c$ & $<$ 204 & $<$ 13.5 &   & $[-80,70]$ \\
Si~IV~1402$^c$ & $<$ 351 & $<$ 14.2 &   & $[-80,70]$ \\
S~IV~748     & $<$ 12 & $<$ 12.6 &   & $[-80,70]$ \\
S~V~786$^c$    & $<$ 77 & $<$ 13.1 &   & $[-80,70]$ \\
S VI 944$^c$   & $<$ 46 & $<$ 13.6 &   & $[-80,70]$ \\
S VI 933    & $<$ 23 & $<$ 12.8 &   & $[-80,70]$ \\
Fe~II~2600  & $<$ 22 & $<$ 12.2 &   & $[-80,70]$ \\
Fe~II~2382  & $<$ 23 & $<$ 12.1 &   & $[-80,70]$ \\

\bottomrule
\end{tabularx}
\end{table}

\begin{figure*}
	\centering
	\includegraphics[scale=0.8, clip=true, trim=0cm 0cm 0cm 1.5cm]{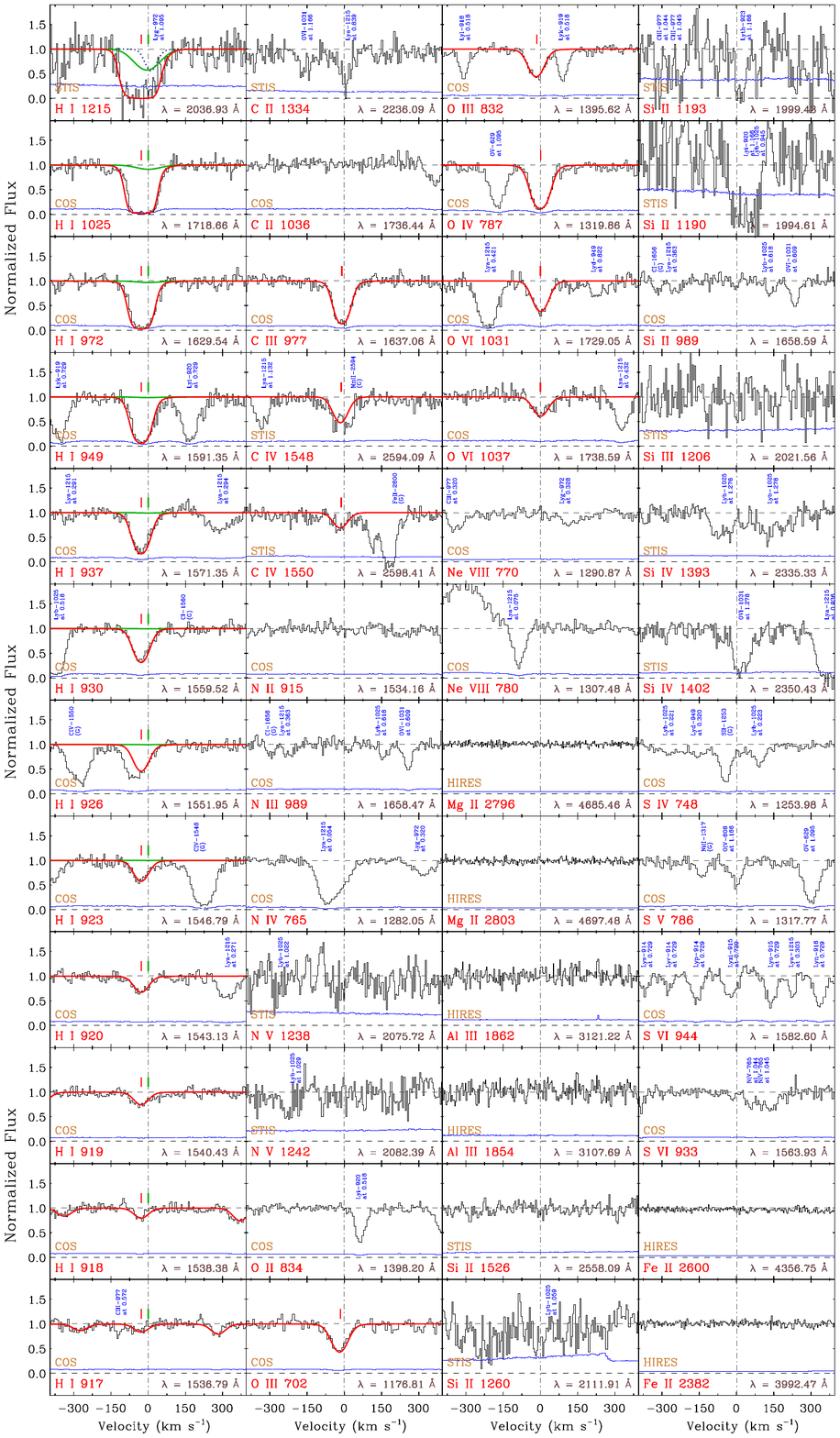}
	\vspace{-1cm}
	\caption{System plot of the z$_{abs}$=0.67556 absorber, with continuum-normalized flux along the Y-axis and the velocity scale relative to the redshift of the absorber along the X-axis. The $v = 0$~{\kms}, marked by the \textit{dashed-dotted} vertical line, indicates the absorber redshift. The $1\sigma$ uncertainty in flux is indicated by the \textit{blue} curve at the bottom of each panel. The \textit{red} curves are the best-fit Voigt profiles. The observed wavelength of each transition is also indicated in the respective panels. Interloping features unrelated to the absorber are also labeled. \label{fig:sysplot0675}}
\end{figure*}


\newcolumntype{Y}{>{\centering\arraybackslash}X}

\begin{table}\centering
\setlength{\tabcolsep}{1.5pt}
\renewcommand{\arraystretch}{1.5}
\caption{Line measurements for the z$_{abs}$ $=$ 0.72885 absorber towards PG~$1522+101$ with the successive columns indicating the equivalent width in the rest-frame of the absorber, the column density measured using the AOD method, or the Voigt profile measured column density and Doppler $b$ parameter. The final column shows the velocity range over which the equivalent width and apparent column densities were integrated, or the velocity centroid for the profile-fitted absorption components. $^c$ - contaminated lines.} 
\label{0728table1}

\begin{tabularx}{\linewidth}{@{}p{4em}YYYY@{}}

\toprule
\multicolumn{1}{c}{Line} & \multicolumn{1}{c}{$\mathit{W_{r}}(m\mathit{A^{o}})$} &  \multicolumn{1}{c}{log[N ($\mathit{cm^{-2}}$)]} & \multicolumn{1}{c}{b (\kms)} & \multicolumn{1}{c}{v (\kms)}  \\ \midrule

H I 1215       & $>$ 565 & $>$ 14.1 &   & $[-70,80]$ \\
H I 1025       & $>$ 434 & $>$ 15.2 &   & $[-70,80]$ \\
H I 972        & 352 $\pm$ 6 & 15.62 $\pm$ 0.32 &   & $[-70,80]$ \\
H I 949        & 297 $\pm$ 6 & 15.80 $\pm$ 0.05 &   & $[-70,80]$ \\
H I 937        & 287 $\pm$ 9 & 16.03 $\pm$ 0.11 &   & $[-70,80]$ \\
H I 930        & 241 $\pm$ 8 & 16.16 $\pm$ 0.08 &   & $[-70,80]$ \\
H I 926        & 206 $\pm$ 9 & 16.26 $\pm$ 0.06 &   & $[-70,80]$ \\
H I 923        & 212 $\pm$ 7 & 16.41 $\pm$ 0.03 &   & $[-70,80]$ \\
H I 920        & 196 $\pm$ 6 & 16.49 $\pm$ 0.03 &   & $[-45,50]$ \\
H I 919        & 182 $\pm$ 5 & 16.56 $\pm$ 0.02 &   & $[-45,50]$ \\
H I 918        & 137 $\pm$ 6 & 16.50 $\pm$ 0.02 &   & $[-45,50]$ \\
H I 917        & 124 $\pm$ 6 & 16.53 $\pm$ 0.02 &   & $[-45,50]$ \\
H I 916.4      & 97 $\pm$ 6 & 16.49 $\pm$ 0.02 &   & $[-45,50]$ \\
H I 915.8      & 82 $\pm$ 7 & 16.49 $\pm$ 0.02 &   & $[-45,50]$ \\
H I 915.3      & 65 $\pm$ 6 & 16.44 $\pm$ 0.02 &   & $[-45,50]$ \\
H I 914.9      & 67 $\pm$ 6 & 16.56 $\pm$ 0.02 &   & $[-45,50]$ \\
H I 914.5      & 31 $\pm$ 7 & 16.25 $\pm$ 0.04 &   & $[-45,50]$ \\
H I 914.2      & 36 $\pm$ 7 & 16.39 $\pm$ 0.03 &   & $[-45,50]$ \\
H I 914.0      & $<$ 15  & $<$ 16.1 &   & $[-45,50]$ \\
H I 913.8      & $<$ 20  & $<$ 16.3 &   & $[-45,50]$ \\
H I 913.6      & $<$ 19.4 & $<$ 16.2 &   & $[-45,50]$ \\
H I 913.4      & $<$ 19.2 & $<$ 16.3 &   & $[-45,50]$ \\
H I 913.3      & $<$ 19.4 & $<$ 16.3 &   & $[-45,50]$ \\
H I 913.2      & $<$ 19.6 & $<$ 16.4 &   & $[-45,50]$ \\
H I 913.1      & $<$ 19.9 & $<$ 16.4 &   & $[-45,50]$ \\
H I 913.0      & $<$ 19.7 & $<$ 16.5 &   & $[-45,50]$ \\
H I 912.9      & $<$ 20.0 & $<$ 16.5 &   & $[-45,50]$ \\
H I 912.8      & $<$ 20.6 & $<$ 16.6 &   & $[-45,50]$ \\
H~I~912.76     & $<$ 20.5 & $<$ 16.6 &   & $[-45,50]$ \\
H~I~912.70     & $<$ 20.4 & $<$ 16.7 &   & $[-45,50]$ \\
H I 912.6      & $<$ 20.5 & $<$ 16.7 &   & $[-45,50]$ \\
H I &  & 16.50 $\pm$ 0.02 & 26 $\pm$ 3 &  2 $\pm$ 1 \\ 

C II 903.9     & $<$ 15.6 & $<$ 12.8 &   & $[-30,30]$ \\
C II 903.6     & $<$ 14.9 & $<$ 13.0 &   & $[-30,30]$ \\
C II 1334      & $<$ 30.9 & $<$ 13.1 &   & $[-30,30]$ \\
C II 1036      & $<$ 38.4 & $<$ 13.5 &   & $[-30,30]$ \\
C III 977      & 80 $\pm$ 6 & 13.24 $\pm$ 0.02 &   & $[-30,30]$ \\
C III &  & 13.34 $\pm$ 0.05 & 17 $\pm$ 3 & 1 $\pm$ 2 \\ 

 \bottomrule
\end{tabularx}
\end{table}

\newcolumntype{Y}{>{\centering\arraybackslash}X}

\begin{table}\centering
\setlength{\tabcolsep}{1.5pt}
\renewcommand{\arraystretch}{1.5}
\caption{Table of line measurements for the z=0.72885 absorber towards PG~$1522+101$, continued from Table \ref{0728table1}. $^c$ - contaminated lines.}
\label{0728table2}

\begin{tabularx}{\linewidth}{@{}p{4em}YYYY@{}}

\toprule
\multicolumn{1}{c}{Line} & \multicolumn{1}{c}{$\mathit{W_{r}}(m\mathit{A^{o}})$} &  \multicolumn{1}{c}{log[N ($\mathit{cm^{-2}}$)]} & \multicolumn{1}{c}{b (\kms)} & \multicolumn{1}{c}{v (\kms)}  \\ \midrule

C~IV~1548$^c$     & $<$ 41 & $<$ 13.1 &   & $[-30,30]$ \\
C~IV~1550$^c$     & $<$ 31 & $<$ 13.2 &   & $[-30,30]$ \\
N II 1083      & $<$ 59 & $<$ 13.7 &   & $[-30,30]$ \\
N III 989      & $<$ 22 & $<$ 13.3 &   & $[-30,30]$ \\
N III 685      & $<$ 11 & $<$ 13.0 &   & $[-30,30]$ \\
N III 684      & $<$ 11 & $<$ 13.3 &   & $[-30,30]$ \\
N IV 765       & $<$ 14 & $<$ 12.6 &   & $[-30,30]$ \\
N~V~1238$^c$      & $<$ 55 & $<$ 13.5 &   & $[-30,30]$ \\
N~V~1242$^c$      & $<$ 84 & $<$ 14.1 &   & $[-30,30]$ \\
O II 834       & $<$ 12.7 & $<$ 13.1 &   & $[-30,30]$ \\
O III 832      & 53 $\pm$ 4 & 14.01 $\pm$ 0.01 &   & $[-30,30]$ \\
O III &  & 14.10 $\pm$ 0.06 & 19 $\pm$ 4 & 2 $\pm$ 2 \\ 

O IV 787    & 114 $\pm$ 3 & 14.51 $\pm$ 0.01 &   & $[-30,30]$ \\
O IV  &  & 14.43 $\pm$ 0.12 & 22 $\pm$ 4 & 1 $\pm$ 4 \\ 

O~VI~1031   & 92 $\pm$ 15 & 13.94 $\pm$ 0.08 &   & $[-100,50]$ \\
O~VI~1037   & 60 $\pm$ 12 & 14.07 $\pm$ 0.10 &   & $[-30,30]$ \\
O VI &  & 14.00 $\pm$ 0.07 & 61 $\pm$ 11 & -21 $\pm$ 8 \\

Ne~VIII~770$^c$   & $<$ 21 & $<$ 13.6 &   & $[-30,30]$ \\
Ne~VIII~780 & $<$ 11 & $<$ 13.6 &   & $[-30,30]$ \\
Mg~II~2796  & $<$ 16 & $<$ 11.5 &   & $[-30,30]$ \\
Mg~II~2803  & $<$ 15 & $<$ 11.8 &   & $[-30,30]$ \\
Al~III~1862 & $<$ 18 & $<$ 12.3 &   & $[-30,30]$ \\
Al~III~1854 & $<$ 21 & $<$ 12.1 &   & $[-30,30]$ \\
Si II 1526  & $<$ 30 & $<$ 13.0 &   & $[-30,30]$ \\
Si II 1260  & $<$ 51 & $<$ 12.4 &   & $[-30,30]$ \\
Si II 1193  & $<$ 56 & $<$ 12.8 &   & $[-30,30]$ \\
Si II 1190  & $<$ 51 & $<$ 13.1 &   & $[-30,30]$ \\
Si II 989   & $<$ 21 & $<$ 13.1 &   & $[-30,30]$ \\
Si~III~1206 & $<$ 55 & $<$ 12.4 &   & $[-30,30]$ \\
Si~IV~1393  & $<$ 26 & $<$ 12.4 &   & $[-30,30]$ \\
Si~IV~1402  & $<$ 24 & $<$ 12.7 &   & $[-30,30]$ \\
S IV 748    & $<$ 11 & $<$ 12.6 &   & $[-30,30]$ \\
S V 786     & $<$ 12 & $<$ 12.1 &   & $[-30,30]$ \\
S~VI~944    & $<$ 16 & $<$ 12.9 &   & $[-30,30]$ \\
S~VI~933$^c$   & $<$ 33 & $<$ 13.0 &   & $[-30,30]$ \\
Fe~II~2600  & $<$ 14 & $<$ 12.0 &   & $[-30,30]$ \\
Fe~II~2382  & $<$ 15 & $<$ 11.9 &   & $[-30,30]$ \\
Fe~II~2344  & $<$ 17 & $<$ 12.4 &   & $[-30,30]$ \\

\bottomrule
\end{tabularx}
\end{table}

\begin{figure*}
	\centering
	\vspace{2cm}
	\includegraphics[scale=0.8]{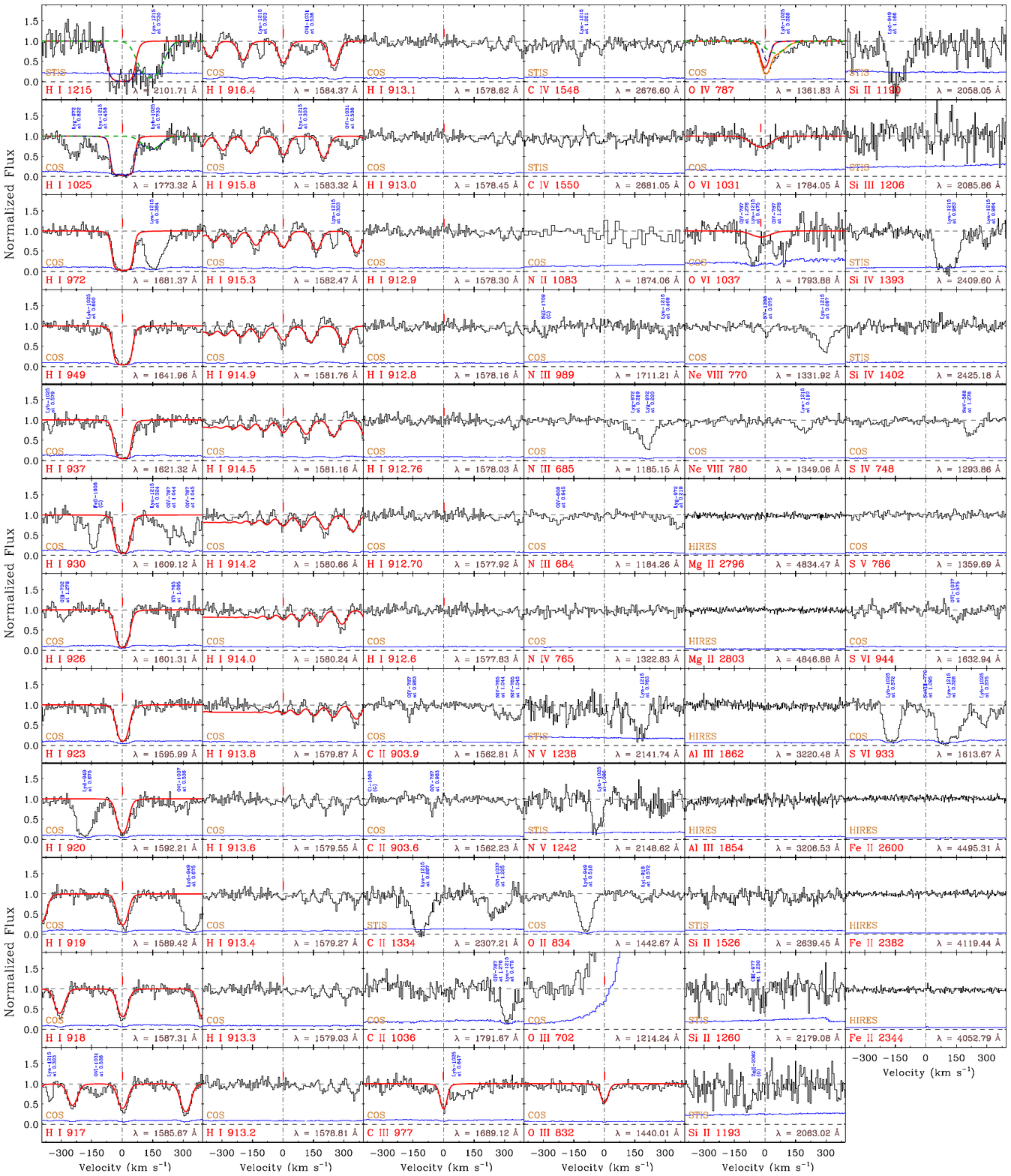}
	\vspace{-6cm}
	\caption{System plot of the z$_{abs}$=0.72885 absorber, with continuum-normalized flux along the Y-axis and the velocity scale relative to the redshift of the absorber along the X-axis. The $v = 0$~{\kms}, marked by the \textit{dashed-dotted} vertical line represents the absorber redshift. The $1\sigma$ uncertainty in flux is the \textit{blue} curve at the bottom of each panel. The \textit{red} curves are the best-fit Voigt profiles. The observed wavelength of each transition is also indicated in the respective panels. The dashed green curves in the {\HI}~1215 {\AA} and {\HI}~1025 {\AA} panels show a fit of the corresponding {\HI} lines at z $\sim 0.730$ and the blue curve shows the composite fit. The {\OIV}~787 {\AA} is severely contaminated by {\HI}~1025 {\AA} at z $\sim 0.327$ and $0.328$. The contamination was accounted for in the fit and is displayed as dashed blue and green curve in the panel.} \label{fig:sysplot0728}
\end{figure*}


\newcolumntype{Y}{>{\centering\arraybackslash}X}

\begin{table}\centering
\setlength{\tabcolsep}{1.5pt}
\renewcommand{\arraystretch}{1.5}
\caption{Line measurements for the z$_{abs}$ $=$ 1.09457 absorber towards PG~$1522+101$ with the successive columns indicating the equivalent width in the rest-frame of the absorber, the column density measured using the AOD method, or the Voigt profile measured column density and Doppler $b$ parameter. The final column shows the velocity range over which the equivalent width and apparent column densities were integrated, or the velocity centroid for the profile-fitted absorption components. (a) - STIS Spectra. continued in Table \ref{1094table2}. $^c$ - contaminated lines.}
\label{1094table1}

\begin{tabularx}{\linewidth}{@{}p{4em}YYYY@{}}

\toprule
\multicolumn{1}{c}{Line} & \multicolumn{1}{c}{${W_{r}}$(m$\angstrom$)} &  \multicolumn{1}{c}{log[N ($\mathit{cm^{-2}}$)]} & \multicolumn{1}{c}{b (\kms)} & \multicolumn{1}{c}{v (\kms)}  \\ \midrule
H I 1215       & 348 $\pm$ 9 & 13.99 $\pm$ 0.25 &   & $[-70,50]$ \\
H I 1025       & 153 $\pm$ 14 & 14.55 $\pm$ 0.18 &   & $[-70,50]$ \\
H I 972        & $<$ 301 & $<$ 15.2 &   & $[-70,50]$ \\
H I 949        & $<$ 80 & $<$ 14.8 &   & $[-70,50]$ \\
H I 937        & $<$ 83 & $<$ 15.1 &   & $[-70,50]$ \\
H I 930        & $<$ 86 & $<$ 15.3 &   & $[-70,50]$ \\
H I 926        & $<$ 62 & $<$ 15.4 &   & $[-70,50]$ \\
H I 923        & $<$ 86 & $<$ 15.7 &   & $[-70,50]$ \\
H I &  & 14.65 $\pm$ 0.15 & 25 $\pm$ 3 & -4 $\pm$ 2 \\ 

He I 584       & $<$ 10 & $<$ 13.0 &   & $[-40,35]$ \\
C II 903.9     & $<$ 66 & $<$ 13.4 &   & $[-40,35]$ \\
C II 903.6     & $<$ 68 & $<$ 13.7 &   & $[-40,35]$ \\
C II 1036      & $<$ 32 & $<$ 13.4 &   & $[-40,35]$ \\
C III 977      & 194 $\pm$ 20 & 13.76 $\pm$ 0.16 &   & $[-40,35]$ \\
C~III~977(a) & $>$ 215 & $>$ 13.6 &   & $[-40,35]$ \\
C~IV~1548      & 179 $\pm$ 5 & 13.91 $\pm$ 0.01 &   & $[-40,35]$ \\
C~IV~1550      & 113 $\pm$ 5 & 13.89 $\pm$ 0.01 &   & $[-40,35]$ \\
C IV &  &  13.91 $\pm$ 0.02 & 18 $\pm$ 1 & -1 $\pm$ 1 \\

N II 1083      & $<$ 25 & $<$ 13.3 &   & $[-40,35]$ \\
N II 915       & $<$ 67 & $<$ 13.7 &   & $[-40,35]$ \\
N III 685      & 62 $\pm$ 3 & 13.85 $\pm$ 0.01 &   & $[-40,35]$ \\
N III 684      & 108 $\pm$ 3 & 14.52 $\pm$ 0.01 &   & $[-40,35]$ \\
N IV 765       & 20 $\pm$ 6 & 12.85 $\pm$ 0.07 &   & $[-40,35]$ \\
N IV &  & 12.96 $\pm$ 0.36 & 17 $\pm$ 2  & 4 $\pm$ 3 \\

N~V~1238       & $<$ 29 & $<$ 13.1 &   & $[-40,35]$ \\
N~V~1242       & $<$ 26 & $<$ 13.3 &   & $[-40,35]$ \\
O II 834       & $<$ 18 & $<$ 13.3 &   & $[-40,35]$ \\
O III 702      & 77 $\pm$ 4 & 14.23 $\pm$ 0.01 &   & $[-55,55]$ \\
O III 832      & 68 $\pm$ 7 & 14.12 $\pm$ 0.03 &   & $[-55,55]$ \\
O III &  & 14.23 $\pm$ 0.06 & 18 $\pm$ 3 & 6 $\pm$ 2  \\ 

O IV 554       & 100 $\pm$ 5 & 14.42 $\pm$ 0.02 &   & $[-55,55]$ \\
O IV 553       & 69 $\pm$ 6 & 14.49 $\pm$ 0.03 &   & $[-55,55]$ \\
O IV 608       & 62 $\pm$ 3 & 14.57 $\pm$ 0.01 &   & $[-55,55]$ \\
O IV 787       & 137 $\pm$ 5 & 14.56 $\pm$ 0.02 &   & $[-55,55]$ \\
O IV &  & 14.69 $\pm$ 0.05 & 18 $\pm$ 3 & 1 $\pm$ 2  \\ 

O V 629        & 127 $\pm$ 4 & 14.09 $\pm$ 0.01 &   & $[-55,55]$ \\
O V &  & 14.63 $\pm$ 0.18 & 18 $\pm$ 2 & 0 $\pm$ 2  \\ 
\bottomrule
\end{tabularx}
\end{table}

\newcolumntype{Y}{>{\centering\arraybackslash}X}

\begin{table}\centering
\setlength{\tabcolsep}{1.5pt}
\renewcommand{\arraystretch}{1.5}
\caption{Table of line measurements for the z=1.09457 absorber towards PG~$1522+101$, continued from Table \ref{1094table1}. $^c$ - contaminated lines.}
\label{1094table2}

\begin{tabularx}{\linewidth}{@{}p{4em}YYYY@{}}

\toprule
\multicolumn{1}{c}{Line} & \multicolumn{1}{c}{$\mathit{W_{r}}$(m$\angstrom$)} &  \multicolumn{1}{c}{log[N ($\mathit{cm^{-2}}$)]} & \multicolumn{1}{c}{b (\kms)} & \multicolumn{1}{c}{v (\kms)}  \\ \midrule
O~VI~1031      & 59 $\pm$ 16 & 13.84 $\pm$ 0.17 &   & $[-55,55]$ \\
O~VI~1037      & 80 $\pm$ 13 & 14.25 $\pm$ 0.10 &   & $[-55,55]$ \\
O VI &  & 14.03 $\pm$ 0.17 & 18 $\pm$ 4 &  12 $\pm$ 8 \\ 

Ne~V~568$^c$ & $<$ 52 & $<$ 14.4 &   & $[-40,35]$ \\
Ne~VI~558      & $<$ 15 & $<$ 13.7 &   & $[-40,35]$ \\
Ne~VIII~770$^c$   & $<$ 32 & $<$ 13.8 &   & $[-40,35]$ \\
Ne~VIII~780    & $<$ 15 & $<$ 13.7 &   & $[-40,35]$ \\
Mg~II~2796     & $<$ 17 & $<$ 11.6 &   & $[-40,35]$ \\
Mg~X~609$^c$      & $<$ 14 & $<$ 13.7 &   & $[-40,35]$ \\
Mg X 624       & $<$ 11 & $<$ 13.8 &   & $[-40,35]$ \\
Al~II~1670     & $<$ 13 & $<$ 11.4 &   & $[-40,35]$ \\
Al~III~1862    & $<$ 19 & $<$ 12.3 &   & $[-40,35]$ \\
Al~III~1854    & $<$ 13 & $<$ 11.9 &   & $[-40,35]$ \\
Si~II~1260     & $<$ 26 & $<$ 12.2 &   & $[-40,35]$ \\
Si~II~1193     & $<$ 29 & $<$ 12.5 &   & $[-40,35]$ \\
Si~II~1190     & $<$ 23 & $<$ 12.7 &   & $[-40,35]$ \\
Si~III~1206    & $<$ 25 & $<$ 12.0 &   & $[-40,35]$ \\
Si~IV~1393     & $<$ 20 & $<$ 12.3 &   & $[-40,35]$ \\
S IV 748       & $<$ 14 & $<$ 12.7 &   & $[-40,35]$ \\
S V 786        & $<$ 16 & $<$ 12.3 &   & $[-40,35]$ \\
S VI 944       & $<$ 69 & $<$ 13.6 &   & $[-40,35]$ \\
S VI 933       & $<$ 75 & $<$ 13.3 &   & $[-40,35]$ \\
Fe~II~2600     & $<$ 15 & $<$ 12.0 &   & $[-40,35]$ \\
Fe~II~2382     & $<$ 15 & $<$ 11.9 &   & $[-40,35]$ \\
Fe~II~2344     & $<$ 17 & $<$ 12.4 &   & $[-40,35]$ \\

\bottomrule
\end{tabularx}
\end{table}

\begin{figure*}
	\centering
	\vspace{-0.5cm}
	\includegraphics[scale=0.8,clip=true, trim = 0cm 0cm 0cm 0.6cm]{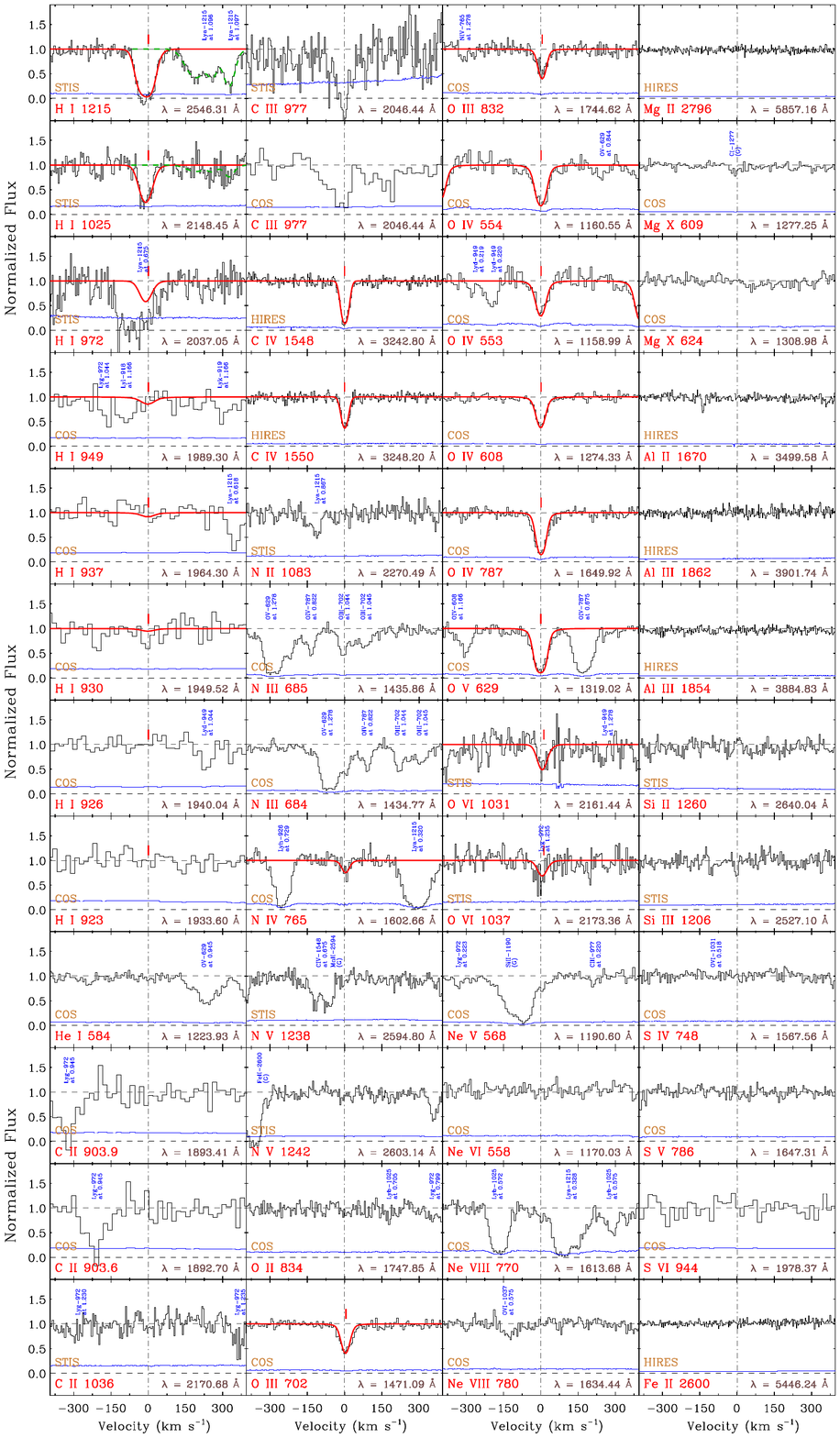}
	\vspace{-1.0cm}
	\caption{System plot of the z$_{abs}$=1.09457 absorber, with continuum-normalized flux along the Y-axis and the velocity scale relative to the redshift of the absorber along the X-axis.  The $1\sigma$ uncertainty in flux is indicated by the \textit{blue} curve at the bottom of each panel. The \textit{red} curves are the best-fit Voigt profiles. The observed wavelength of each transition is also indicated in the respective panels. The dashed green curves in the {\HI}~1215~{\AA}~and~{\HI}~1025~{\AA}~panels show a three component fit of the corresponding {\HI} lines at z $\sim 1.096$. The absorption seen at the location of {\CII}~$1334$ in the FOS data is Galactic {\MgII}~$2796$. The {\CII} ion is a non-detection as seen from the {\CII}~$1036$ panel. Similarly, the strong absorption in the {\NeV}~$568$ panel is Galactic {\SiII}~$1190$.} \label{fig:sysplot1094}
\end{figure*}


\newcolumntype{Y}{>{\centering\arraybackslash}X}

\begin{table}\centering
\setlength{\tabcolsep}{1.5pt}
\renewcommand{\arraystretch}{1.5}
\caption{Line measurements for the z$_{abs}$ $=$ 1.16592 absorber towards PG~$1522+101$ with the successive columns indicating the equivalent width in the rest-frame of the absorber, the column density measured using the AOD method, or the Voigt profile measured column density and Doppler $b$ parameter. The final column shows the velocity range over which the equivalent width and apparent column densities were integrated, or the velocity centroid for the profile-fitted absorption components. $^c$ - contaminated lines.} 
\label{1166table1}

\begin{tabularx}{\linewidth}{@{}p{4em}YYYY@{}}

\toprule
\multicolumn{1}{c}{Line} & \multicolumn{1}{c}{$\mathit{W_{r}}(m\mathit{A^{o}})$} &  \multicolumn{1}{c}{log[N ($\mathit{cm^{-2}}$)]} & \multicolumn{1}{c}{b (\kms)} & \multicolumn{1}{c}{v (\kms)}  \\ \midrule
H I 1215       & $>$ 541 & $>$ 14.3 &   & $[-100,65]$ \\
H I 1025       & $>$ 374 & $>$ 14.9 &   & $[-100,65]$ \\
H I 972        & $>$ 274 & $>$ 15.2 &   & $[-100,65]$ \\
H I 949        & $>$ 304 & $>$ 15.6 &   & $[-100,65]$ \\
H I 937        & $>$ 280 & $>$ 15.9 &   & $[-100,65]$ \\
H I 930        & $>$ 187 & $>$ 15.8 &   & $[-100,65]$ \\
H I 918        & 115 $\pm$ 24 & 16.34 $\pm$ 0.08 &   & $[-35,70]$ \\
H I 917        & $<$ 73 & $<$ 16.1 &   & $[-35,70]$ \\
H I 916.4      & $<$ 74 & $<$ 16.2 &   & $[-35,70]$ \\
H I 915.8      & $<$ 75 & $<$ 16.3 &   & $[-35,70]$ \\
H I &  & 16.52 $\pm$ 0.11 & 21 $\pm$ 1 & 5 $\pm$ 2 \\
       & & 14.69 $\pm$ 0.16 & 29 $\pm$ 3 & -27 $\pm$ 6 \\
       
He I 584    & 75 $\pm$ 2 & 14.15 $\pm$ 0.01 &   & $[-30,50]$ \\
He I 537    & 32 $\pm$ 4 & 14.36 $\pm$ 0.03 &   & $[-35,30]$ \\
He I &  & 14.48 $\pm$ 0.17 & 14 $\pm$ 3 & 6 $\pm$ 2 \\ 

C~II~1036$^c$& $<$ 37 & $<$ 13.6 &   & $[-30,50]$ \\
C III 977      & 114 $\pm$ 17 & 13.33 $\pm$ 0.50 &   & $[-30,50]$ \\
C III &  &  13.81 $\pm$ 0.70 & 14 $\pm$ 2.0 & 9 $\pm$ 11 \\

C~IV~1548      & 97 $\pm$ 5 & 13.53 $\pm$ 0.01 &   & $[-30,50]$ \\
C~IV~1550      & 49 $\pm$ 5 & 13.47 $\pm$ 0.01 &   & $[-30,50]$ \\
C IV &  &  13.45 $\pm$ 0.10 & 11 $\pm$ 2 & 1 $\pm$ 1 \\
     & & 12.93 $\pm$ 0.33 & 23 $\pm$ 14 & -24 $\pm$ 15 \\
     
O II 834       & $<$ 58 & $<$ 13.8 &   & $[-30,50]$ \\
O III 702      & 53 $\pm$ 4 & 14.06 $\pm$ 0.01 &   & $[-30,50]$ \\
O III 832      & 88 $\pm$ 22 & 14.28 $\pm$ 0.12 &   & $[-30,50]$ \\
O III &  & 14.18 $\pm$ 0.14 & 12 $\pm$ 2  & 6 $\pm$ 3  \\ 

O IV 554       & 100 $\pm$ 2 & 14.47 $\pm$ 0.01 &   & $[-30,50]$ \\
O IV 553       & 55 $\pm$ 3 & 14.36 $\pm$ 0.01 &   & $[-50,50]$ \\
O IV 608       & 56 $\pm$ 4 & 14.51 $\pm$ 0.01 &   & $[-50,50]$ \\
O IV 787       & 101 $\pm$ 6 & 14.38 $\pm$ 0.02 &   & $[-50,50]$ \\
O IV &  & 14.36 $\pm$ 0.12 & 10 $\pm$ 1  & 9 $\pm$ 2 \\ 
     & & 13.98 $\pm$ 0.12 & 22 $\pm$ 7 & -19 $\pm$ 6 \\

O V 629        & 106 $\pm$ 4 & 13.95 $\pm$ 0.01 &   & $[-85,45]$ \\
O V &  & 13.94 $\pm$ 0.17 & 10 $\pm$ 1  & 4 $\pm$ 3 \\ 
    & & 13.67 $\pm$ 0.17 & 22 $\pm$ 7 & -27 $\pm$ 7 \\
    
O~VI~1031      & 104 $\pm$ 13 & 14.03 $\pm$ 0.07 &   & $[-85,45]$ \\
O~VI~1037      & 54 $\pm$ 13 & 14.02 $\pm$ 0.09 &   & $[-85,45]$ \\
O VI &  & 13.61 $\pm$ 0.20 & 10 $\pm$ 2 & 4 $\pm$ 3 \\ 
     & & 13.78 $\pm$ 0.14 & 23 $\pm$ 7 & -27 $\pm$ 7 \\
\bottomrule
\end{tabularx}
\end{table}

\newcolumntype{Y}{>{\centering\arraybackslash}X}

\begin{table}\centering
\setlength{\tabcolsep}{1.5pt}
\renewcommand{\arraystretch}{1.5}
\caption{Table of line measurements for the z=1.16592 absorber towards PG1522+101, continued from Table \ref{1166table1}. $^c$ - contaminated lines.}
\label{1166table2}

\begin{tabularx}{\linewidth}{@{}p{4em}YYYY@{}}

\toprule
\multicolumn{1}{c}{Line} & \multicolumn{1}{c}{$\mathit{W_{r}}(m\mathit{A^{o}})$} &  \multicolumn{1}{c}{log[N ($\mathit{cm^{-2}}$)]} & \multicolumn{1}{c}{b (\kms)} & \multicolumn{1}{c}{v (\kms)}  \\ \midrule
Ne~V~568$^c$      & $<$ 16 & $<$ 13.7 &   & $[-30,50]$ \\
Ne~VI~558$^c$     & $<$ 21 & $<$ 14.0 &   & $[-30,50]$ \\
Ne~VIII~770    & $<$ 16 & $<$ 13.4 &   & $[-30,50]$ \\
Ne~VIII~780    & $<$ 18 & $<$ 13.8 &   & $[-30,50]$ \\
Mg~II~1240     & $<$ 21 & $<$ 16.0 &   & $[-30,50]$ \\
Mg X 609       & $<$ 13 & $<$ 13.6 &   & $[-30,50]$ \\
Al~II~1670     & $<$ 11 & $<$ 11.3 &   & $[-30,50]$ \\
Al~III~1862    & $<$ 13 & $<$ 12.2 &   & $[-30,50]$ \\
Al~III~1854    & $<$ 12 & $<$ 11.8 &   & $[-30,50]$ \\
Si II 1260     & $<$ 24 & $<$ 12.1 &   & $[-30,50]$ \\
Si II 1193     & $<$ 21 & $<$ 12.4 &   & $[-30,50]$ \\
Si~II~1190$^c$     & $<$ 31 & $<$ 13.0 &   & $[-30,50]$ \\
Si~III~1206    & 75 $\pm$ 7 & 12.66 $\pm$ 0.02 &   & $[-30,50]$ \\
Si III &  & 12.80 $\pm$ 0.40 & 11 & 8 $\pm$ 7 \\ 

Fe~II~2600     & $<$ 19 & $<$ 12.1 &   & $[-30,50]$ \\
Fe~II~2382     & $<$ 16 & $<$ 12.0 &   & $[-30,50]$ \\
Fe~II~2344     & $<$ 17 & $<$ 12.4 &   & $[-30,50]$ \\

\bottomrule
\end{tabularx}
\end{table}

\begin{figure*}
	\centering
	\includegraphics[scale=0.8, clip=true, trim= 0.5cm 0cm 0cm 1.5cm]{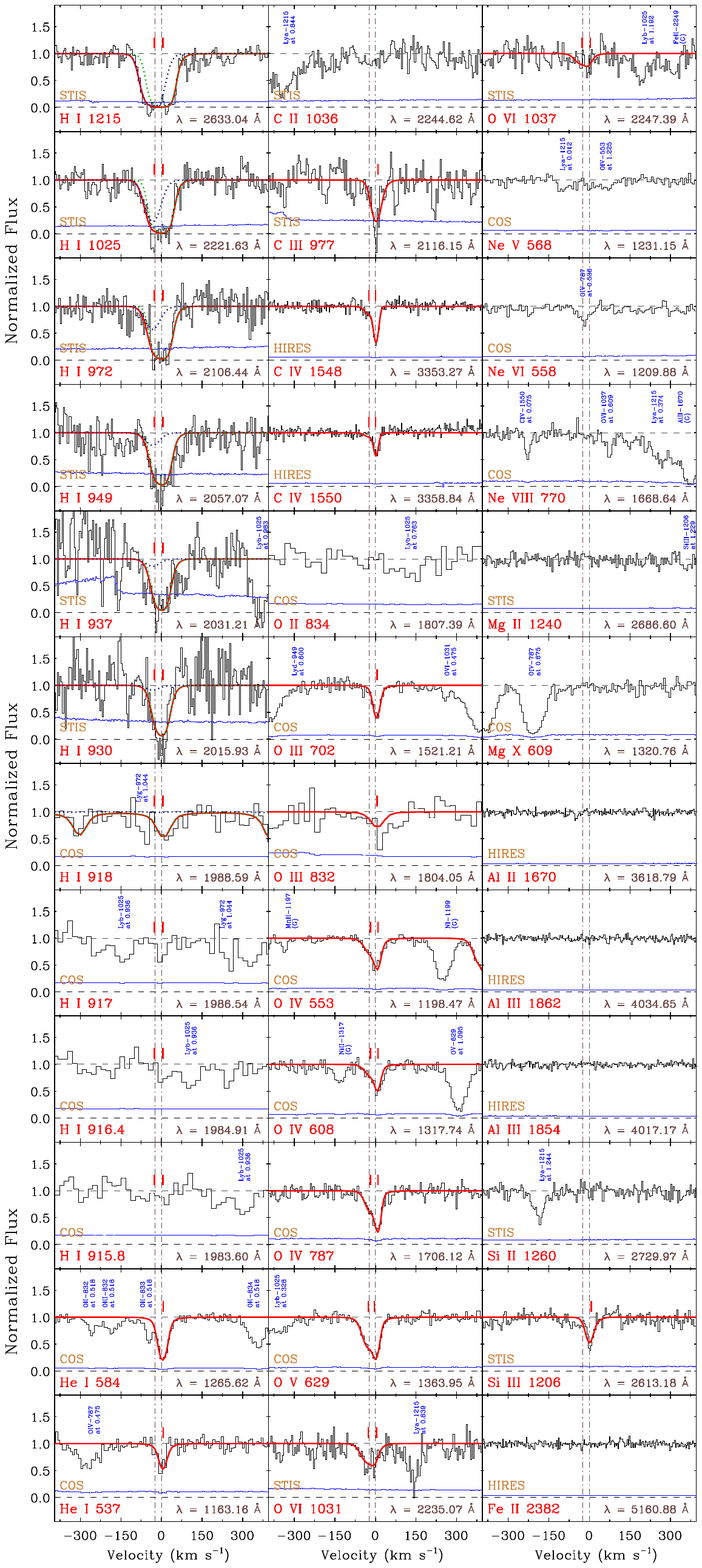}
	\vspace{-1cm}
	\caption{System plot of the z$_{abs}$=1.16592 absorber, with continuum-normalized flux along the Y-axis and the velocity scale relative to the redshift of the absorber along the X-axis. The $v = 0$~{\kms}, marked by the \textit{dashed-dotted} vertical line, indicates the absorber redshift. The $1\sigma$ uncertainty in flux is indicated by the \textit{blue} curve at the bottom of each panel. The \textit{red} curves are the best-fit Voigt profiles. The observed wavelength of each transition is also indicated in the respective panels.\label{fig:sysplot1166}}
\end{figure*}


\newcolumntype{Y}{>{\centering\arraybackslash}X}

\begin{table}\centering
\setlength{\tabcolsep}{1.5pt}
\renewcommand{\arraystretch}{1.5}
\caption{Line measurements for the z$_{abs}$ $=$ 1.27768 absorber towards PG1522+101 with the successive columns indicating the corresponding equivalent width in the rest-frame of the absorber, the column density measured through the AOD method or Voigt profile fitting and the Doppler parameters obtained through profile fitting. The final column shows the velocity range over which the equivalent width and apparent column densities were integrated, or the centroid for the profile-fitted absorption components. Continued in Table \ref{1277table2}. $^c$ - contaminated lines.}
\label{1277table1}

\begin{tabularx}{\linewidth}{@{}p{4em}YYYY@{}}

\toprule
\multicolumn{1}{c}{Line} & \multicolumn{1}{c}{$\mathit{W_{r}}(m\mathit{A^{o}})$} &  \multicolumn{1}{c}{log[N ($\mathit{cm^{-2}}$)]} & \multicolumn{1}{c}{b (\kms)} & \multicolumn{1}{c}{v (\kms)}  \\ \midrule
H I 1215    & $>$ 502 & $>$ 14 &   & $[-45,115]$ \\
H I 1025    & 178 $\pm$ 12 & 14.49 $\pm$ 0.04 &   & $[-45,115]$ \\
H I 972$^c$& $<$ 284  & $<$ 15.3  &   & $[-45,115]$ \\
H I 949     & 60 $\pm$ 16 & 14.84 $\pm$ 0.14 &   & $[-45,115]$ \\
H I 937     & $<$ 64 & $<$ 15 &   & $[-45,115]$ \\
H I 930$^c$& $<$ 90 & $<$ 15.5 &   & $[-45,115]$ \\
H I 926$^c$& $<$ 136 & $<$ 16.0 &   & $[-45,115]$ \\
H I 923$^c$& $<$ 331 & $<$ 16.4 &   & $[-45,115]$ \\
H I & & 14.31 $\pm$ 0.09 & 25 $\pm$ 5 & -7 $\pm$ 6 \\
    & & 14.09 $\pm$ 0.11 & 24 $\pm$ 6 & 56 $\pm$ 7 \\

He I 584    & $<$ 10 & $<$ 13.0 &   & $[-30,60]$ \\
He I 537    & $<$ 9 & $<$ 13.6 &   & $[-30,60]$ \\
C II 903.9  & $<$ 45 & $<$ 13.2 &   & $[-30,60]$ \\
C II 903.6  & $<$ 46 & $<$ 13.5 &   & $[-30,60]$ \\
C II 1334   & $<$ 25 & $<$ 13.0 &   & $[-30,60]$ \\
C II 1036   & $<$ 27 & $<$ 13.3 &   & $[-30,60]$ \\
C III 977   & 74 $\pm$ 12 & 13.17 $\pm$ 0.09 &   & $[-30,60]$ \\
C III & & 13.10 $\pm$ 0.16 & 10 $\pm$ 1 & -8 $\pm$ 5 \\ 
      & & 12.84 $\pm$ 0.21 & 10 $\pm$ 2 & 40 $\pm$ 8 \\
      
C~IV~1548   & 173 $\pm$ 4 & 13.82 $\pm$ 0.01 &   & $[-30,60]$ \\
C~IV~1550   & 107 $\pm$ 4 & 13.82 $\pm$ 0.01 &   & $[-30,60]$ \\
C IV  & & 13.61 $\pm$ 0.02 & 10 $\pm$ 1 & 0 $\pm$ 1 \\ 
      & & 13.42 $\pm$ 0.02 & 10 $\pm$ 2 & 39 $\pm$ 1 \\

N II 1083   & $<$ 29 & $<$ 13.4 &   & $[-30,60]$ \\
N II 915    & $<$ 50 & $<$ 13.6 &   & $[-30,60]$ \\
N III 989   & $<$ 33 & $<$ 13.5 &   & $[-30,60]$ \\
N III 685   & $<$ 12 & $<$ 13.0 &   & $[-30,60]$ \\
N III 684   & $<$ 12 & $<$ 13.3 &   & $[-30,60]$ \\
N IV 765    & 37 $\pm$ 5 & 13.11 $\pm$ 0.03 &   & $[-30,60]$ \\
N IV  & & 13.01 $\pm$ 0.10 & 9 $\pm$ 1 & -2 $\pm$ 2 \\ 
      & & 12.49 $\pm$ 0.23 & 9 $\pm$ 2 & 38 $\pm$ 7 \\
      
N V 1238    & $<$ 13 & $<$ 12.7 &   & $[-30,60]$ \\
N V 1242    & $<$ 14 & $<$ 13.1 &   & $[-30,60]$ \\
O III 702   & 29 $\pm$ 4.47 & 13.74 $\pm$ 0.03 &   & $[-30,60]$ \\
O III 832   & $<$ 54 & $<$ 13.9 &   & $[-30,60]$ \\
O III & & 13.66 $\pm$ 0.11 & 9 $\pm$ 1 & -5 $\pm$ 4 \\ 
      & & 13.25 $\pm$ 0.22 & 9 $\pm$ 2 & 32 $\pm$ 7 \\
\bottomrule
\end{tabularx}
\end{table}

\newcolumntype{Y}{>{\centering\arraybackslash}X}

\begin{table}\centering
\setlength{\tabcolsep}{1.5pt}
\renewcommand{\arraystretch}{1.5}
\caption{Table of line measurements for the z=1.27768 absorber towards PG1522+101, continued from Table \ref{1277table1}.$^c$ - contaminated lines.}
\label{1277table2}

\begin{tabularx}{\linewidth}{@{}p{4em}YYYY@{}}

\toprule
\multicolumn{1}{c}{Line} & \multicolumn{1}{c}{$\mathit{W_{r}}(m\mathit{A^{o}})$} &  \multicolumn{1}{c}{log[N ($\mathit{cm^{-2}}$)]} & \multicolumn{1}{c}{b (\kms)} & \multicolumn{1}{c}{v (\kms)}  \\ \midrule
O IV 554    & 80 $\pm$ 2 & 14.28 $\pm$ 0.00 &   & $[-30,60]$ \\
O IV 608    & 55 $\pm$ 3 & 14.49 $\pm$ 0.01 &   & $[-30,60]$ \\
O IV 787    & 119 $\pm$ 10 & 14.48 $\pm$ 0.10 &   & $[-30,60]$ \\
O IV  & & 14.44 $\pm$ 0.07 & 9 $\pm$ 1 & 1 $\pm$ 1 \\ 
      & & 14.23 $\pm$ 0.07 & 8 $\pm$ 2 & 39 $\pm$ 1 \\

O V 629     & 144 $\pm$ 2 & 14.19 $\pm$ 0.01 &   & $[-30,60]$ \\
O V   & & 14.69 $\pm$ 0.30 & 12 $\pm$ 2 & 0 $\pm$ 1 \\ 
      & & 14.03 $\pm$ 0.19 & 10 $\pm$ 2 & 40 $\pm$ 1 \\

O~VI~1031   & $>$ 220 & $>$ 14.57 &   & $[-20,75]$ \\
O~VI~1037   & $>$ 281 & $>$ 15.11 &   & $[-20,75]$ \\
O~VI        & & 15.55 $\pm$ 0.67 & 10 & -1 $\pm$ 4 \\ 
            & & 14.53 $\pm$ 0.89 & 9 $\pm$ 3 & 39 $\pm$ 5 \\
      

Ne V 568    & 33 $\pm$ 3 & 14.08 $\pm$ 0.01 &   & $[-20,75]$ \\
Ne V  & & 14.02 $\pm$ 0.11 & 8 $\pm$ 1 & 3 $\pm$ 3 \\ 
      & & 13.64 $\pm$ 0.16 & 8 $\pm$ 2 & 38 $\pm$ 5 \\

Ne~VI~558   & 40 $\pm$ 2 & 14.26 $\pm$ 0.01 &   & $[-20,75]$ \\
Ne VI & & 14.13 $\pm$ 0.09 & 8 $\pm$ 1 & 9 $\pm$ 3 \\ 
      & & 13.86 $\pm$ 0.12 & 8 $\pm$ 2 & 39 $\pm$ 4 \\ 

Ne~VIII~770 & $<$ 17 & $<$ 13.5 &   & $[-30,60]$ \\
Ne~VIII~780 & $<$ 26 & $<$ 13.9 &   & $[-30,60]$ \\
Mg X 609    & $<$ 10 & $<$ 13.5 &   & $[-30,60]$ \\
Mg X 624    & $<$ 9 & $<$ 13.8 &   & $[-30,60]$ \\
Al~II~1670  & $<$ 12 & $<$ 11.4 &   & $[-30,60]$ \\
Al~III~1862 & $<$ 14 & $<$ 12.2 &   & $[-30,60]$ \\
Al~III~1854 & $<$ 13 & $<$ 11.9 &   & $[-30,60]$ \\
Si~II~1260  & $<$ 18 & $<$ 12.0 &   & $[-30,60]$ \\
Si~II~1020  & $<$ 28 & $<$ 14.2 &   & $[-30,60]$ \\
Si~III~1206 & $<$ 23 & $<$ 12.0 &   & $[-30,60]$ \\
Si~IV~1393  & $<$ 22 & $<$ 12.4 &   & $[-30,60]$ \\
Si~IV~1402  & $<$ 19 & $<$ 12.6 &   & $[-30,60]$ \\
S IV 1062   & $<$ 26 & $<$ 13.8 &   & $[-30,60]$ \\
S IV 748    & $<$ 16 & $<$ 12.8 &   & $[-30,60]$ \\
S V 786     & $<$ 32 & $<$ 12.6 &   & $[-30,60]$ \\
S VI 944    & $<$ 39 & $<$ 13.3 &   & $[-20,75]$ \\
S VI 933    & 63 $\pm$ 13 & 13.33 $\pm$ 0.11 &   & $[-20,75]$ \\
S VI  & & 13.13 $\pm$ 0.10 & 6 $\pm$ 3 & -14 $\pm$ 10 \\ 
      & & 13.05 $\pm$ 0.11 & 6 $\pm$ 3 & 18 $\pm$ 12 \\ 

Fe~II~2382  & $<$ 17 & $<$ 12.0 &   & $[-30,60]$ \\
Fe~II~2344  & $<$ 16 & $<$ 12.4 &   & $[-30,60]$ \\

\bottomrule
\end{tabularx}
\end{table}

\begin{figure*}
	\centering
	\includegraphics[scale=0.8, clip=true, trim= 0cm 0cm 0cm 1cm]{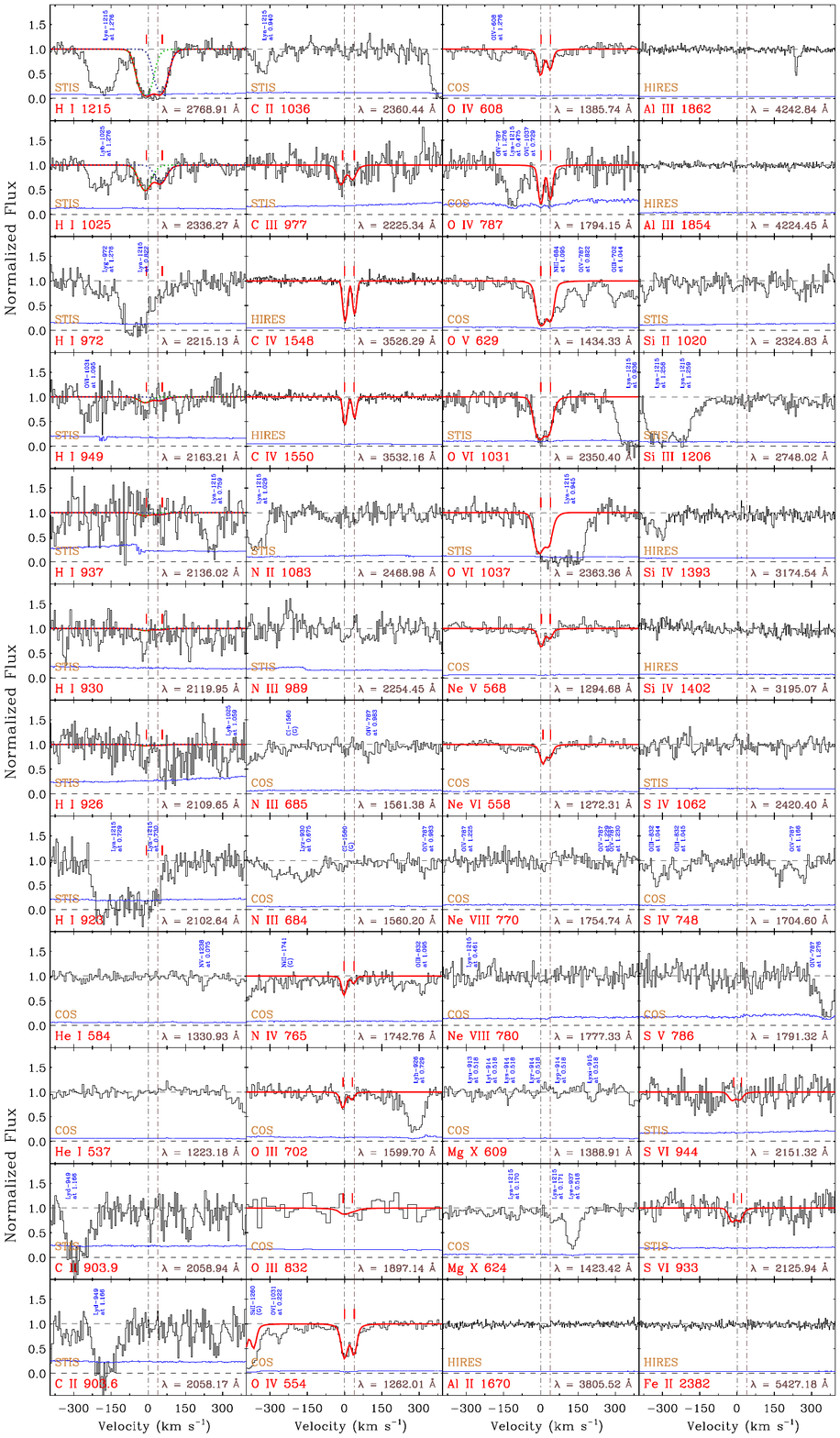}
	\vspace{-1cm}
	\caption{System plot of the z$_{abs}$=1.27768 absorber, with continuum-normalized flux along the Y-axis and the velocity scale relative to the redshift of the absorber along the X-axis. The $v = 0$~{\kms}, marked by the \textit{dashed-dotted} vertical line, indicates the absorber redshift. The $1\sigma$ uncertainty in flux is indicated by the \textit{blue} curve at the bottom of each panel. The \textit{red} curves are the best-fit Voigt profiles. The observed wavelength of each transition is also indicated in the respective panels.\label{fig:sysplot1277}}
\end{figure*}

\bsp	
\label{lastpage}
\end{document}